\documentclass[iop,numberedappendix]{emulateapj}

\usepackage{graphicx,amsmath,natbib}
\usepackage{here}
\usepackage{epsf}
\usepackage{epstopdf}
\usepackage{xspace}
\usepackage{hyperref}

\usepackage[normalem]{ulem} 
\usepackage{cancel} 

\usepackage{color}

\newcommand{\um}{\ensuremath{\mu\rm{m}}\xspace}

\newcommand{\sigfir}{\ensuremath{\Sigma_{\rm{FIR}}}\xspace}

\newcommand{\lfir}{\ensuremath{L_{\rm{FIR}}}\xspace}

\newcommand{\Tdust}{\ensuremath{T_{\rm{dust}}}\xspace}
\newcommand{\Mdust}{\ensuremath{M_{\rm{dust}}}\xspace}
\newcommand{\Lz}{\ensuremath{\lambda_0}\xspace}
\newcommand{\msol}{\ensuremath{\rm{M}_\odot}\xspace}
\newcommand{\lsol}{\ensuremath{\rm{L}_\odot}\xspace}

\newcommand{\sopf}{\ensuremath{S_\mathrm{1.4mm}}\xspace}
\newcommand{\sopo}{\ensuremath{S_\mathrm{1.1mm}}\xspace}
\newcommand{\stwo}{\ensuremath{S_\mathrm{2mm}}\xspace}
\newcommand{\ses}{\ensuremath{S_\mathrm{870\um}}\xspace}
\newcommand{\siges}{\ensuremath{\sigma_\mathrm{870\um}}\xspace}
\newcommand{\mues}{\ensuremath{\mu_\mathrm{870\um}}\xspace}

\newcommand{\reff}{\ensuremath{r_\mathrm{eff}}\xspace}
\newcommand{\etal}{et~al.\xspace}
\newcommand{\cii}{[C{\scriptsize II}]\xspace}
\newcommand{\arc}{\ensuremath{''}\xspace}

\def\Arizona{1}
\def\Diego{2}
\def\ESOGarching{3}
\def\Cambridge{4}
\def\KICPChicago{5}
\def\AAUChicago{6}
\def\Dal{7}
\def\Davis{8}
\def\UFlorida{9}
\def\UCL{10}
\def\Stanford{11}
\def\UCLA{12}
\def\MPIfR{13}
\def\Illinois{14}
\def\Oxford{15}

\begin{document}

\title{ALMA Imaging and Gravitational Lens Models of South Pole Telescope-- \\
	   Selected Dusty, Star-Forming Galaxies at High Redshifts}


\shortauthors{J. S. Spilker, et al.}

\author{
J.~S.~Spilker$^{\Arizona}$,
D.~P.~Marrone$^{\Arizona}$,   
M.~Aravena$^{\Diego}$,
M.~B\'ethermin$^{\ESOGarching}$,
M.~S.~Bothwell$^{\Cambridge}$,
J.~E.~Carlstrom$^{\KICPChicago,\AAUChicago}$, 
S.~C.~Chapman$^{\Dal}$,
T.~M.~Crawford$^{\KICPChicago,\AAUChicago}$, 
C.~de~Breuck$^{\ESOGarching}$,
C.~D.~Fassnacht$^{\Davis}$,
A.~H.~Gonzalez$^{\UFlorida}$, 
T.~R.~Greve$^{\UCL}$,	
Y.~Hezaveh$^{\Stanford}$,
K.~Litke$^{\Arizona}$,
J.~Ma$^{\UFlorida}$, 
M.~Malkan$^{\UCLA}$,
K.~M.~Rotermund$^{\Dal}$,
M.~Strandet$^{\MPIfR}$, 
J.~D.~Vieira$^{\Illinois}$,
A.~Wei\ss$^{\MPIfR}$,
N.~Welikala$^{\Oxford}$,
}

\altaffiltext{\Arizona}{Steward Observatory, University of Arizona, 933 North Cherry Avenue, Tucson, AZ 85721, USA; \href{mailto:jspilker@as.arizona.edu}{jspilker@as.arizona.edu}}
\altaffiltext{\Diego}{N\'ucleo de Astronom\'{\i}a, Facultad de Ingenier\'{\i}a, Universidad Diego Portales, Av. Ej\'ercito 441, Santiago, Chile}
\altaffiltext{\ESOGarching}{European Southern Observatory, Karl Schwarzschild Stra\ss e 2, 85748 Garching, Germany}
\altaffiltext{\Cambridge}{Cavendish Laboratory, University of Cambridge, JJ Thompson Ave, Cambridge CB3 0HA, UK}
\altaffiltext{\KICPChicago}{Kavli Institute for Cosmological Physics, University of Chicago, 5640 South Ellis Avenue, Chicago, IL 60637, USA}
\altaffiltext{\AAUChicago}{Department of Astronomy and Astrophysics, University of Chicago, 5640 South Ellis Avenue, Chicago, IL 60637, USA}
\altaffiltext{\Dal}{Dalhousie University, Halifax, Nova Scotia, Canada}
\altaffiltext{\Davis}{Department of Physics,  University of California, One Shields Avenue, Davis, CA 95616, USA}
\altaffiltext{\UFlorida}{Department of Astronomy, University of Florida, Gainesville, FL 32611, USA}
\altaffiltext{\UCL}{Department of Physics and Astronomy, University College London, Gower Street, London WC1E 6BT, UK}
\altaffiltext{\Stanford}{Kavli Institute for Particle Astrophysics and Cosmology, Stanford University, Stanford, CA 94305, USA}
\altaffiltext{\UCLA}{Department of Physics and Astronomy, University of California, Los Angeles, CA 90095-1547, USA}
\altaffiltext{\MPIfR}{Max-Planck-Institut f\"{u}r Radioastronomie, Auf dem H\"{u}gel 69 D-53121 Bonn, Germany}
\altaffiltext{\Illinois}{Department of Astronomy and Department of Physics, University of Illinois, 1002 West Green St., Urbana, IL 61801}
\altaffiltext{\Oxford}{Department of Physics, Oxford University, Denis Wilkinson Building, Keble Road, Oxford, OX1 3RH, UK}

\begin{abstract}

The South Pole Telescope has discovered one hundred gravitationally lensed, 
high-redshift, dusty, star-forming galaxies (DSFGs). We present 0.5\arc resolution 
870\,\um Atacama Large Millimeter/submillimeter Array imaging of a sample of 47 
DSFGs spanning $z=1.9-5.7$, and construct gravitational lens models of these 
sources.  Our visibility-based lens modeling incorporates
several sources of residual interferometric calibration uncertainty, 
allowing us to properly account for noise in the observations.
At least 70\% of the sources are strongly lensed by foreground galaxies 
($\mues > 2$), with a median magnification $\mues = 6.3$, extending
to $\mues > 30$.  We compare the intrinsic size distribution of the strongly lensed
sources to a similar number of unlensed DSFGs and find no significant differences
in spite of a bias between the magnification and intrinsic source size.
This may indicate that the true size distribution of DSFGs is relatively narrow.
We use the source sizes to constrain the wavelength at
which the dust optical depth is unity and find 
this wavelength to be correlated
with the dust temperature.  This correlation leads to discrepancies in dust mass
estimates of a factor of 2 compared to estimates using a single value for this
wavelength.  We investigate the relationship between the \cii line
and the far-infrared luminosity and find that the same correlation between
the \cii/\lfir ratio and \sigfir found for low-redshift star-forming galaxies applies
to high-redshift galaxies and extends at least two orders of magnitude higher in
\sigfir.  This lends further credence to the claim that the compactness
of the IR-emitting region is the controlling parameter in establishing the
``\cii deficit.''

\end{abstract}

\keywords{galaxies: high-redshift --- galaxies: ISM --- 
galaxies: star formation}

\section{Introduction} \label{intro}

With the Atacama Large Millimeter/submillimeter Array (ALMA) now in full operation,
our understanding of dust-enshrouded star formation at high redshifts is
advancing more rapidly than ever before.  The most intense star formation in
the universe takes place in dusty, star-forming galaxies (DSFGs), 
at high redshifts ($z>1$), creating new
stars at rates of $>100-1000$\,\msol\,yr$^{-1}$ (see a recent review by
\citealt{casey14}).  The otherwise high UV luminosity
from massive young stars in these galaxies is almost entirely reprocessed 
by interstellar dust, which 
absorbs the short-wavelength radiation and re-radiates it at far-infrared (FIR)
and (sub)millimeter wavelengths.  Although DSFGs represent a significant contribution
to the comoving star formation rate density out to at least $z=4$ 
\citep[e.g.,][]{chapman05,casey13}, producing
a realistic population of DSFGs has long been a challenge for theoretical models
of galaxy evolution \citep[e.g.,][]{baugh05,dave10,hayward13,narayanan15}.

Observations of these galaxies benefit from a strongly negative
``K-correction'' at submillimeter wavelengths \citep[e.g.,][]{blain93}, in which the 
dimming due to increased cosmological distance is countered by the rapidly rising
dust spectral energy distribution (SED) at fixed observing wavelength. 
DSFGs were initially discovered in low-resolution ($>10$\arc) 850\,\um deep images
\citep{smail97,barger98,hughes98}, and high-resolution follow-up studies at
submillimeter wavelengths remain challenging, as are observations at other 
wavelengths that do not benefit from the negative K-correction.
One fairly straightforward method of gaining resolution is to target a sample
of gravitationally lensed galaxies, such as those discovered by the South Pole
Telescope \citep{carlstrom11,vieira10,mocanu13} or the \textit{Herschel} Space
Observatory \citep{negrello10,wardlow13}.  Follow-up observations of these galaxies at 
FIR/submillimeter wavelengths, where they are brightest, with interferometers such
as ALMA and the Submillimeter Array have
shown that the bulk of the brightest objects are consistent with strong
gravitational lensing \citep[e.g.,][]{hezaveh13,vieira13,bussmann13}.
Lensed samples offer the opportunity to study
DSFGs at higher resolution and using fainter observational diagnostics than
otherwise possible \citep[e.g.,][]{swinbank10,fu12,bothwell13b,spilker14}.

Taking advantage of gravitational lensing requires careful modeling to understand
its effects. In this paper, we present lens models of a sample of 47 DSFGs
discovered in South Pole Telescope data and observed by ALMA at $\sim0.5$''
resolution.  \citet{hezaveh13} presented models of four sources which were spatially
resolved at the $\sim1.5$\arc resolution of the first data acquired for 
this project; here we
expand this work to include the completed dataset, including all sources and
array configurations.  As in \citet{hezaveh13}, our models are performed
in the Fourier plane native to the interferometer, and marginalized over several
common calibration uncertainties.  The resulting intrinsic source properties
span a large range in luminosity, and we use these derived properties to
explore the intrinsic size distribution of DSFGs, their dust SEDs, and
the relation between the \cii fine structure line and the FIR luminosity.

In Section~\ref{obs}, we describe the selection criteria and ALMA observations.
Section~\ref{lensmodels} describes our gravitational lens modeling technique, 
with the results of these models detailed in Section~\ref{results}.
In Section~\ref{discussion} we use these models to address selected topics
of interest, including the intrinsic size distribution of DSFGs and the
relationship between the \cii fine structure line and the FIR luminosity. We 
conclude in Section~\ref{conclusions}. Throughout this work, we assume
a flat WMAP9 $\Lambda$CDM cosmology, $h=0.693$, $\Omega_m = 0.286$, and
$\Omega_\Lambda = 0.713$ \citep{hinshaw13}.  We define the far-infrared 
luminosity \lfir to be integrated over rest-frame $42.5-122.5$\,\um
\citep{helou88}.

\section{Sample Selection and Observations} \label{obs}

The selection criteria used to generate the SPT DSFG sample are described in detail
by \citet{weiss13}. Briefly, sources were selected to have dust-like spectral
indices between 1.4 and 2\,mm (i.e., \sopf/\stwo$>1.7$; \citealt{mocanu13}). 
Further selection criteria remove synchrotron-dominated
and low-redshift ($z<0.1$) contaminant sources. 
Redshifts for some of the SPT DSFGs are presented in \citet{strandet16}.  
Optical and near-infrared spectroscopic redshifts
of the foreground lenses, where available, will be presented in Rotermund \etal, 
in prep.  Finally, we make use of optical and infrared 
imaging data obtained from
a variety of facilities, including the \textit{Hubble} Space Telescope,
Very Large Telescope, Magellan-Baade telescope, and \textit{Spitzer}/IRAC.

To refine the coarse SPT positions, each source was observed at higher spatial
resolution to improve the positional accuracy, typically at 870\,\um using 
the Large Apex BOlometer CAmera (LABOCA)
or at 1.3\,mm using the Submillimeter Array (SMA). From this catalog, we selected 
47 bright sources which could be placed into four
groups of targets that lie within 15$^\circ$ of each other on the sky in order
to share calibrator sources.  The targets are listed
in Table~\ref{tab:targets}.  In Figure~\ref{fig:selection} we compare the objects
in the subsample observed by ALMA with all SPT sources and with the 
\textit{Herschel}-selected objects observed by \citet{bussmann13,bussmann15}.

These 47 SPT sources were observed by ALMA at 870\,\um as part of 
Cycle 0 program 2011.0.00958.S (PI D. Marrone). The ALMA observations were 
carried out in eight sessions from November 2011 to 
August 2012 and are summarized in Table~\ref{tab:obstable}. Given the limited 
number of antennas available at the beginning
of Cycle 0 (minimum 14), each group of sources was observed with two different array
configurations, corresponding to approximately 0.5 and 1.5\arc resolution,
to provide better sampling of the $uv$ plane. Over the series of observations
the number of antennas increased (up to 25), providing greater sensitivity in
later observations.  Additional sources with precisely known positions from
the International Celestial Reference Frame (ICRF; \citealt{ma98}) were observed 
to verify the astrometric and antenna baseline solutions.  Each source was 
observed for 60--90\,s per array configuration.
The total observing time for all calibrators and science targets was 9.4 hours.

Four basebands, each processing 2\,GHz of telescope bandwidth, were centered near
336.8, 338.8, 348.8, and 350.8\,GHz.  The correlator was configured to provide
128 channels of 15.6\,MHz width for each baseband.
Bandpass calibration was performed by observing a bright quasar at the beginning
of each track. Time-dependent amplitude and phase variations were calibrated
using several quasars near (typically within $<5^\circ$ of) the science targets.
The flux scale was determined at the beginning of each track using an available 
solar system object or quasar with
a recently determined flux density, as detailed in Table~\ref{tab:obstable}. This flux
scale is estimated to be correct to within 10\%, although we allow an amplitude
re-scaling between the two observations of each group of sources in our modeling
(see Section~\ref{lensmodels}).  We estimate the noise
on each visibility measurement by calculating the scatter after differencing 
successive visibilities on the same baseline, baseband, and polarization.  
After calibration, the data from each track 
were combined and imaged using Briggs weighting (robust parameter $= -0.5$).
This weighting represents a compromise which somewhat favors higher resolution 
at the expense of sensitivity.

In four objects (SPT0125-47, SPT0125-50, SPT2103-60, SPT2354-58), we 
serendipitously detected a spectral feature in the ALMA data. As we consider
only models of the continuum emission in this work, for these sources, we exclude
the spectral window containing the spectral line.  

Another four objects (SPT0550-53, SPT0551-50, SPT2351-57, SPT2353-50)
appear to be lensed by galaxy groups or clusters. 
\textit{HST} imaging shows numerous galaxies in the vicinity of the 870\,\um emission.
Images of these sources are
shown in Appendix~\ref{app:clusters}. The ALMA measurements show only single images,
and the ALMA field of view does not encompass the expected locations of counterimages.
For these sources,
the lensing geometry cannot be constrained by the ALMA data alone.  Beyond counting
them among the sources identified as lensed, we ignore
them for the remainder of this paper.

Images of the sources we model in this paper, overlaid on the best-available
near-IR or optical imaging, are shown in Fig.~\ref{fig:images}.

\begin{figure}[htb]%
\includegraphics[width=\columnwidth]{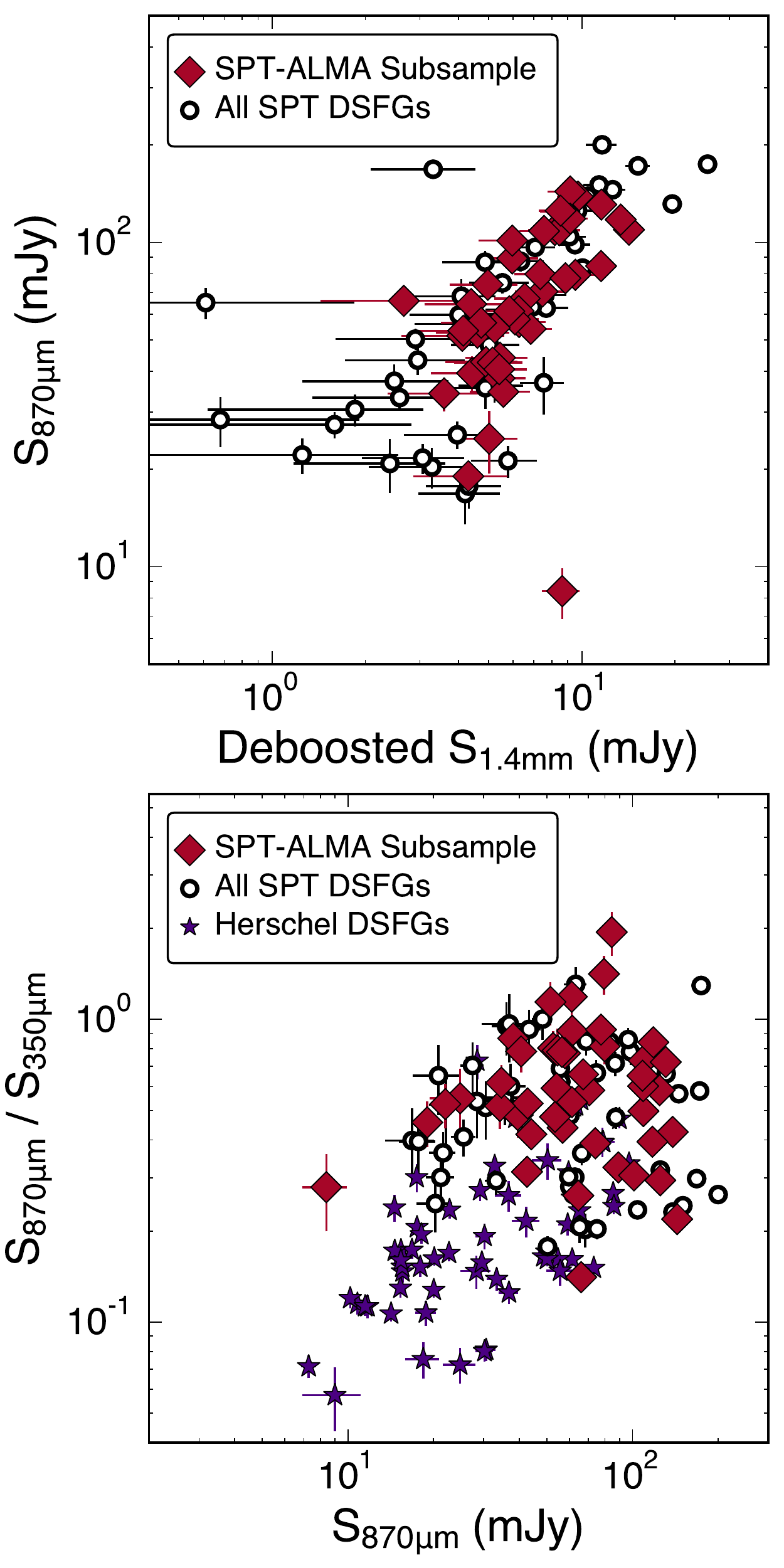}%
\caption{
Comparison of the subsample of SPT sources observed by ALMA to all SPT sources
and the \textit{Herschel}-selected samples of \citet{bussmann13,bussmann15}.
Note that \ses shown in this figure is derived from single-dish LABOCA measurements
for the SPT sources.  Single-dish photometry is not available for the
\textit{Herschel} sources, so these points are derived from interferometric
(SMA or ALMA) observations only and may underestimate the true total
flux density; see Section~\ref{fluxcomp}.
\textit{Top:} The subsample of SPT sources observed by ALMA was selected to have
high \sopf, and spans most of the range of \ses seen in the full sample.
\textit{Bottom:} Flux density -- FIR color diagram for SPT- and \textit{Herschel}-
selected DSFGs \citep{bussmann13,bussmann15}. The SPT sources are redder on average,
and at higher redshift \citep[e.g.,][]{weiss13,bethermin15b},
largely due to their longer selection wavelength.
}%
\label{fig:selection}%
\end{figure}

\begin{deluxetable*}{llllccrr} 
\tablecaption{Observed Target Summary\label{tab:targets}} 
\startdata 
\tableline 
\\ 
Short Name & IAU Name & RA$_\mathrm{ALMA}$ & Dec$_\mathrm{ALMA}$ & $S_\mathrm{LABOCA}$ & $S_\mathrm{ALMA}^\mathrm{a}$ & $z_L$ & $z_S$ \\ 
           &          &                    &                    &  mJy              &  mJy                        &    &       \\ 
\tableline 
\vspace{-0.07in} 
\\
SPT0020-51 & SPT-S J002023-5146.3 & 00:20:23.45 & -51:46:34.80 & 70 $\pm$ 8 & 77 $\pm$ 8 & 0.693 & ... \\
SPT0027-50 & SPT-S J002706-5007.3 & 00:27:06.84 & -50:07:19.00 & 138 $\pm$ 16 & 126 $\pm$ 13 & ... & ... \\
SPT0103-45 & SPT-S J010312-4538.9 & 01:03:11.57 & -45:38:51.90 & 124 $\pm$ 14 & 105 $\pm$ 11 & 0.740 & 3.0917 \\
SPT0109-47 & SPT-S J010949-4702.1 & 01:09:49.91 & -47:02:09.50 & 109 $\pm$ 14 & 82 $\pm$ 9 & 0.669 & ... \\
SPT0113-46 & SPT-S J011308-4617.7 & 01:13:09.03 & -46:17:56.90 & 79 $\pm$ 11 & 54 $\pm$ 6 & ... & 4.2328 \\
SPT0125-47 & SPT-S J012507-4723.8 & 01:25:07.18 & -47:23:55.50 & 144 $\pm$ 17 & 144 $\pm$ 15 & 0.305 & 2.5148 \\
SPT0125-50 & SPT-S J012549-5038.3 & 01:25:48.41 & -50:38:17.40 & 109 $\pm$ 14 & 81 $\pm$ 9 & 0.510 & 3.9593 \\
SPT0128-51 & SPT-S J012809-5129.7 & 01:28:09.87 & -51:29:43.80 & 19 $\pm$ 3 & 17 $\pm$ 4 & ... & ... \\
SPT0202-61 & SPT-S J020258-6121.2 & 02:02:58.86 & -61:21:13.20 & 109 $\pm$ 13 & 81 $\pm$ 9 & ... & ... \\
SPT0243-49 & SPT-S J024308-4915.6 & 02:43:09.07 & -49:15:33.00 & 84 $\pm$ 10 & 57 $\pm$ 7 & ... & 5.6991 \\
SPT0245-63 & SPT-S J024544-6320.7 & 02:45:44.23 & -63:20:44.30 & 61 $\pm$ 8 & 48 $\pm$ 6 & ... & ... \\
SPT0300-46 & SPT-S J030004-4621.4 & 03:00:04.21 & -46:21:25.30 & 57 $\pm$ 8 & 58 $\pm$ 7 & ... & 3.5954 \\
SPT0319-47 & SPT-S J031931-4724.6 & 03:19:32.37 & -47:24:33.20 & 67 $\pm$ 9 & 57 $\pm$ 7 & ... & 4.5164 \\
SPT0345-47 & SPT-S J034510-4725.7 & 03:45:10.97 & -47:25:40.90 & 89 $\pm$ 11 & 92 $\pm$ 10 & 0.364 & 4.2958 \\
SPT0346-52 & SPT-S J034640-5205.0 & 03:46:41.19 & -52:05:05.50 & 131 $\pm$ 15 & 123 $\pm$ 13 & ... & 5.6559 \\
SPT0348-62 & SPT-S J034841-6220.9 & 03:48:41.55 & -62:20:55.80 & 52 $\pm$ 7 & 40 $\pm$ 5 & 0.378 & ... \\
SPT0403-58 & SPT-S J040331-5850.1 & 04:03:32.28 & -58:50:06.70 & 40 $\pm$ 6 & 50 $\pm$ 6 & ... & ... \\
SPT0404-59 & SPT-S J040446-5949.2 & 04:04:45.82 & -59:49:09.90 & 25 $\pm$ 6 & 14 $\pm$ 4 & 1.10 & ... \\
SPT0418-47 & SPT-S J041839-4751.9 & 04:18:39.27 & -47:51:50.10 & 108 $\pm$ 15 & 102 $\pm$ 11 & 0.263 & 4.2248 \\
SPT0441-46 & SPT-S J044143-4605.5 & 04:41:44.13 & -46:05:29.50 & 80 $\pm$ 12 & 100 $\pm$ 11 & 0.882 & 4.4771 \\
SPT0452-50 & SPT-S J045246-5018.5 & 04:52:45.51 & -50:18:40.60 & 43 $\pm$ 6 & 64 $\pm$ 7 & 1.218 & 2.0104 \\
SPT0459-58 & SPT-S J045901-5805.3 & 04:59:00.47 & -58:05:17.00 & 53 $\pm$ 8 & 63 $\pm$ 7 & ... & 4.8564 \\
SPT0459-59 & SPT-S J045913-5942.4 & 04:59:12.62 & -59:42:21.20 & 61 $\pm$ 8 & 68 $\pm$ 8 & 0.938 & 4.7993 \\
SPT0529-54 & SPT-S J052903-5436.6 & 05:29:03.37 & -54:36:40.30 & 118 $\pm$ 14 & 115 $\pm$ 12 & 0.140 & 3.3689 \\
SPT0532-50 & SPT-S J053250-5047.1 & 05:32:51.27 & -50:47:09.50 & 118 $\pm$ 14 & 172 $\pm$ 18 & 1.15 & 3.3988 \\
SPT0538-50 & SPT-S J053816-5030.8 & 05:38:16.83 & -50:30:52.00 & 125 $\pm$ 13 & 146 $\pm$ 15 & 0.404 & 2.7817 \\
SPT0550-53$^{\mathrm{b}}$ & SPT-S J055002-5356.6 & 05:50:01.08 & -53:56:41.20 & 53 $\pm$ 8 & 55 $\pm$ 6 & 0.85 & 3.1280 \\
SPT0551-50$^{\mathrm{b}}$ & SPT-S J055138-5058.0 & 05:51:38.97 & -50:58:03.30 & 74 $\pm$ 10 & 84 $\pm$ 9 & 0.365 & 3.1638 \\
SPT2031-51 & SPT-S J203100-5112.3 & 20:30:59.33 & -51:12:26.40 & 64 $\pm$ 7 & 53 $\pm$ 6 & 0.624 & ... \\
SPT2048-55 & SPT-S J204823-5520.7 & 20:48:23.47 & -55:20:43.70 & 54 $\pm$ 7 & 56 $\pm$ 7 & ... & ... \\
SPT2052-56 & SPT-S J205239-5611.9 & 20:52:40.87 & -56:11:57.50 & 22 $\pm$ 3 & 15 $\pm$ 4 & ... & ... \\
SPT2103-60 & SPT-S J210330-6032.8 & 21:03:31.55 & -60:32:46.40 & 78 $\pm$ 10 & 62 $\pm$ 7 & 0.76 & 4.4357 \\
SPT2132-58 & SPT-S J213244-5803.1 & 21:32:43.54 & -58:02:54.00 & 58 $\pm$ 8 & 57 $\pm$ 7 & ... & 4.7677 \\
SPT2134-50 & SPT-S J213403-5013.4 & 21:34:03.85 & -50:13:27.10 & 101 $\pm$ 12 & 86 $\pm$ 9 & 0.776 & 2.7799 \\
SPT2146-55 & SPT-S J214654-5507.9 & 21:46:54.13 & -55:07:52.10 & 54 $\pm$ 7 & 49 $\pm$ 6 & ... & 4.5672 \\
SPT2146-56 & SPT-S J214644-5617.0 & 21:46:44.58 & -56:17:00.90 & 8 $\pm$ 2 & 4 $\pm$ 3 & 0.673 & ... \\
SPT2147-50 & SPT-S J214719-5035.9 & 21:47:19.62 & -50:35:59.00 & 61 $\pm$ 8 & 54 $\pm$ 6 & 0.845 & 3.7602 \\
SPT2300-51$^{\mathrm{c}}$ & SPT-S J230012-5157.4 & 23:00:12.48 & -51:57:23.70 & 20 $\pm$ 3 & 4 $\pm$ 3 & ... & ... \\
SPT2311-54 & SPT-S J231124-5450.6 & 23:11:24.26 & -54:50:32.80 & 44 $\pm$ 5 & 40 $\pm$ 5 & 0.44 & 4.2796 \\
SPT2319-55 & SPT-S J231921-5557.9 & 23:19:22.20 & -55:57:57.80 & 38 $\pm$ 5 & 36 $\pm$ 5 & 0.91 & 5.2928 \\
SPT2340-59 & SPT-S J234010-5943.3 & 23:40:09.57 & -59:43:30.40 & 34 $\pm$ 5 & 35 $\pm$ 5 & 0.113 & 3.8641 \\
SPT2349-50 & SPT-S J234942-5053.6 & 23:49:42.70 & -50:53:33.20 & 43 $\pm$ 5 & 43 $\pm$ 6 & 0.450 & 2.6480 \\
SPT2349-56 & SPT-S J234942-5638.2 & 23:49:42.70 & -56:38:18.90 & 56 $\pm$ 10 & 22 $\pm$ 4 & ... & 4.3002 \\
SPT2351-57$^{\mathrm{b}}$ & SPT-S J235150-5722.3 & 23:51:51.03 & -57:22:16.40 & 35 $\pm$ 5 & 32 $\pm$ 5 & 0.589 & 5.8113 \\
SPT2353-50$^{\mathrm{b}}$ & SPT-S J235338-5010.2 & 23:53:39.50 & -50:10:04.60 & 41 $\pm$ 6 & 35 $\pm$ 5 & 0.697 & 5.5764 \\
SPT2354-58 & SPT-S J235434-5815.1 & 23:54:34.58 & -58:15:06.50 & 66 $\pm$ 8 & 58 $\pm$ 7 & 0.428 & 1.8671 \\
SPT2357-51 & SPT-S J235718-5153.7 & 23:57:16.85 & -51:53:51.50 & 53 $\pm$ 8 & 36 $\pm$ 5 & ... & 3.0700
\enddata 
\tablecomments{Positions listed correspond to the ALMA phase center. 
Source redshifts are given in \citet{weiss13} and 
Strandet \etal (submitted). Lens redshifts are given in Rotermund \etal, in prep. 
Note that both LABOCA and ALMA flux densities are measured at 870\,\um.} 
\tablenotetext{a}{Total flux density in the ALMA image; see Section~\ref{fluxcomp}. 
Flux densities include 10\% absolute calibration uncertainties.} 
\tablenotetext{b}{Source appears to be lensed by a large group or cluster and 
cannot be modeled; these sources are shown in Appendix~\ref{app:clusters}.} 
\tablenotetext{c}{SPT2300-51 is undetected by ALMA; the listed ALMA flux is derived 
from a $\sim$4$\sigma$ source outside the primary beam half-power radius.} 
\end{deluxetable*}

\begin{deluxetable*}{llclccc}
\tablecaption{Observational Summary \label{tab:obstable}}
\startdata
\tableline
\\
Source Group             & Date         & Total Time$^\mathrm{a}$ (h) & Flux Calibrator & N$_\mathrm{ant}$ & $\sigma^\mathrm{b}$ (mJy)   & Beam Size$^\mathrm{a}$ \\
\tableline
\vspace{-0.07in}
\\
SPT0202-61 -- SPT0418-47 & 28 Nov. 2011 & 1.4            & J0403-360       & 14               & 0.60                        & 0.8''$\times$2.2'' \\
                         & 04 June 2012 & 1.4            & Neptune         & 18               & 0.29                        & 0.5''$\times$0.6'' \\
SPT0441-46 -- SPT0551-50 & 16 Nov. 2011 & 0.8            & Callisto        & 16               & 0.72                        & 1.3''$\times$1.5'' \\
                         & 15 June 2012 & 0.9            & Callisto        & 20               & 0.64                        & 0.5''$\times$0.7'' \\
SPT2031-51 -- SPT2147-50 & 06 May 2012  & 0.9            & Neptune         & 17               & 0.42                        & 0.4''$\times$0.5'' \\
						 & 14 Aug. 2012 & 0.9            & Neptune         & 22               & 0.31                        & 0.5''$\times$0.6'' \\
SPT2300-51 -- SPT0128-51 & 22 May 2012  & 1.4            & Neptune         & 19               & 0.40                        & 0.4''$\times$0.5'' \\
						 & 13 Aug. 2012 & 1.7            & Neptune         & 25               & 0.21                        & 0.5''$\times$0.6''
\enddata
\tablenotetext{a}{Total observation time includes overheads.}
\tablenotetext{b}{Sensitivity and beam size are averages for all science targets in each track.}
\end{deluxetable*}

\section{Visibility-Based Lens Modeling} \label{lensmodels}

When modeling the effects of gravitational lensing, many methods perform the fitting
procedure directly on observed images of the lensed emission. However, 
ALMA does not directly image the sky emission; rather, it measures the Fourier
components of the sky emission at a range of two-dimensional spatial frequencies.
Inverting these visibilities leads to correlated noise in the resulting images which
can introduce bias into later measurements. Instead, a better option is to 
model the visibilities directly, where the noise and measurement are well 
understood.  Modeling in the $uv$ plane also allows us to model and account
for residual calibration errors, including improper antenna delay calibrations and
mismatched absolute flux scales from observations taken on different days. Our
lens modeling procedure is based on the work of \citet{hezaveh13} (see also
\citealt{bussmann12,bussmann13} for a similar technique).

The lens mass profile is represented by one or more Singular Isothermal Ellipsoids
(SIEs), with lensing deflections derived by \citet{kormann94}. The SIE is 
parameterized by its two-dimensional location relative to the phase center 
($x_L$, $y_L$), the lens strength in the form of the angular Einstein
radius $\theta_{E,L}$, ellipticity $e_L$, and position angle of the major axis
$\phi_L$ in degrees east of north. In some cases, the data also favor the
existence of an external tidal shear ($\gamma$), with deflections calculated as in 
\citet{keeton00} (we have redefined the shear position angle, $\phi_\gamma$, to match
the convention used here for $\phi_L$).  Background source 
emission and any unlensed sources are represented as one or more
unresolved point sources (with position $x_S$ and $y_S$, and flux density
\ses as free parameters) or S\'{e}rsic profiles (\citealt{sersic68}; 
with position $x_S$ and $y_S$,
flux density \ses, S\'{e}rsic index $n_S$, half-light radius \reff, axis ratio 
$b_S$/$a_S$, and position angle $\phi_S$ as free parameters). Note that a S\'{e}rsic 
index $n=4$ corresponds to a 
\citet{devaucouleurs53} profile, $n=1$ an exponential disk, and $n=0.5$ a
Gaussian light profile (in \citealt{hezaveh13}, all sources were modeled
as circularly symmetric Gaussian profiles).  
For lensed sources, we define the location of the source
to be relative to the primary lens in the model, while for unlensed sources
it is defined relative to the ALMA phase center. Within the framework we have 
developed, any of these
lens and source parameters may be held fixed during fitting, and loose flat
priors may be used. We use available optical/NIR imaging to guide
the models (e.g., a single lens vs. multiple lenses), but the positions of
galaxies identified in these images are not otherwise used, except for singly-imaged
sources for which the ALMA data alone are not sufficiently constraining.

To reproduce the information present in our high signal-to-noise ratio measurements,
and to represent realistic calibration uncertainties, our modeling must be more
flexible than that used in previous work \citep[e.g.,][]{bussmann13,hezaveh13,bussmann15}.
For example, because we are jointly modeling multiple datasets taken several months
apart (see Table~\ref{tab:obstable}), small differences in absolute 
calibration or atmospheric conditions between
epochs could be translated into false shifts in parameters.
To address this possibility, we allow for a
multiplicative amplitude re-scaling factor and an astrometric offset
between the two tracks. 
We also calibrate uncorrected antenna-based phase errors using the procedure described
in \citet{hezaveh13}. These phase errors may be attributed to uncompensated 
atmospheric delays or imprecisely known antenna positions.  These phase
errors are generally small except in the two Nov. 2011 tracks, which were observed
prior to antenna baseline solutions being incorporated into the reduction pipeline.
The phase errors and astrometric shifts derived from this procedure 
are consistent with those found for the ICRF sources that we
added to our observations to test the calibration and astrometry of the data.

We employ a Markov Chain Monte Carlo (MCMC) fitting procedure, using the
\texttt{emcee} \citep{foremanmackey13} code to sample the posterior probability
function. At each point in parameter space, we generate a model image from a given
set of lens and source parameters, including the flux scaling and astrometric 
offsets mentioned above. We then invert this image to the Fourier plane and 
measure the modeled visibilities at the $uv$ coordinates of each dataset. The 
quality of fit is calculated using the $\chi^2$ metric. When comparing models
of the same source with different numbers of free parameters, we use the 
Deviance Information Criterion (DIC; \citealt{spiegelhalter02}) for model selection.
The DIC determines, for example, whether including an additional source-plane
component is justified.

The code used to generate all the models in this work, along with example usage
scripts, is available at \url{https://github.com/jspilker/visilens}.

\section{Results} \label{results}

Images of each system along with the best-fit image- and source-plane models are
shown in Fig.~\ref{fig:images}.  These models are briefly described in 
Appendix~\ref{app:notes}. Summaries of the properties of the lenses and sources are
provided in Tables~\ref{tab:lenses} and \ref{tab:sources}, respectively.

\begin{figure*}[!tbp]%
\begin{centering}
\includegraphics[width=0.495\textwidth]{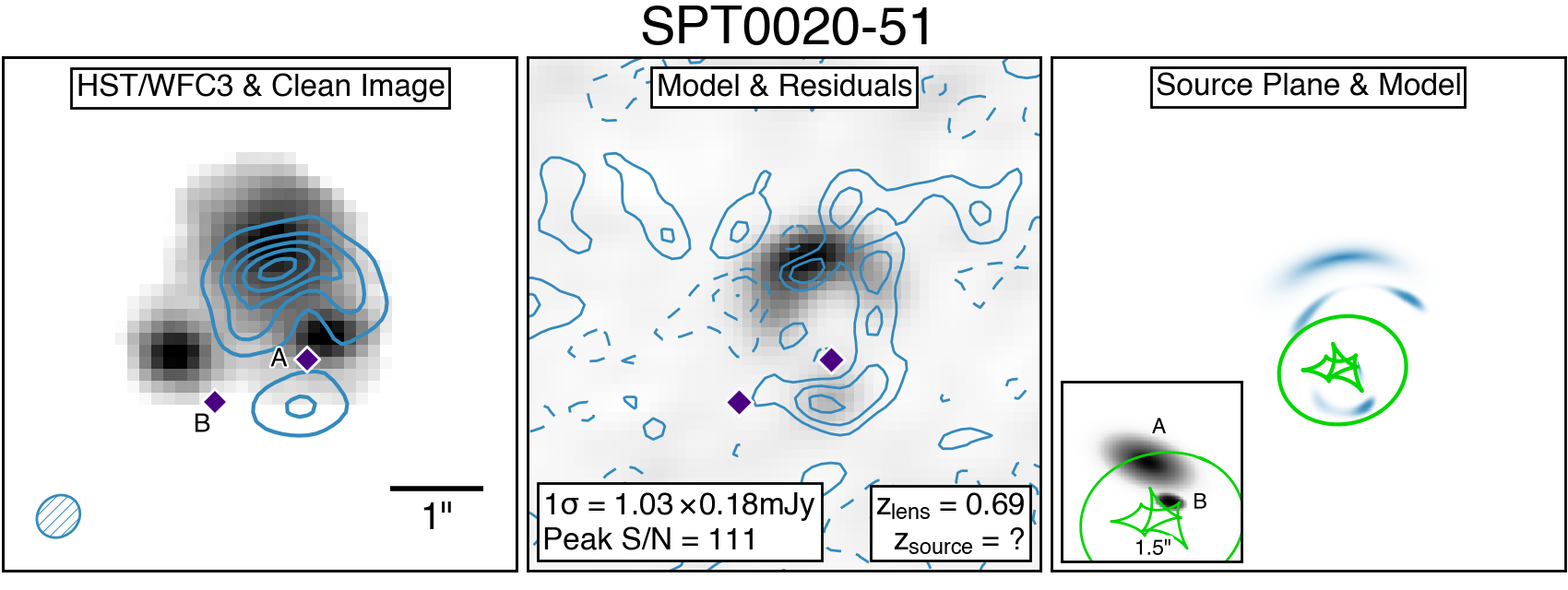}
\includegraphics[width=0.495\textwidth]{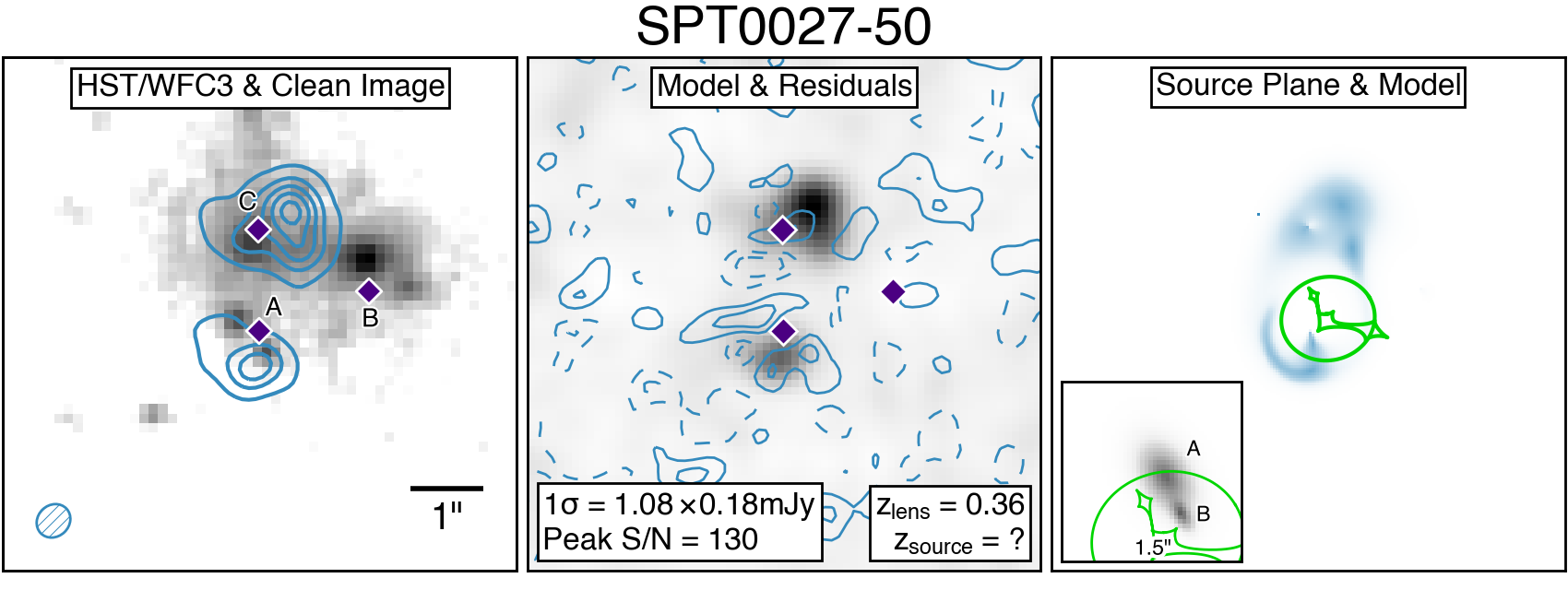}
\includegraphics[width=0.495\textwidth]{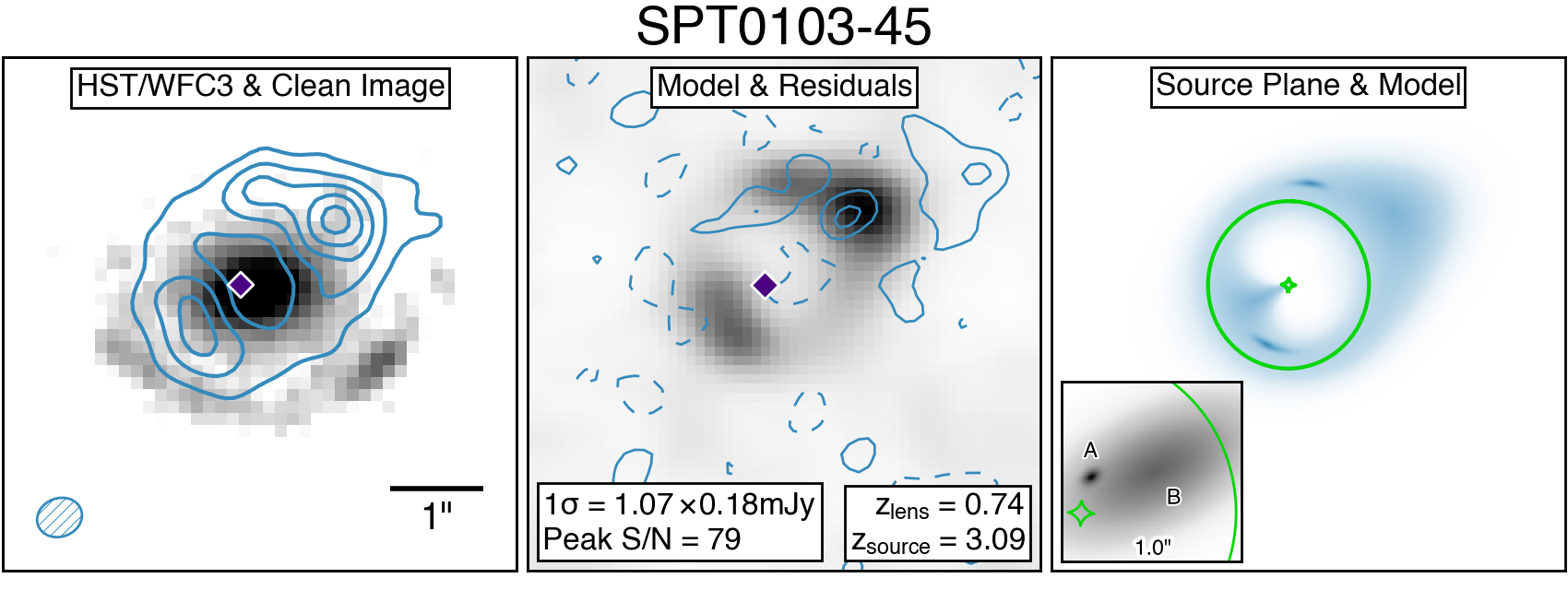}
\includegraphics[width=0.495\textwidth]{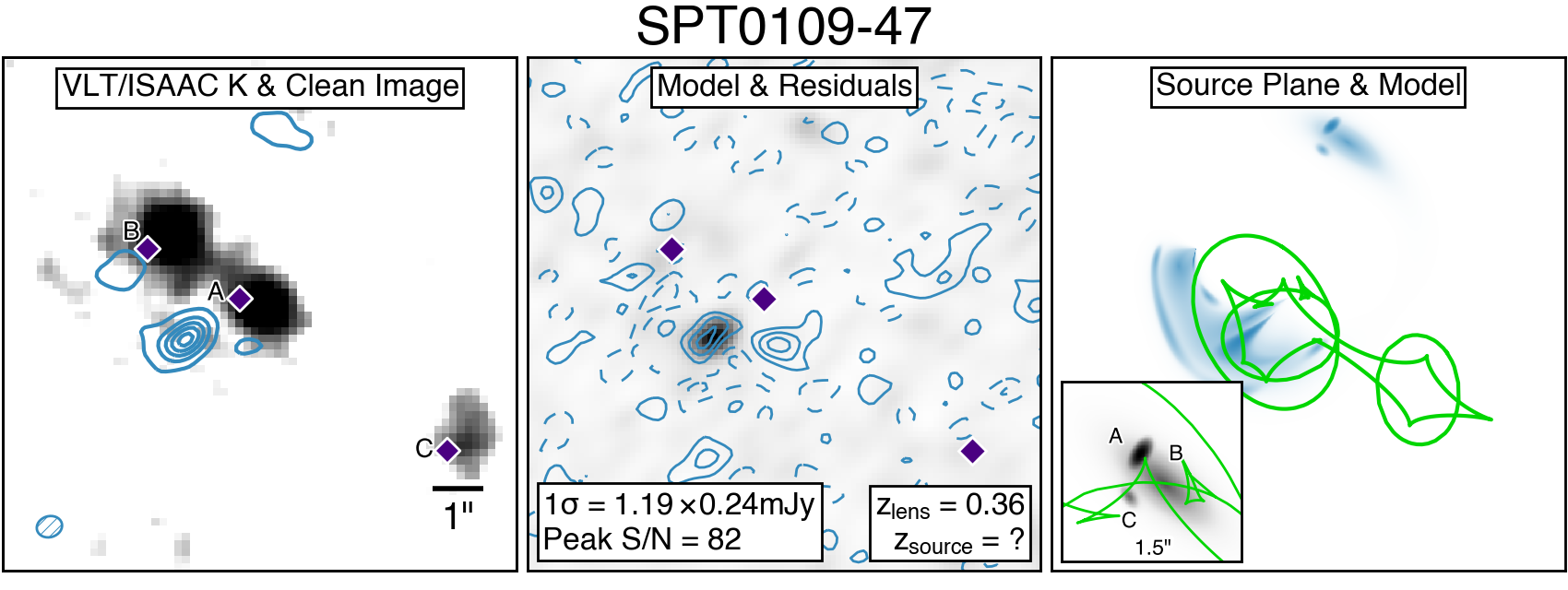}
\includegraphics[width=0.495\textwidth]{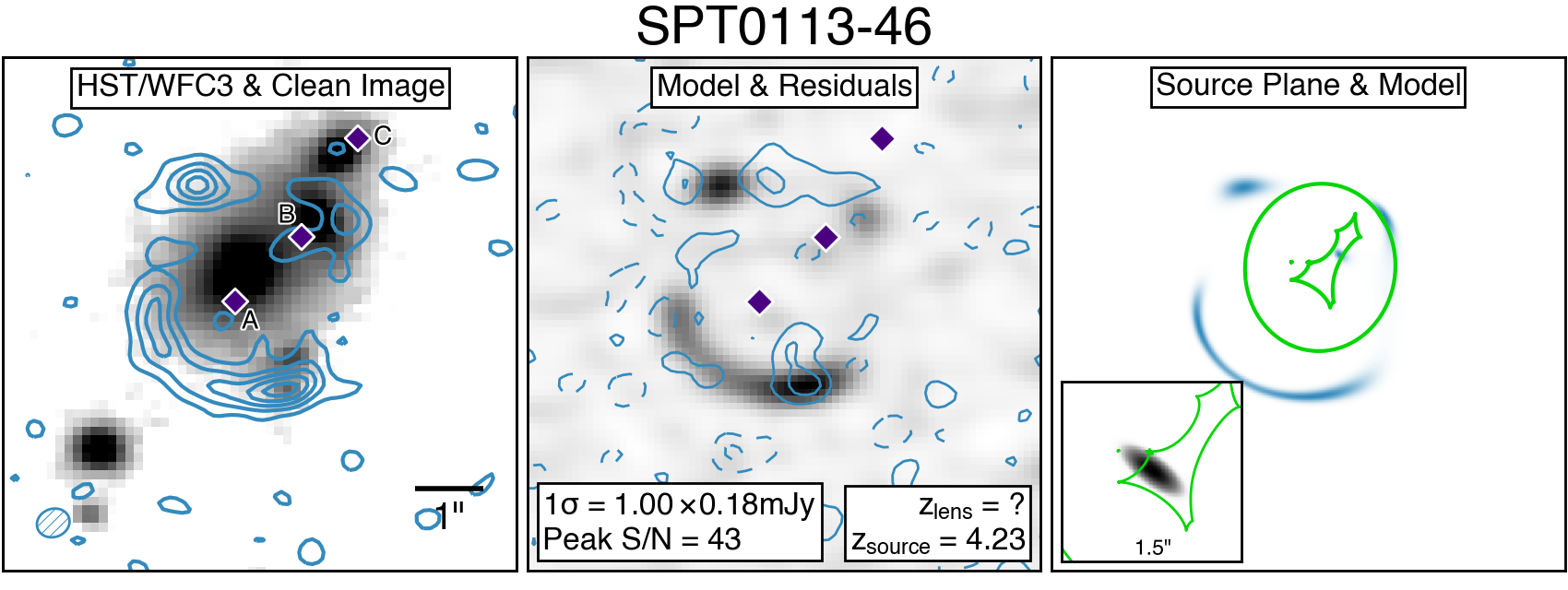}
\includegraphics[width=0.495\textwidth]{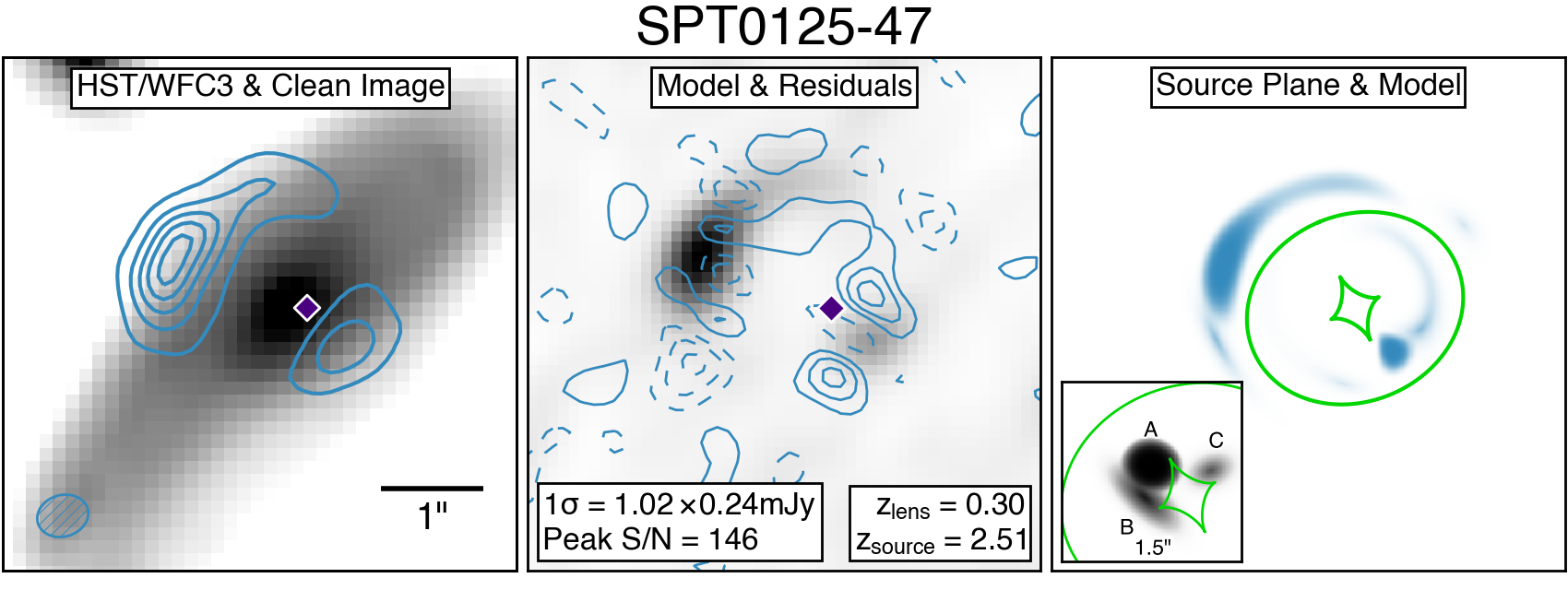}
\includegraphics[width=0.495\textwidth]{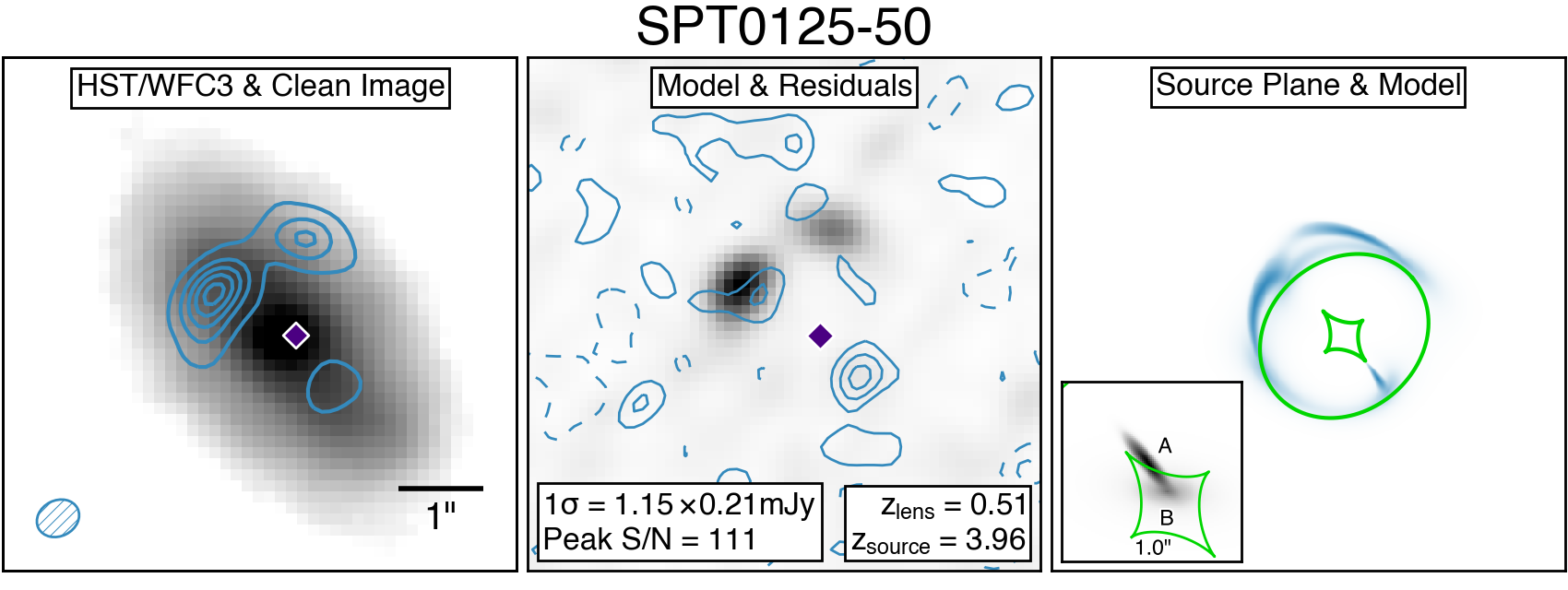}
\includegraphics[width=0.495\textwidth]{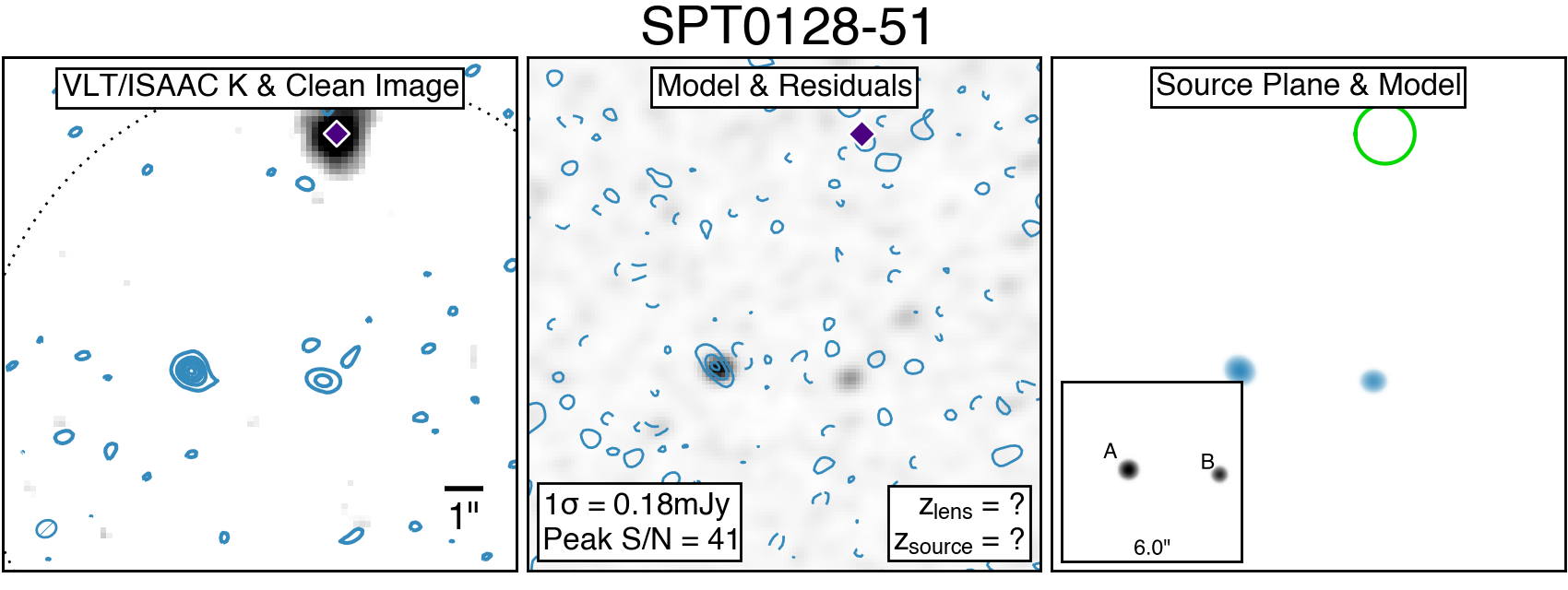}
\includegraphics[width=0.495\textwidth]{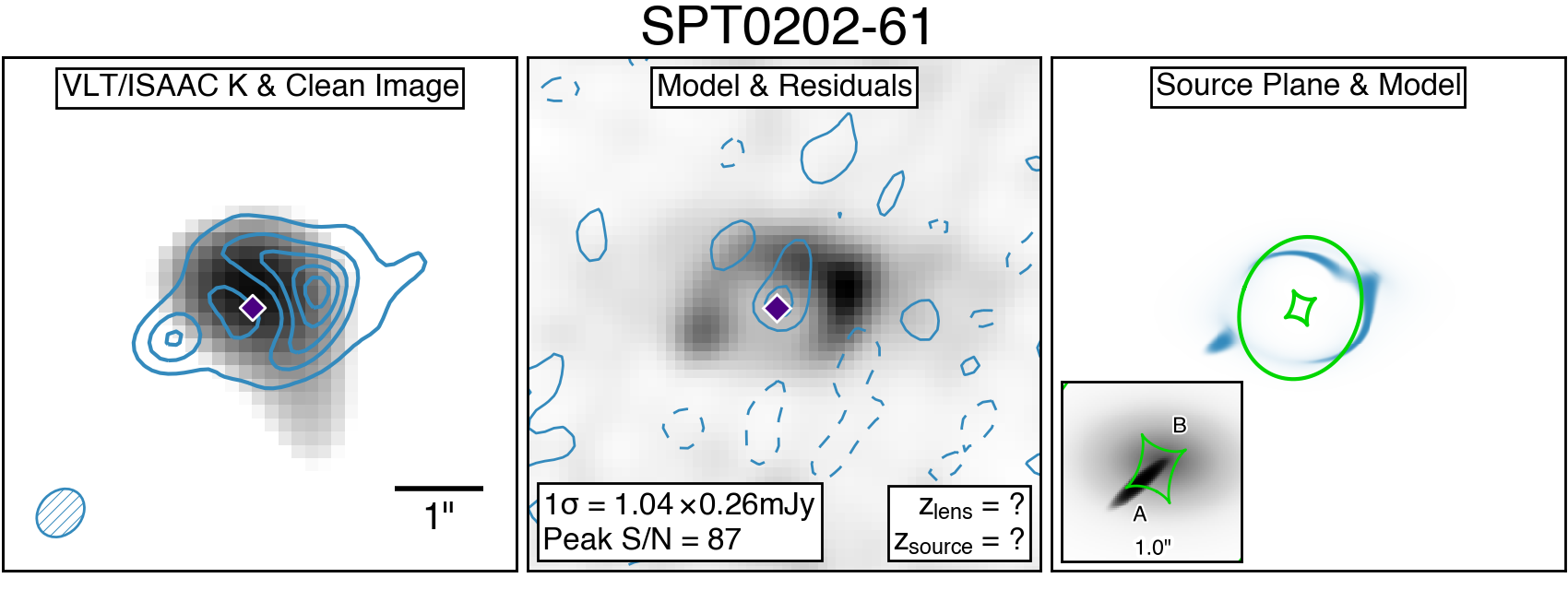}
\includegraphics[width=0.495\textwidth]{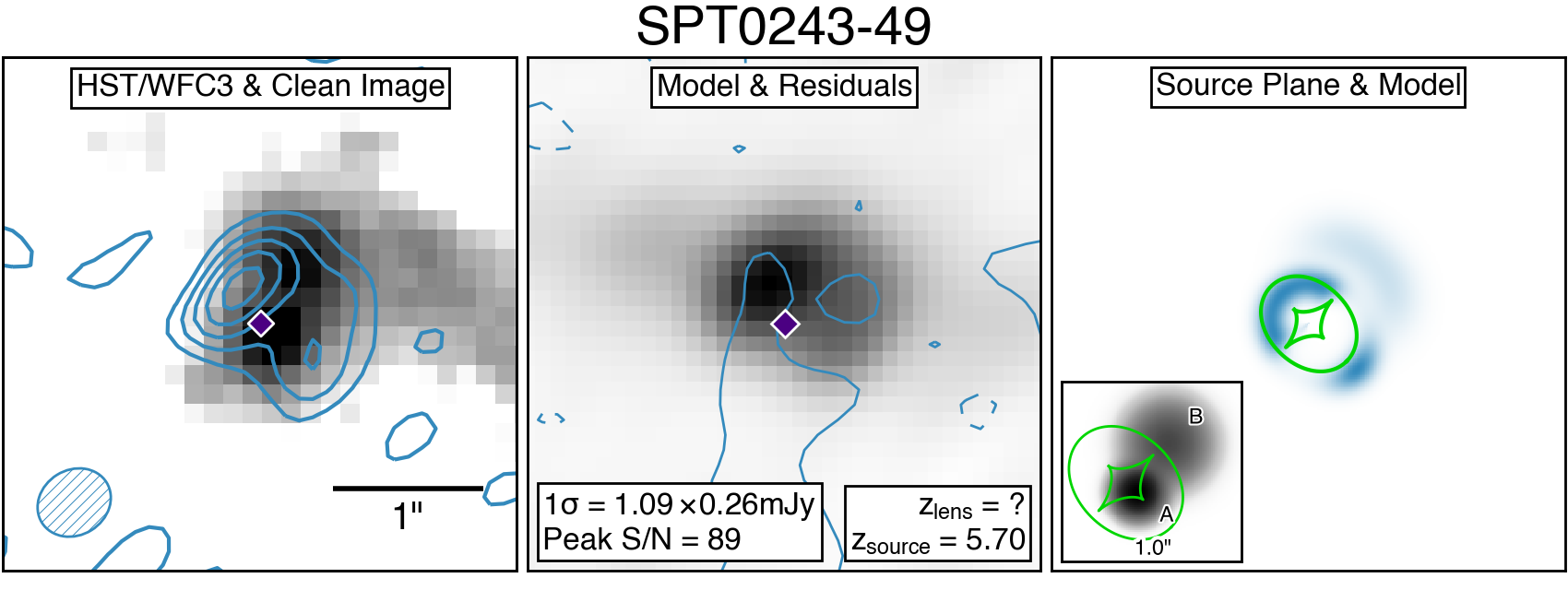}
\includegraphics[width=0.495\textwidth]{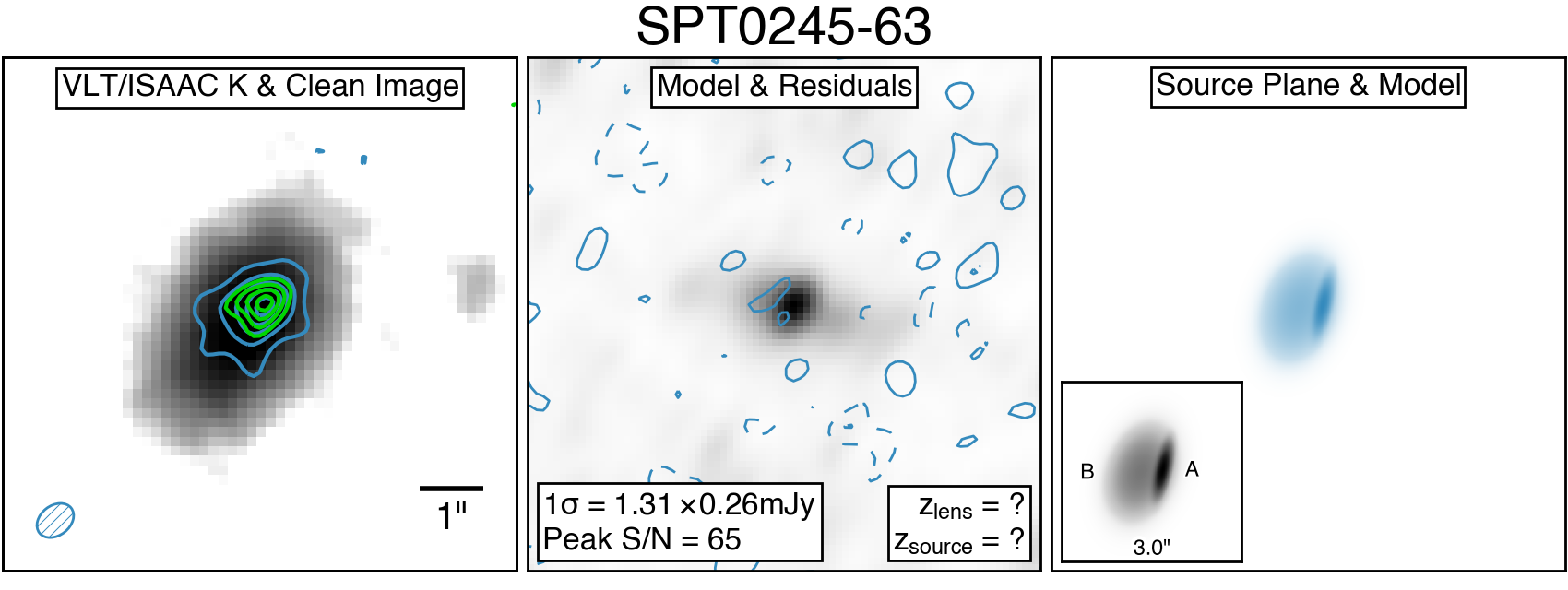}
\includegraphics[width=0.495\textwidth]{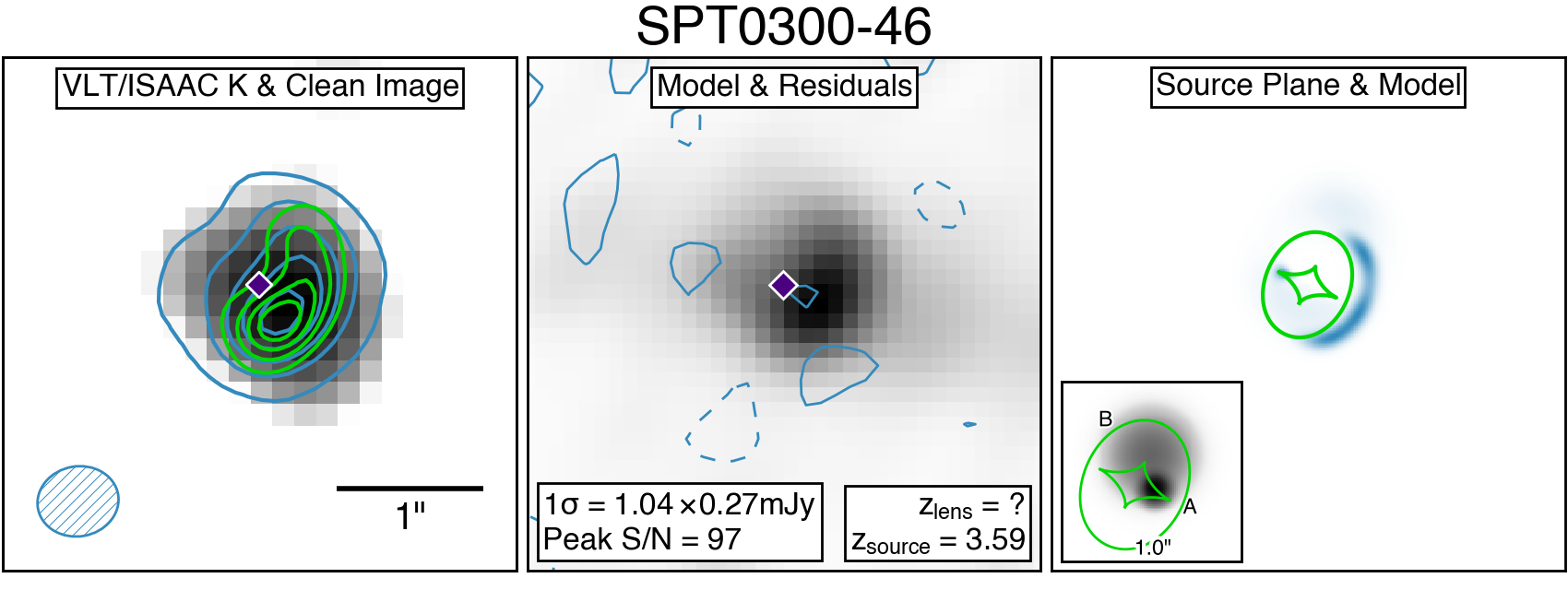}
\end{centering}
\caption{
Images and lens models for all sources modeled in this work. 
\textit{Left:} ALMA
870\,\um emission (blue contours) overlaid on the best-available optical/NIR image
(greyscale) for each source. Contours are drawn at 10, 30, ... percent of the peak 
value. 
The synthesized beam is indicated in the lower left corner.
For some objects, we also show images of the 870\,\um emission which
highlight the resolved structure present in the data (green contours; see 
Appendix~\ref{app:notes} for details).
Greyscale images are logarithmically scaled to emphasize the objects
detected. Fitted lens positions are shown with navy diamonds; sources
with multiple lenses are labeled as in Table~\ref{tab:lenses}. 
In panels with a large field-of-view, the ALMA primary beam half-power radius is
indicated with a dotted line; for the other objects, the primary beam correction at
the center of the image is given in the middle panel as the scale factor before
the noise level in mJy.
\textit{Middle:} Model dirty images (greyscale), with residual contours 
(blue) in steps of $\pm$2, 4, ...$\sigma$.  
\textit{Right:} Fully resolved best-fit model images (blue), with caustics shown
in green.  The inset of each panel shows a zoomed-in view of the source-plane
emission, where the size of the inset is given in the lower-center of each panel.
Multiple sources are labeled as in Table~\ref{tab:sources}.
\label{fig:images}} \addtocounter{figure}{-1}
\end{figure*}

\begin{figure*}[!tbp]%
\begin{centering}
\includegraphics[width=0.495\textwidth]{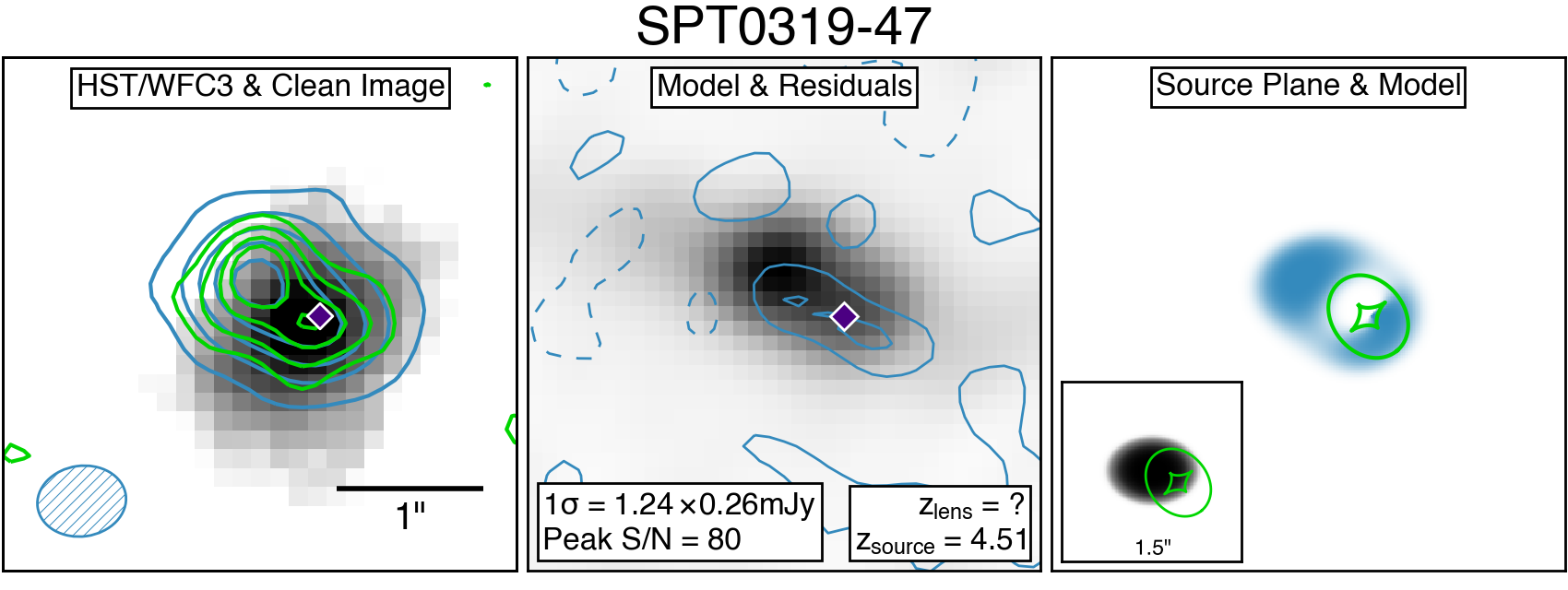}
\includegraphics[width=0.495\textwidth]{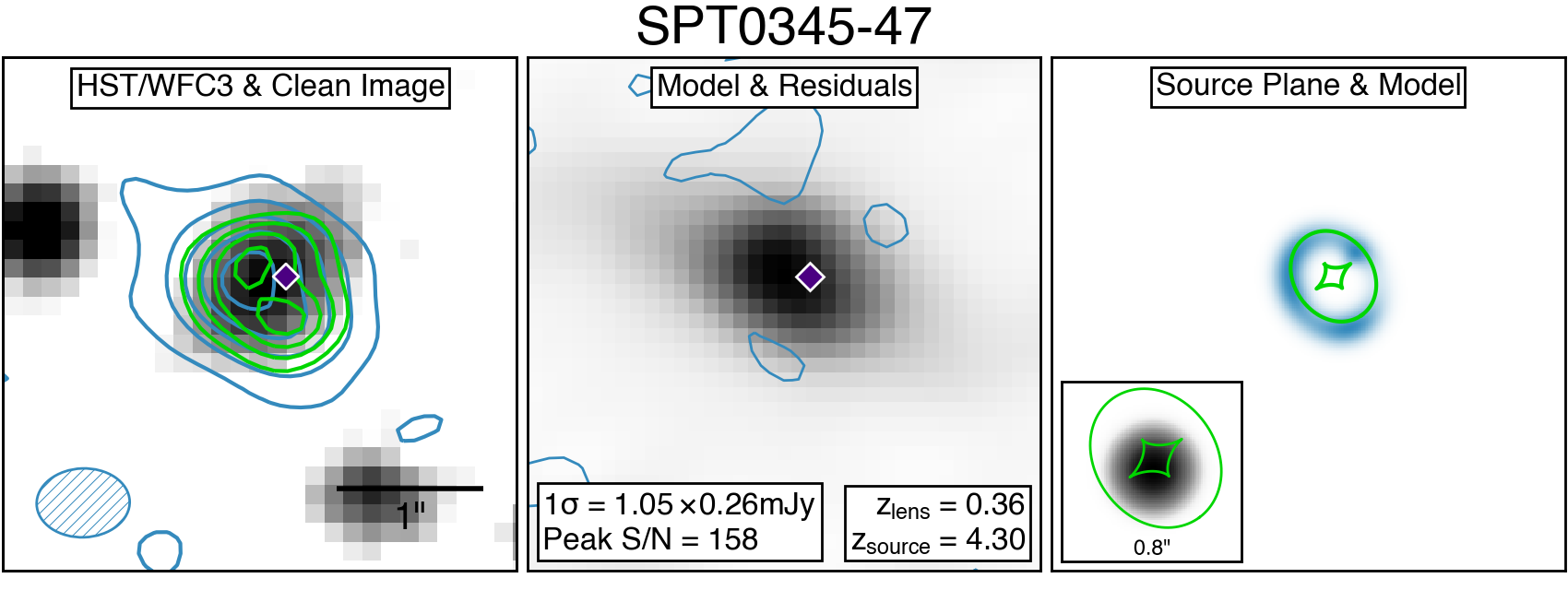}
\includegraphics[width=0.495\textwidth]{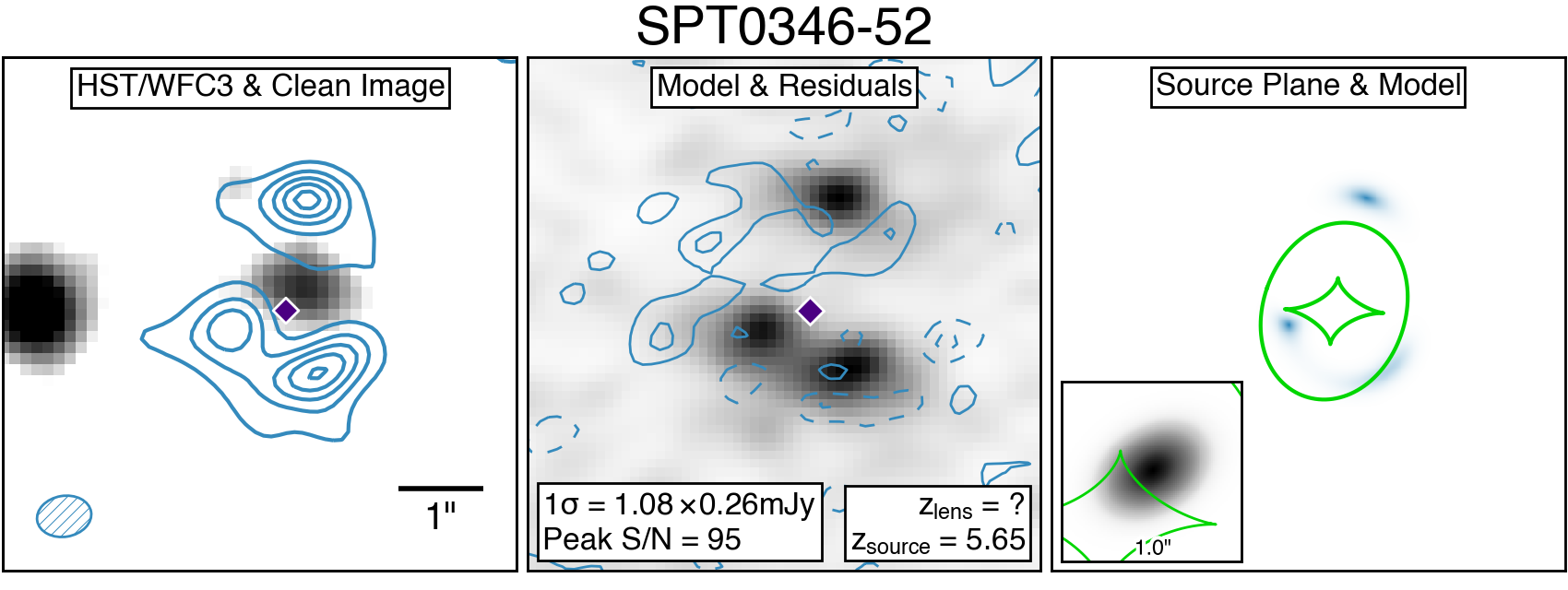}
\includegraphics[width=0.495\textwidth]{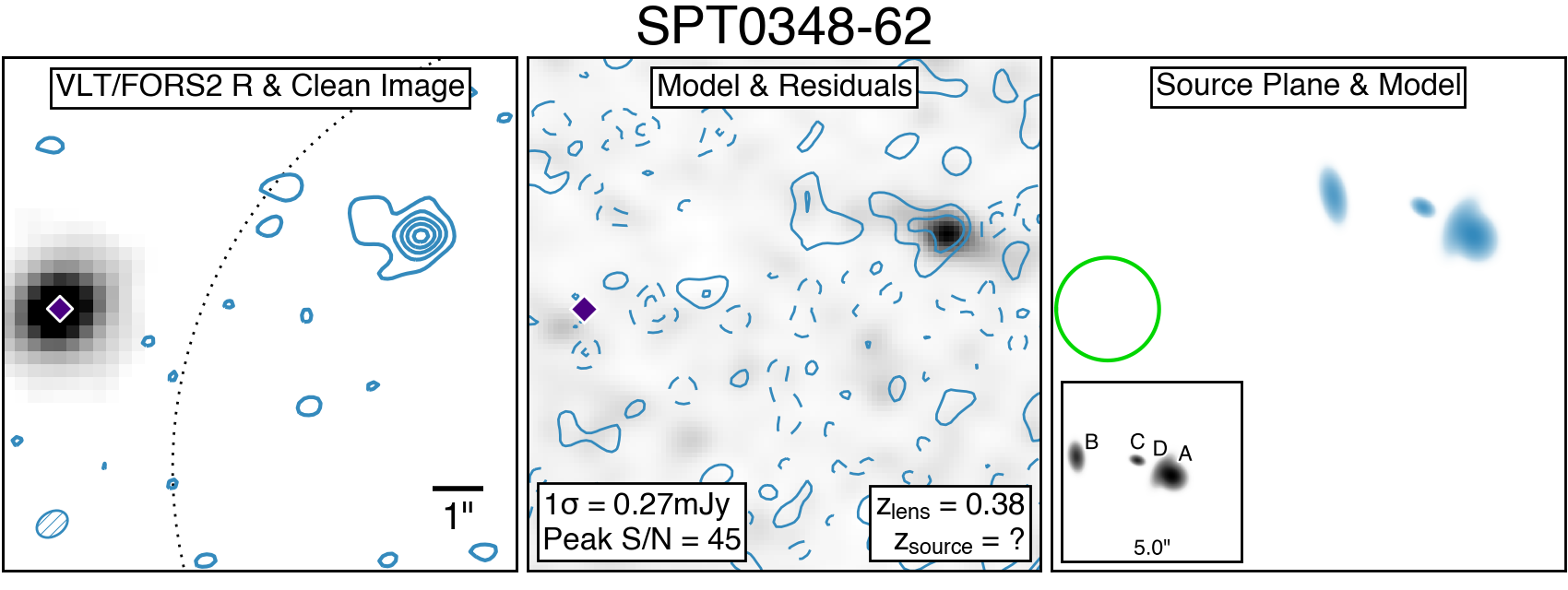}
\includegraphics[width=0.495\textwidth]{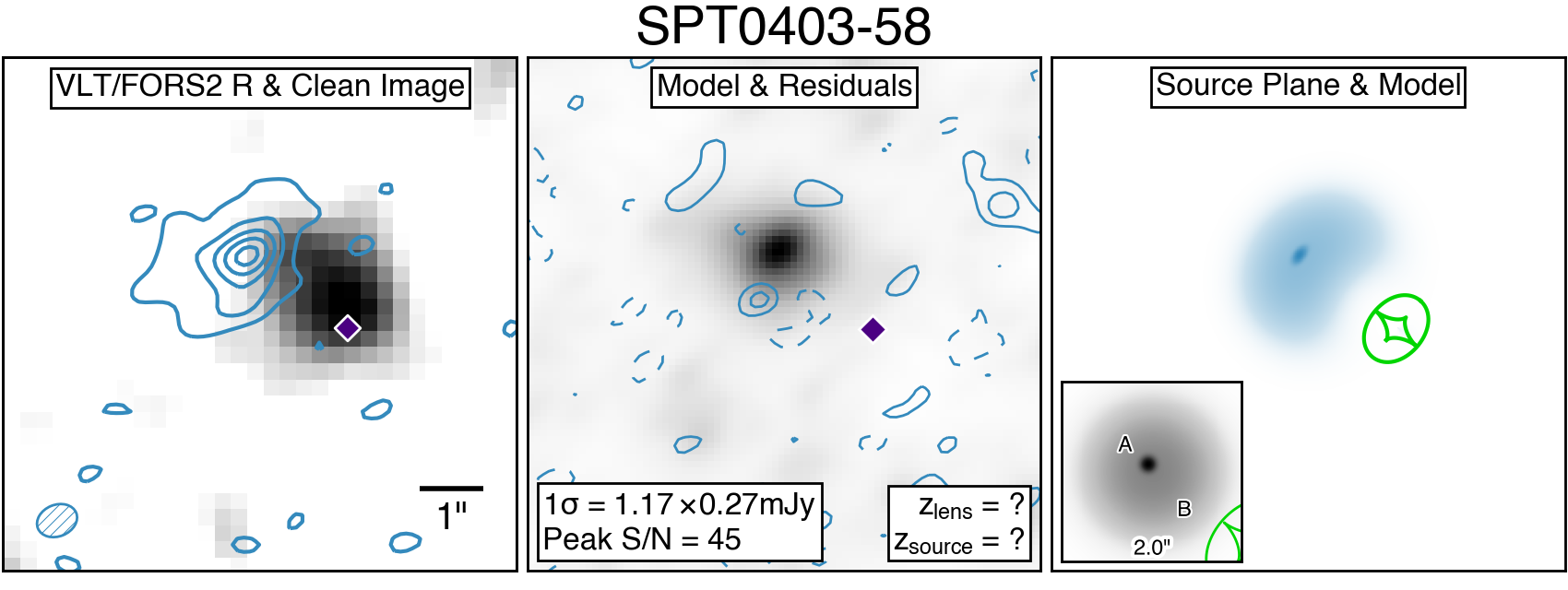}
\includegraphics[width=0.495\textwidth]{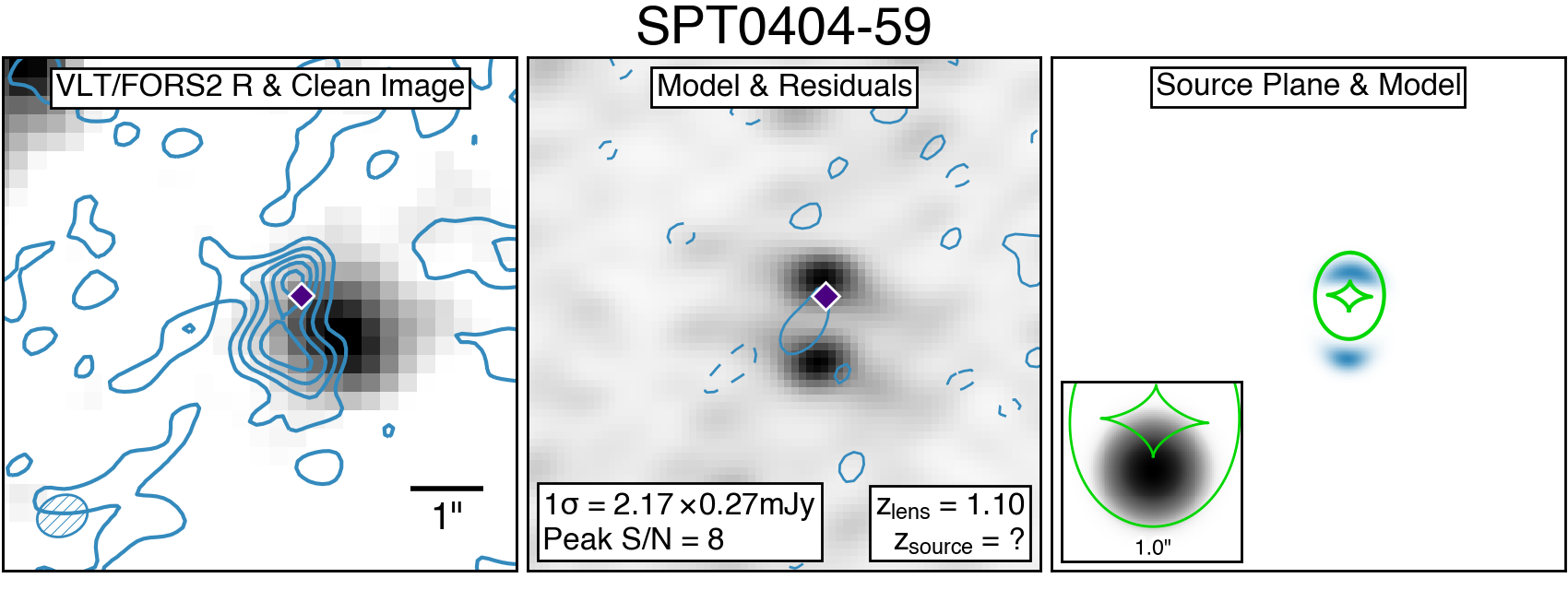}
\includegraphics[width=0.495\textwidth]{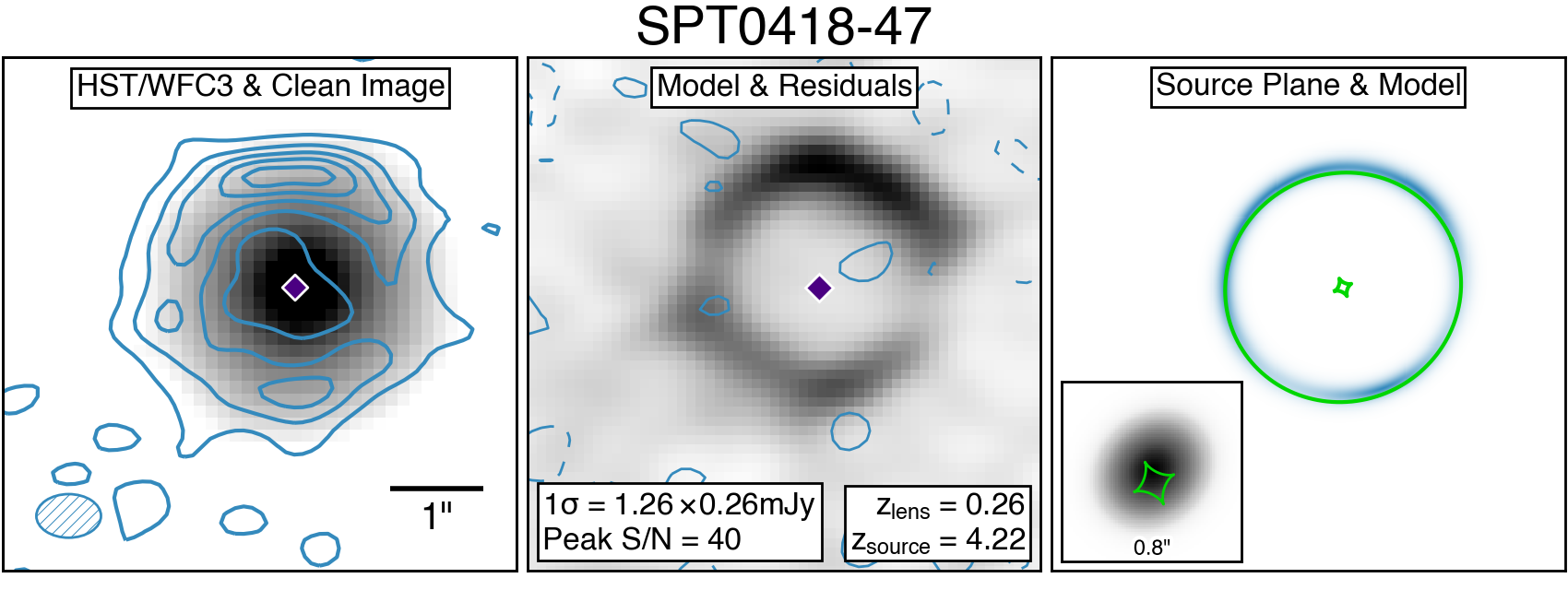}
\includegraphics[width=0.495\textwidth]{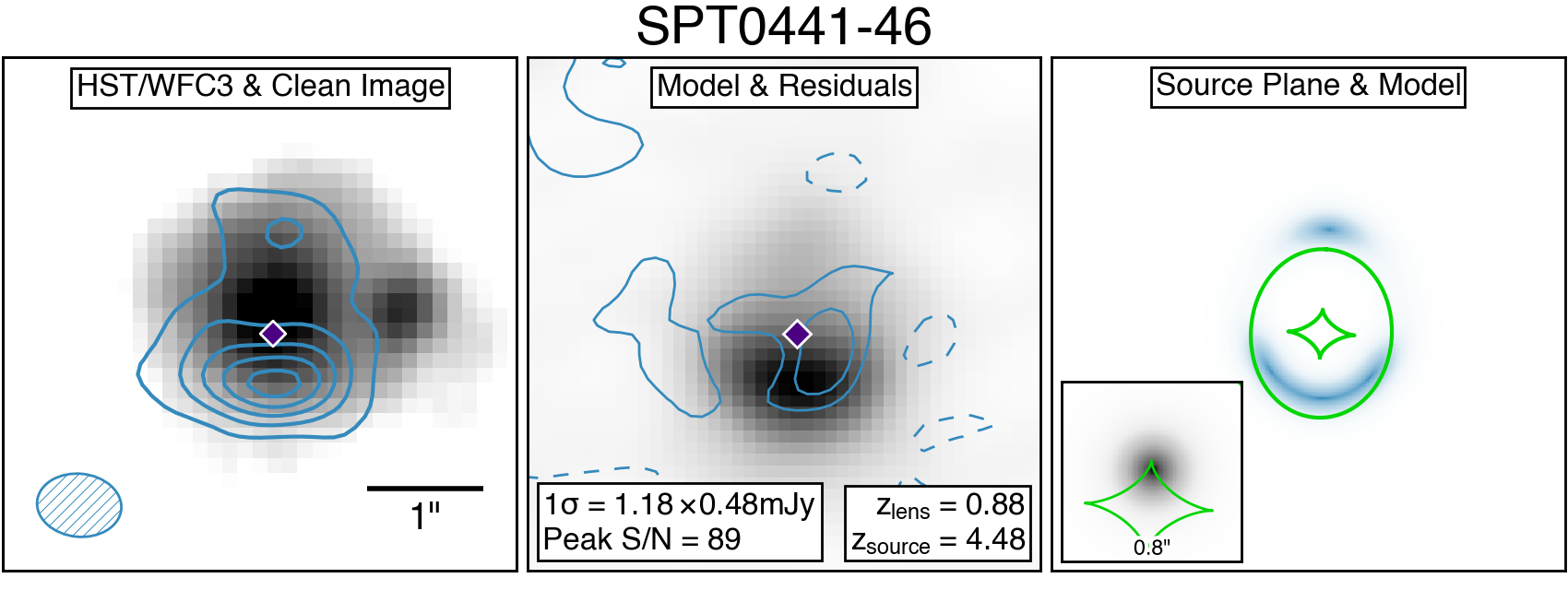}
\includegraphics[width=0.495\textwidth]{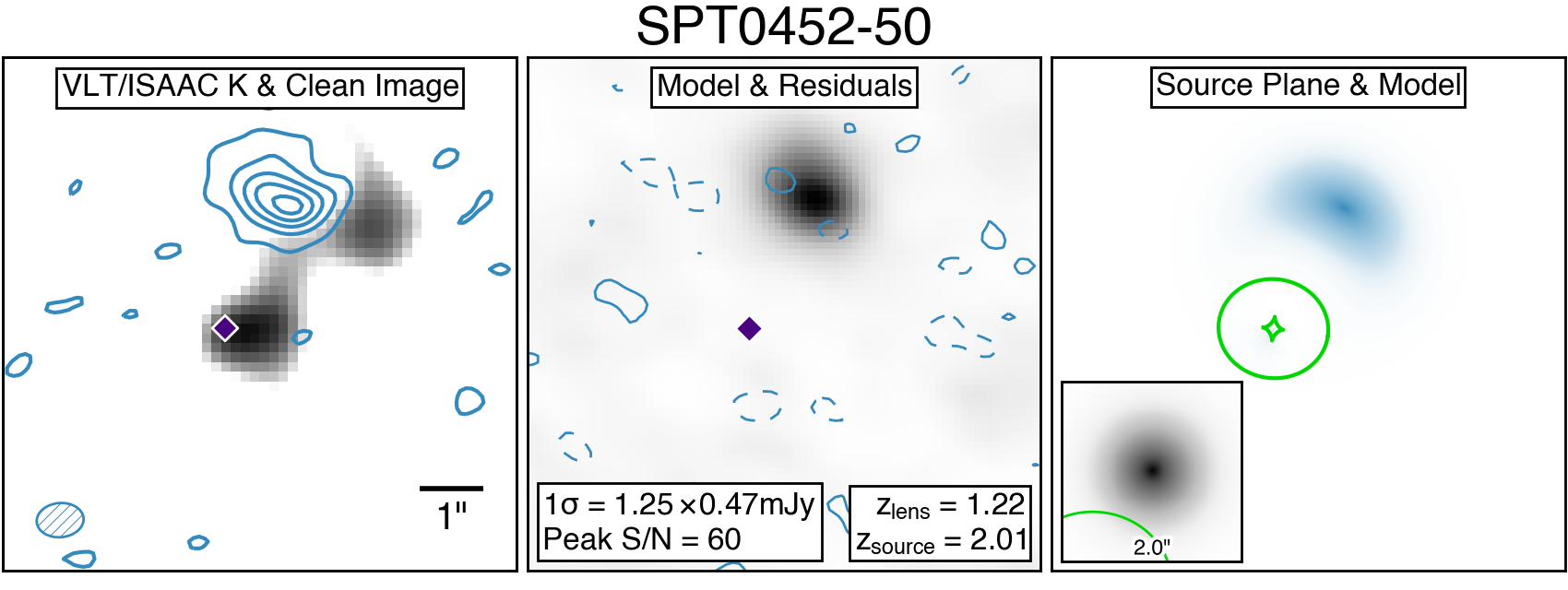}
\includegraphics[width=0.495\textwidth]{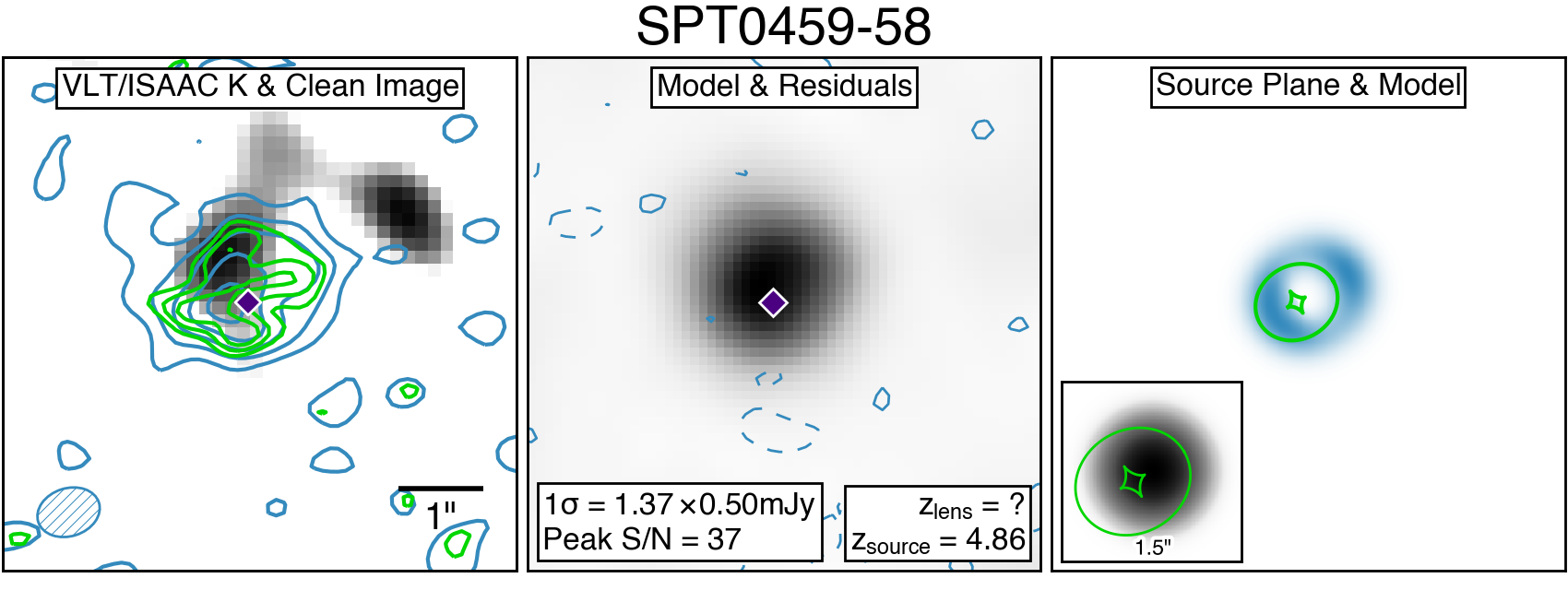}
\includegraphics[width=0.495\textwidth]{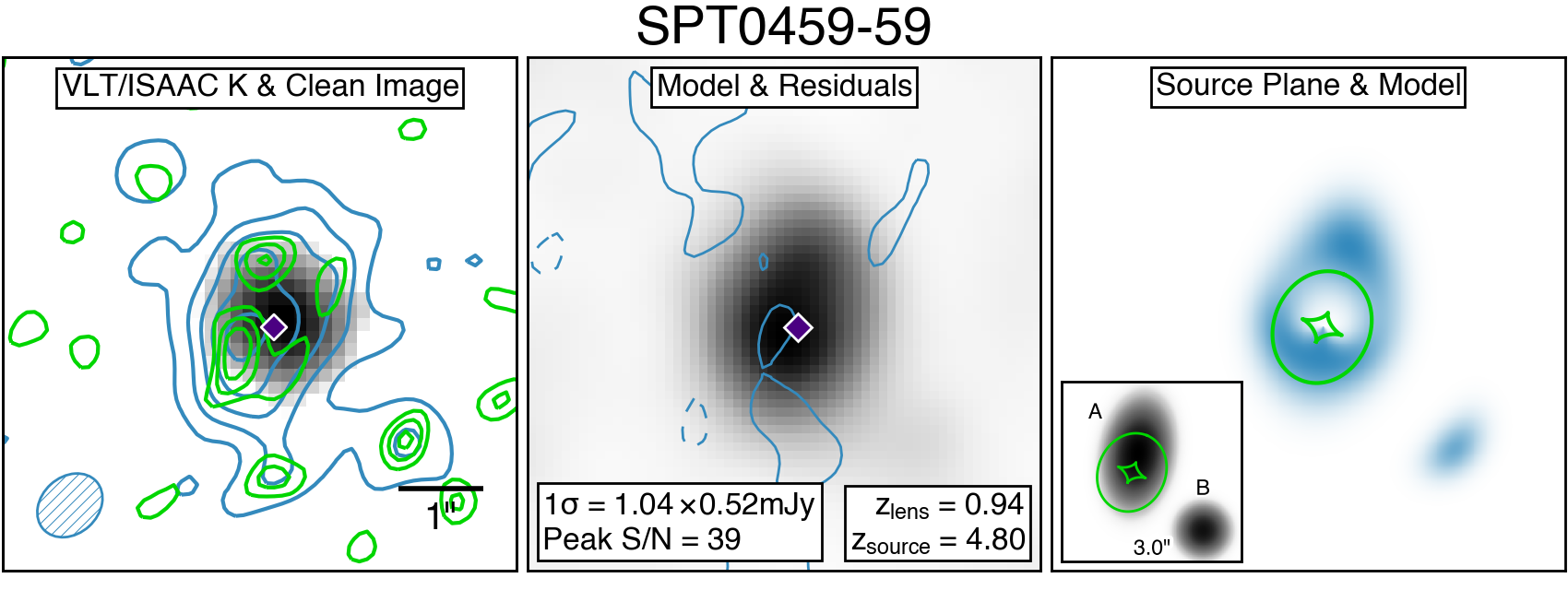}
\includegraphics[width=0.495\textwidth]{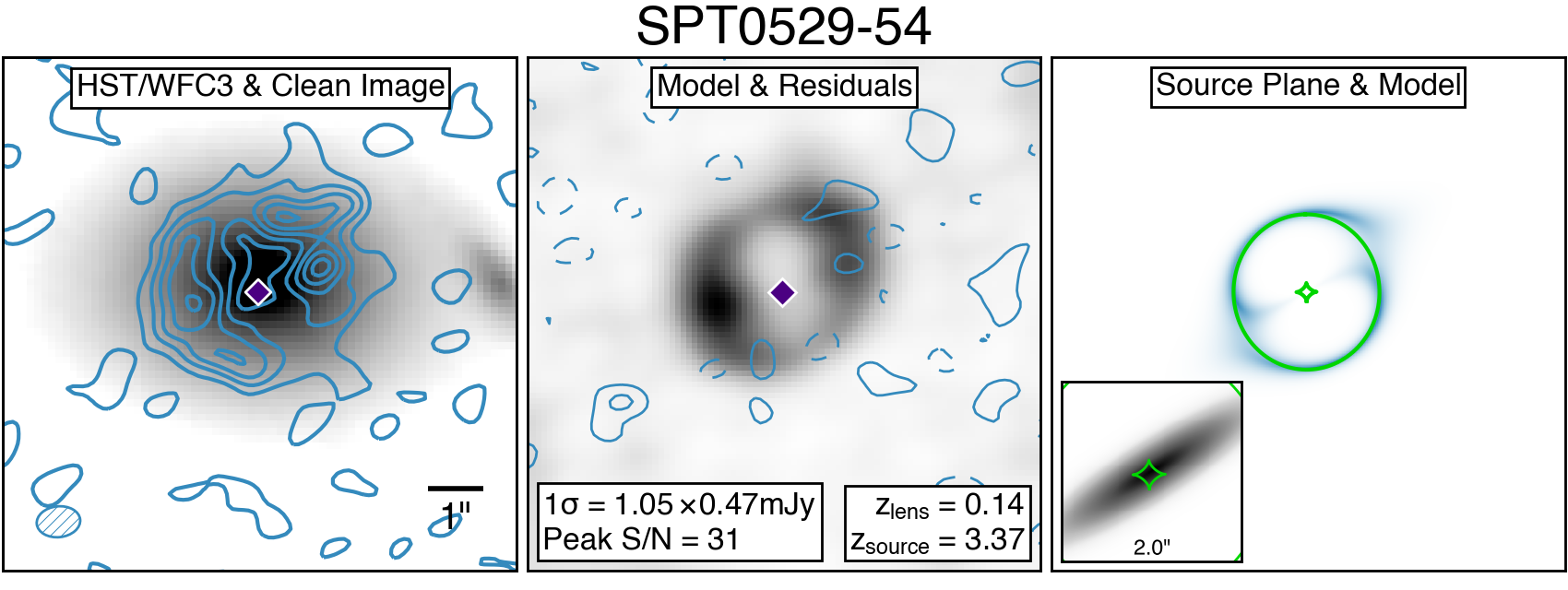}

\end{centering}
\caption{ Continued.  \label{fig:images1}} \addtocounter{figure}{-1}
\end{figure*}

\begin{figure*}[!tbp]%
\begin{centering}
\includegraphics[width=0.495\textwidth]{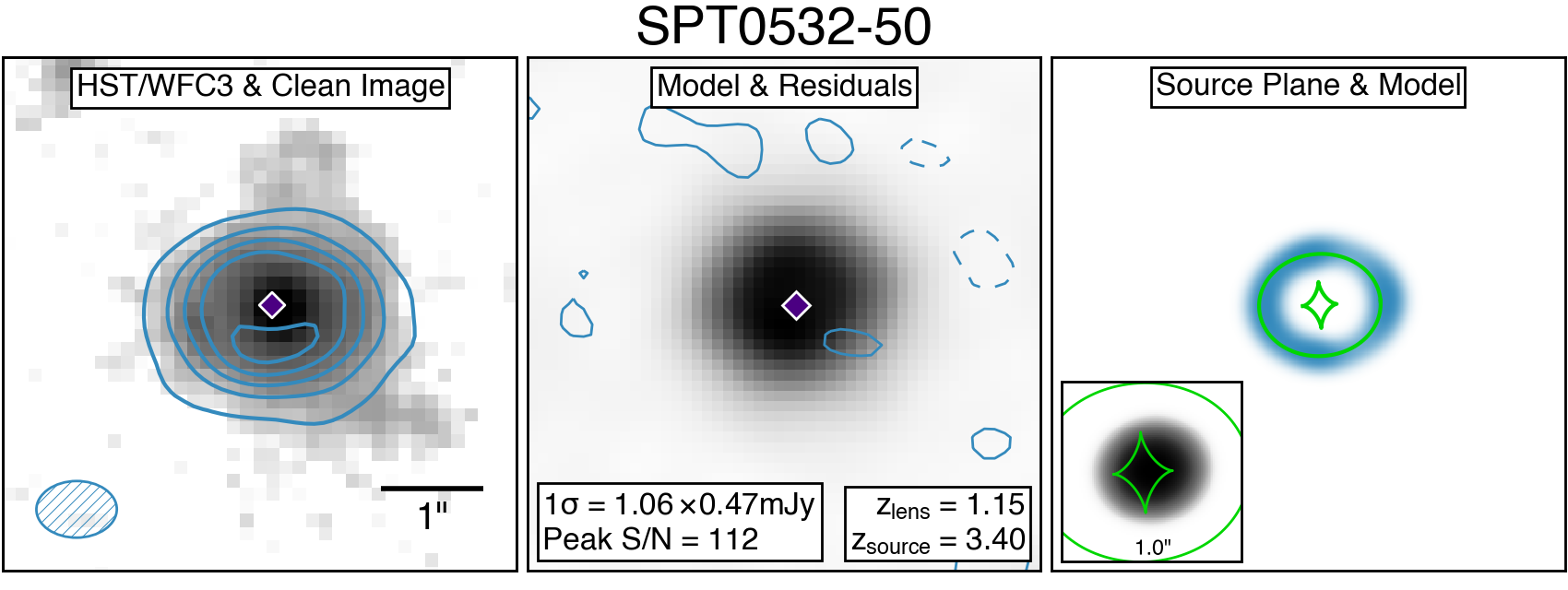}
\includegraphics[width=0.495\textwidth]{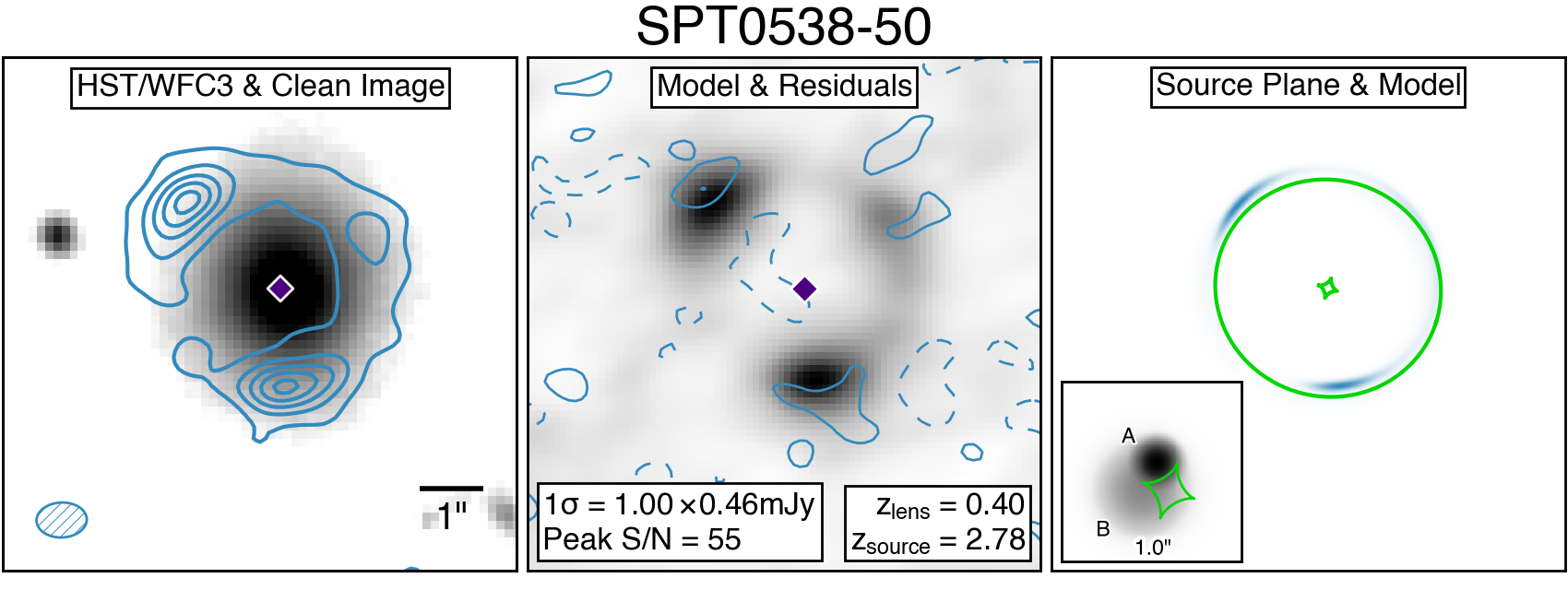}
\includegraphics[width=0.495\textwidth]{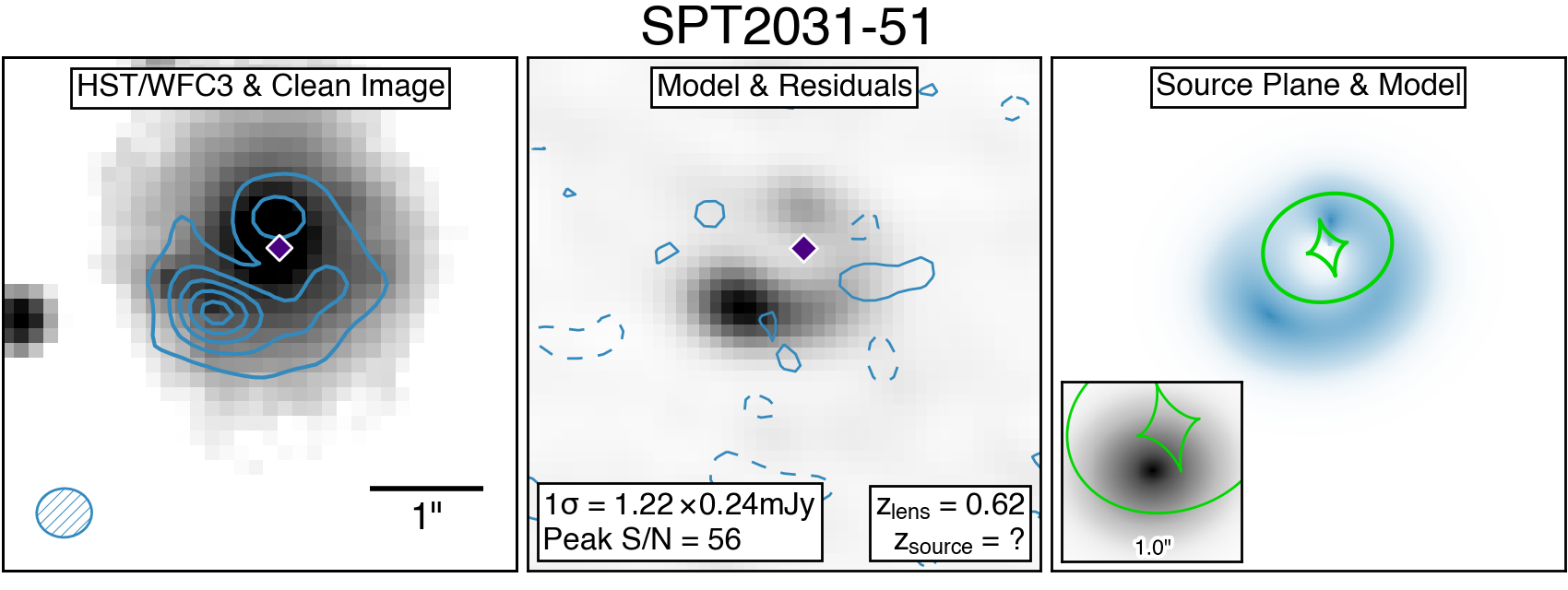}
\includegraphics[width=0.495\textwidth]{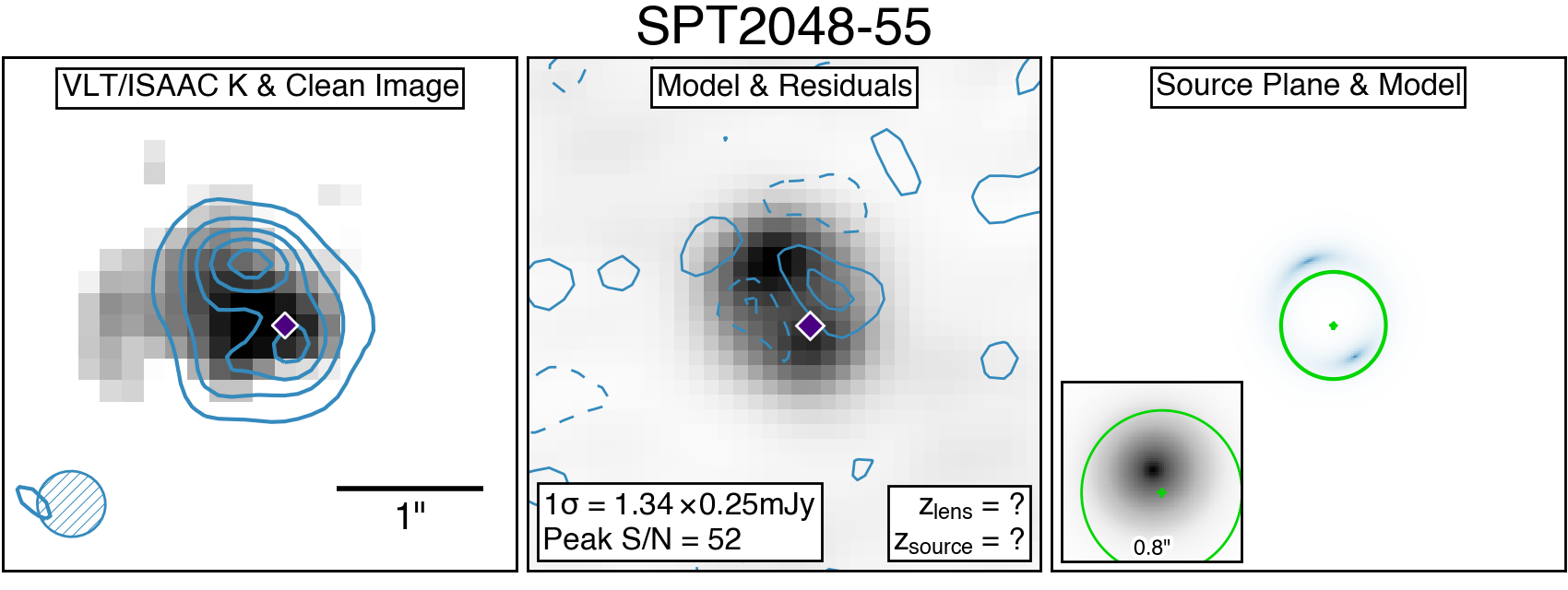}
\includegraphics[width=0.495\textwidth]{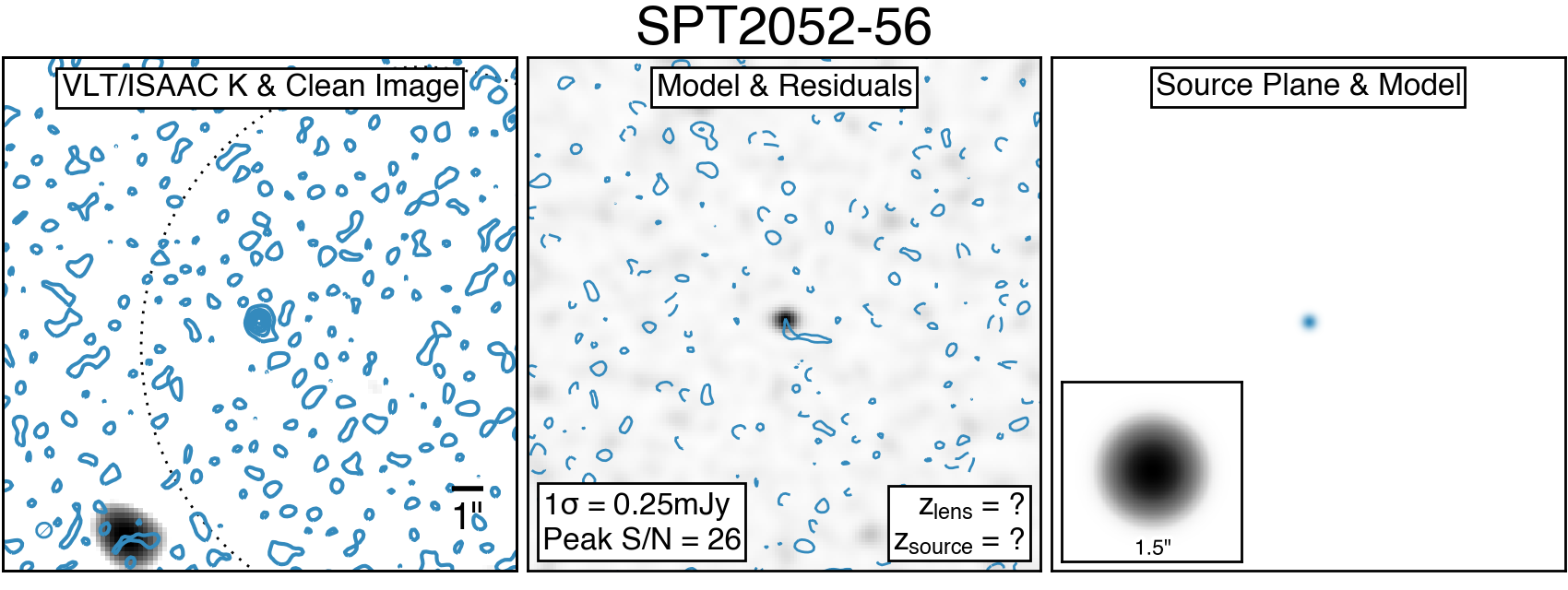}
\includegraphics[width=0.495\textwidth]{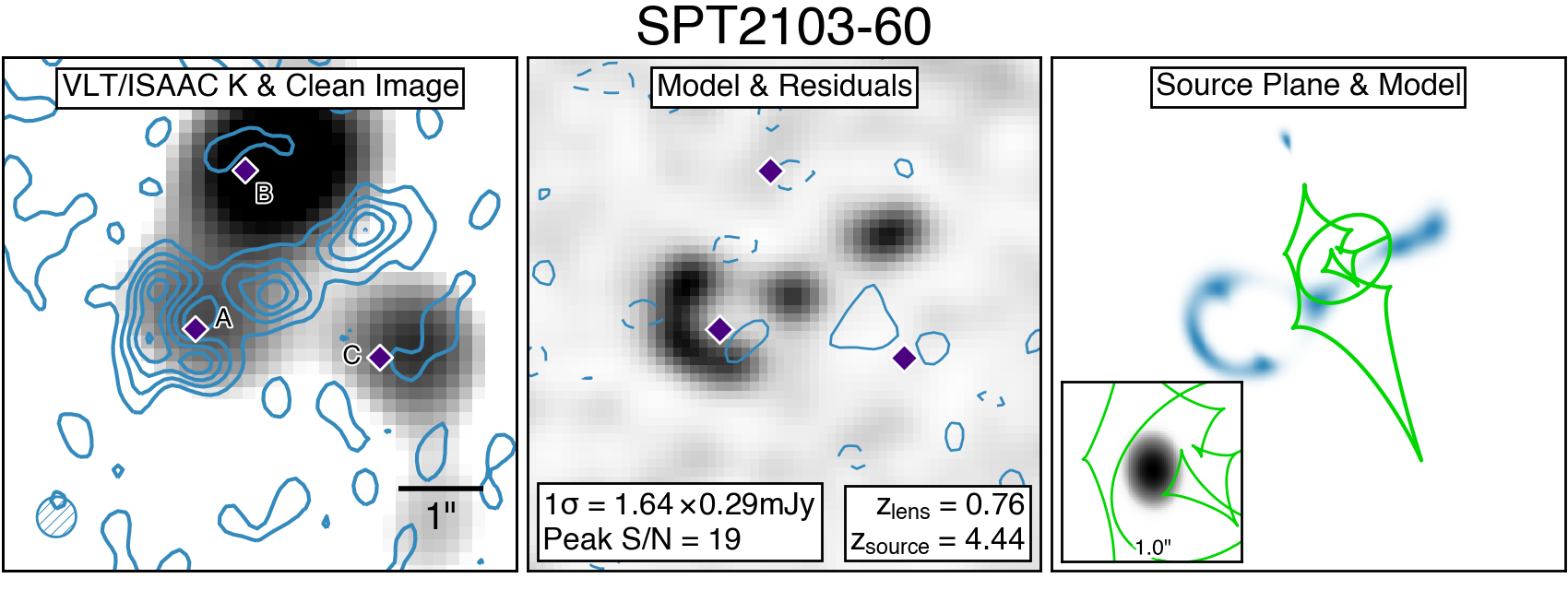}
\includegraphics[width=0.495\textwidth]{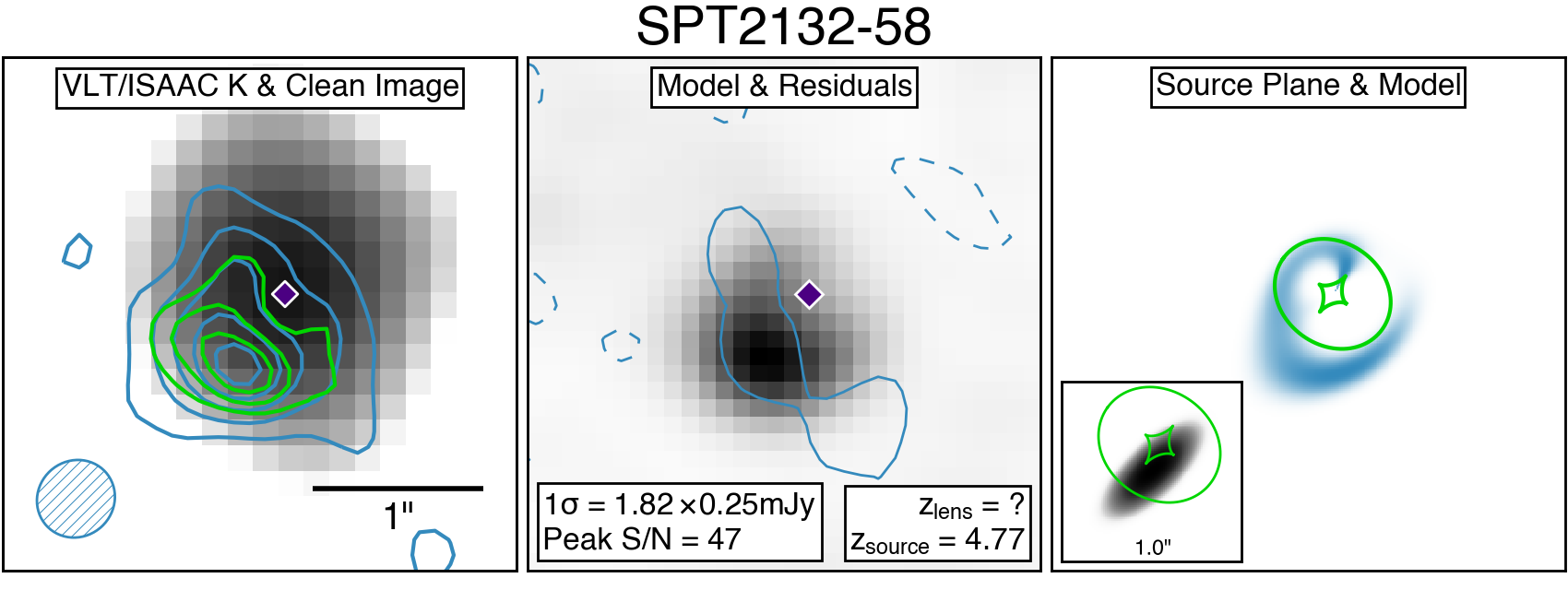}
\includegraphics[width=0.495\textwidth]{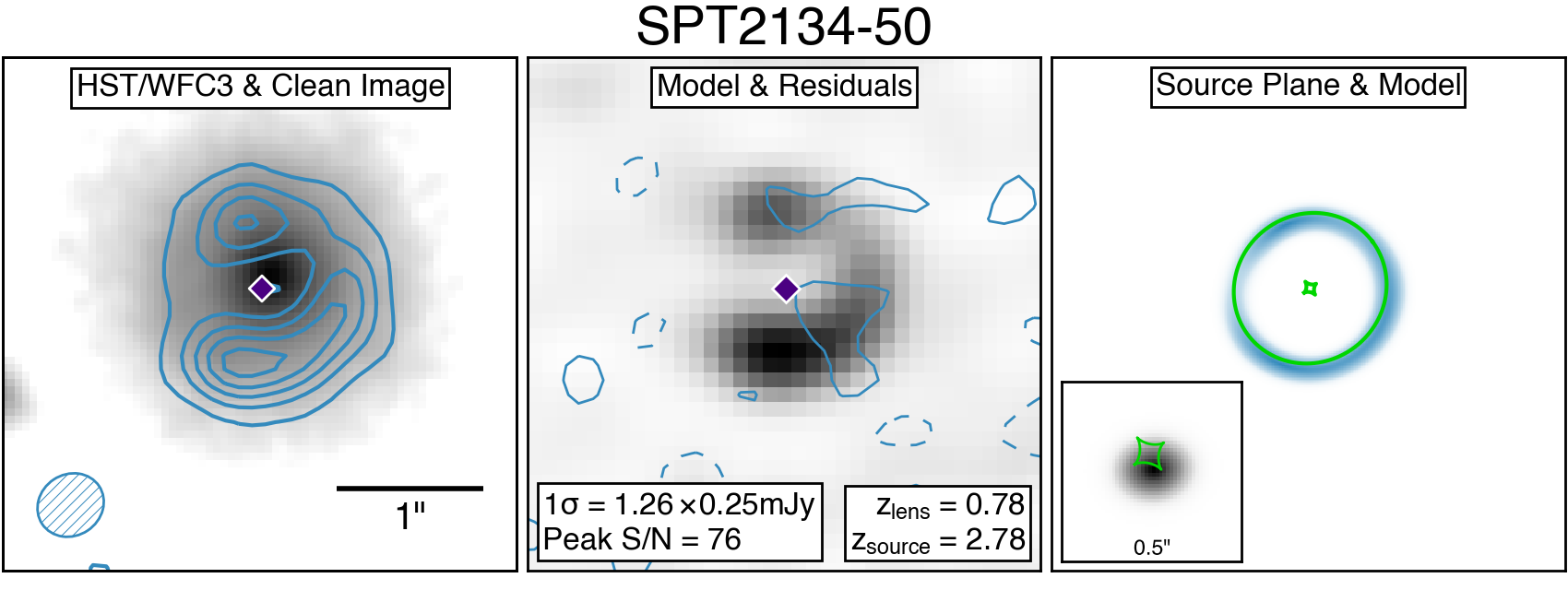}
\includegraphics[width=0.495\textwidth]{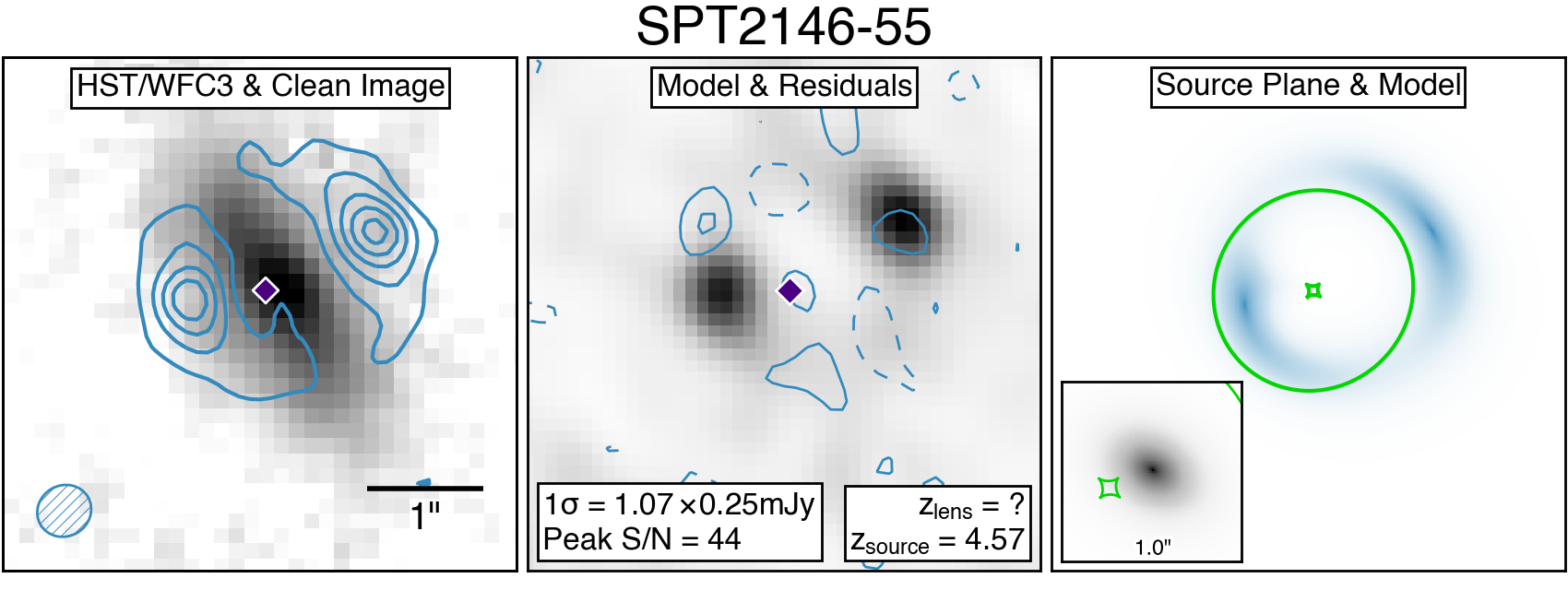}
\includegraphics[width=0.495\textwidth]{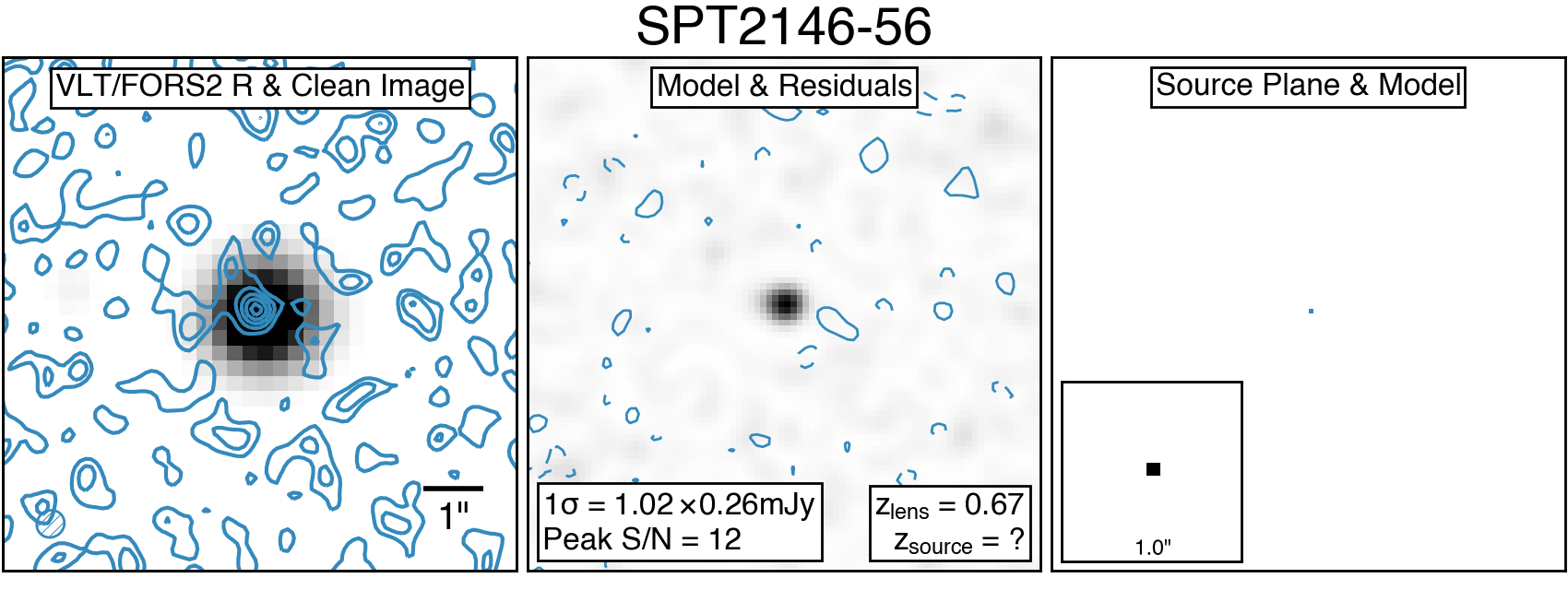}
\includegraphics[width=0.495\textwidth]{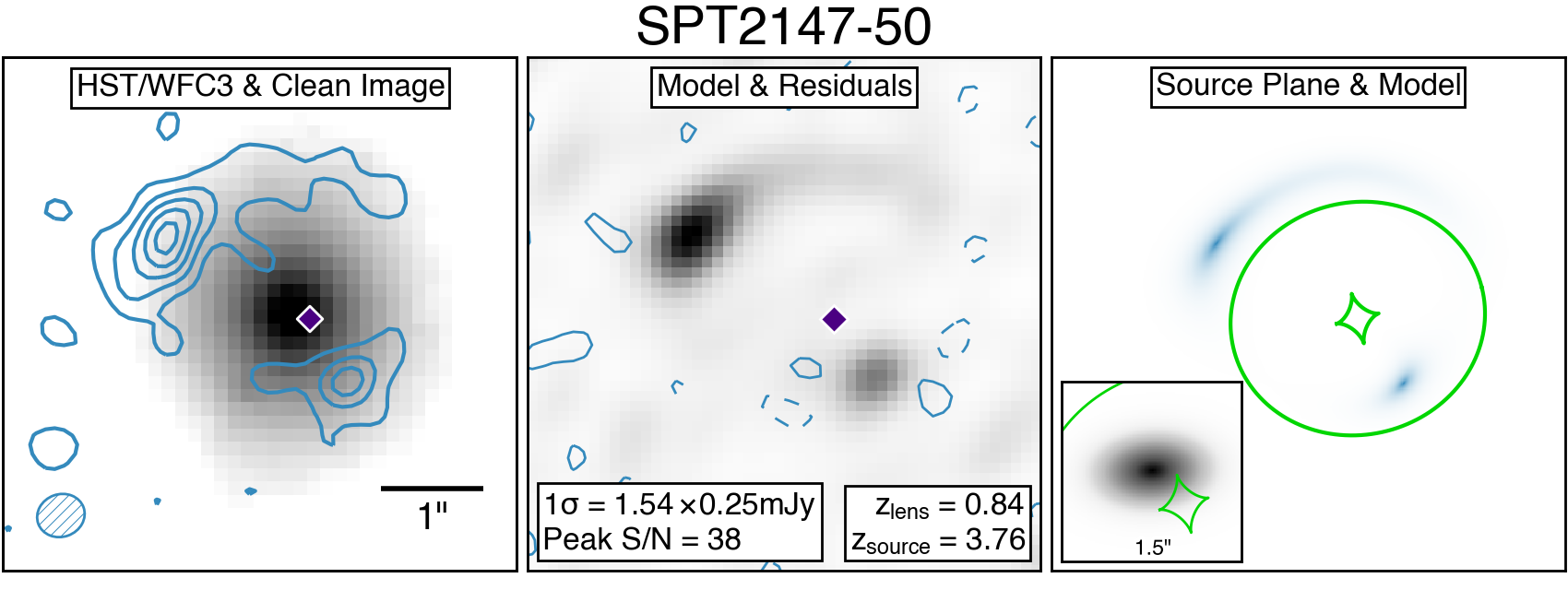}
\includegraphics[width=0.495\textwidth]{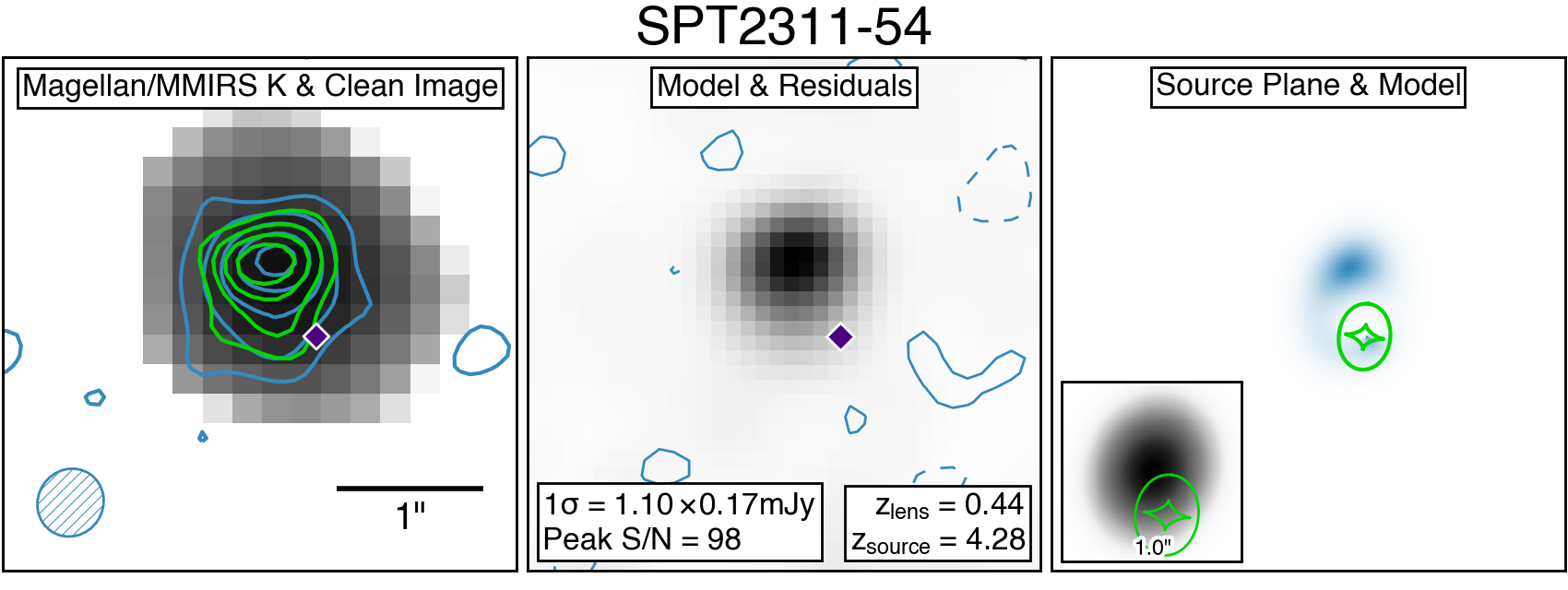}

\end{centering}
\caption{ Continued.  \label{fig:images2}} \addtocounter{figure}{-1}
\end{figure*}

\begin{figure*}[htb]%
\begin{centering}
\includegraphics[width=0.495\textwidth]{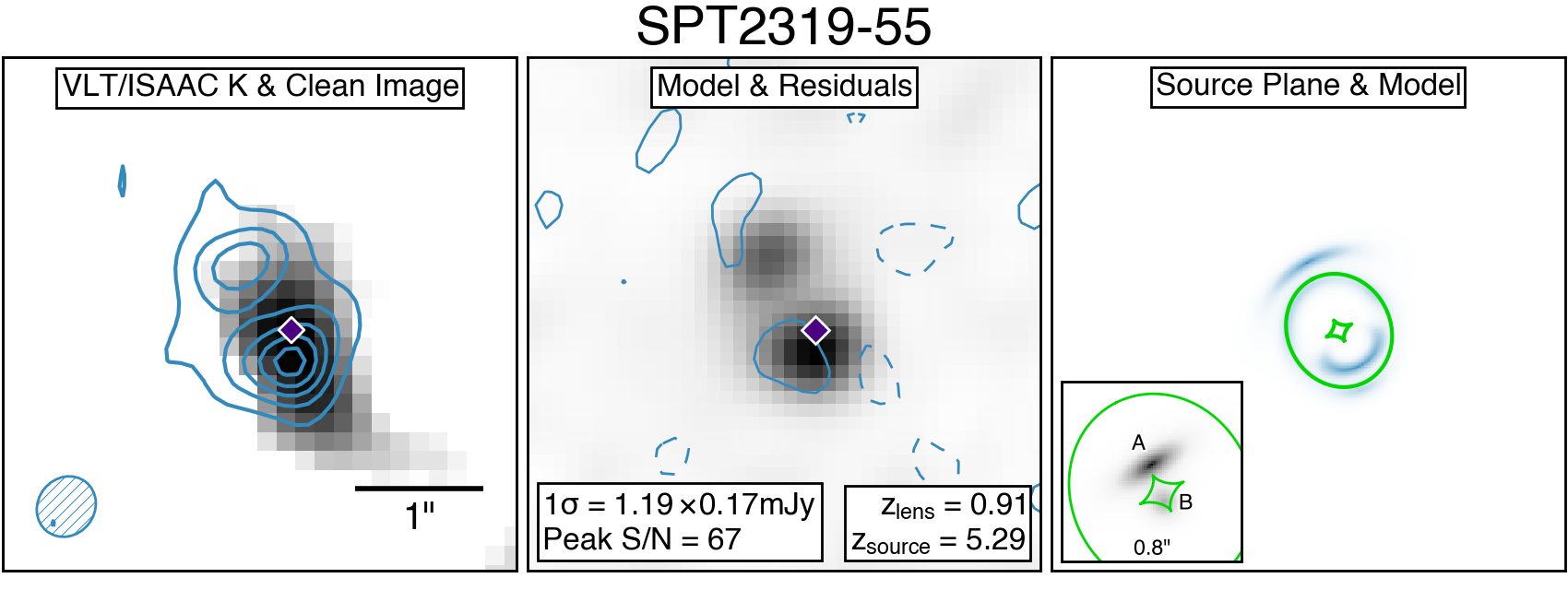}
\includegraphics[width=0.495\textwidth]{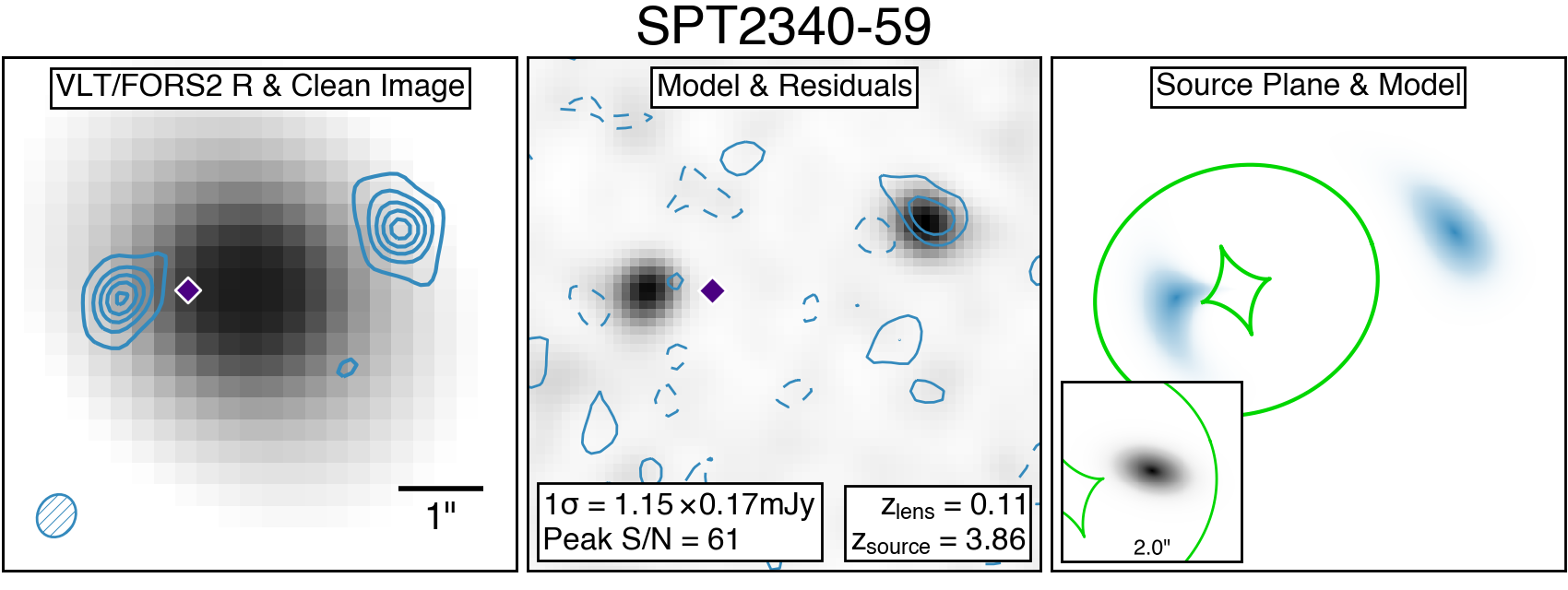}
\includegraphics[width=0.495\textwidth]{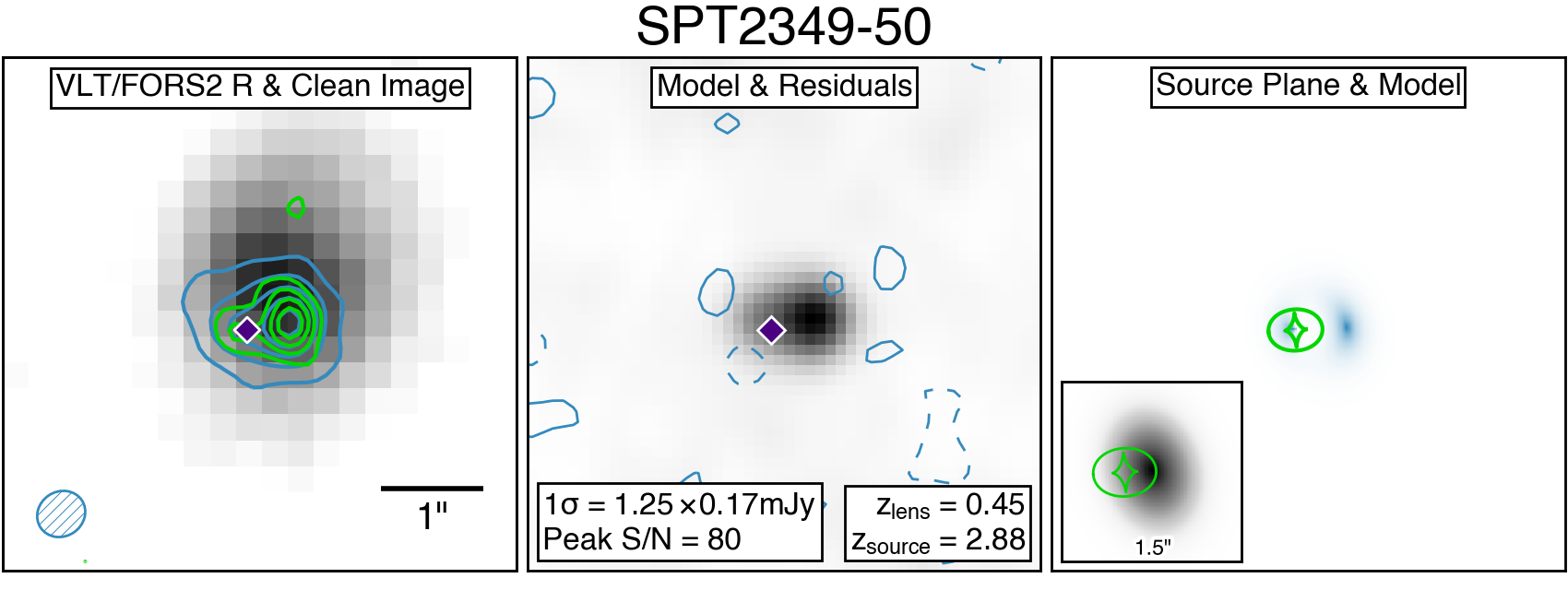}
\includegraphics[width=0.495\textwidth]{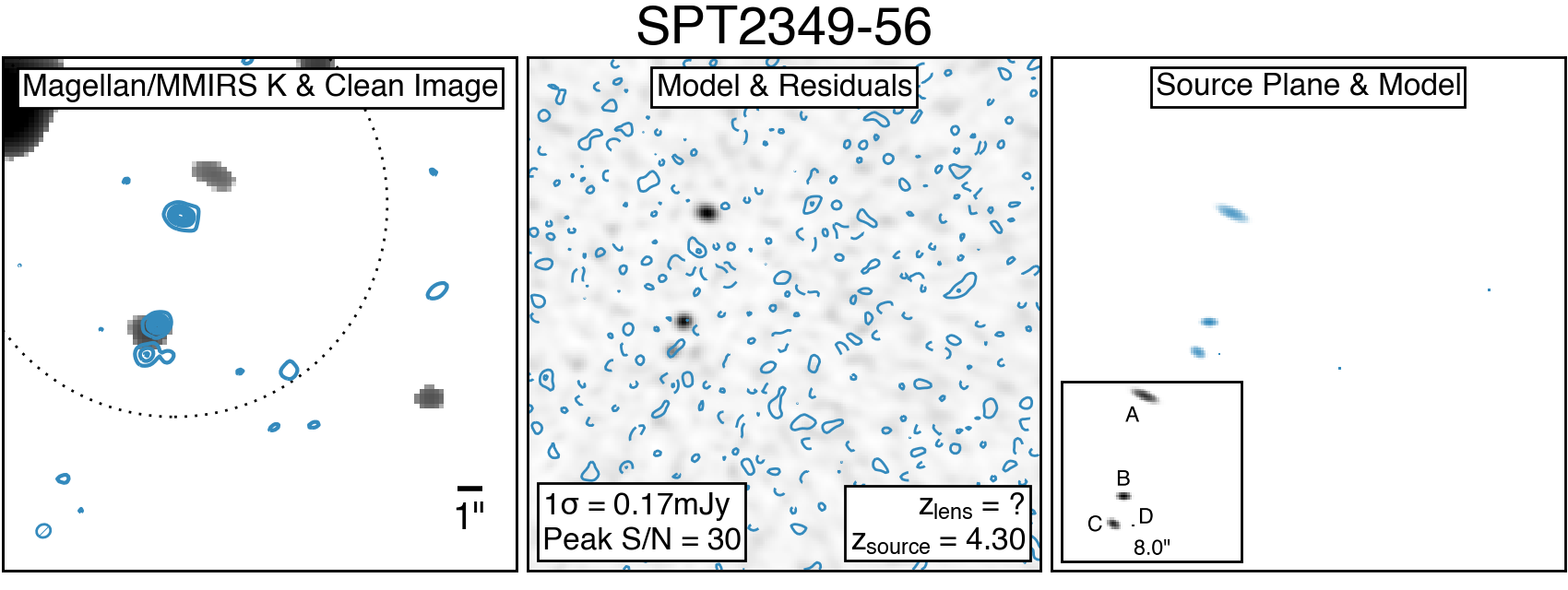}
\includegraphics[width=0.495\textwidth]{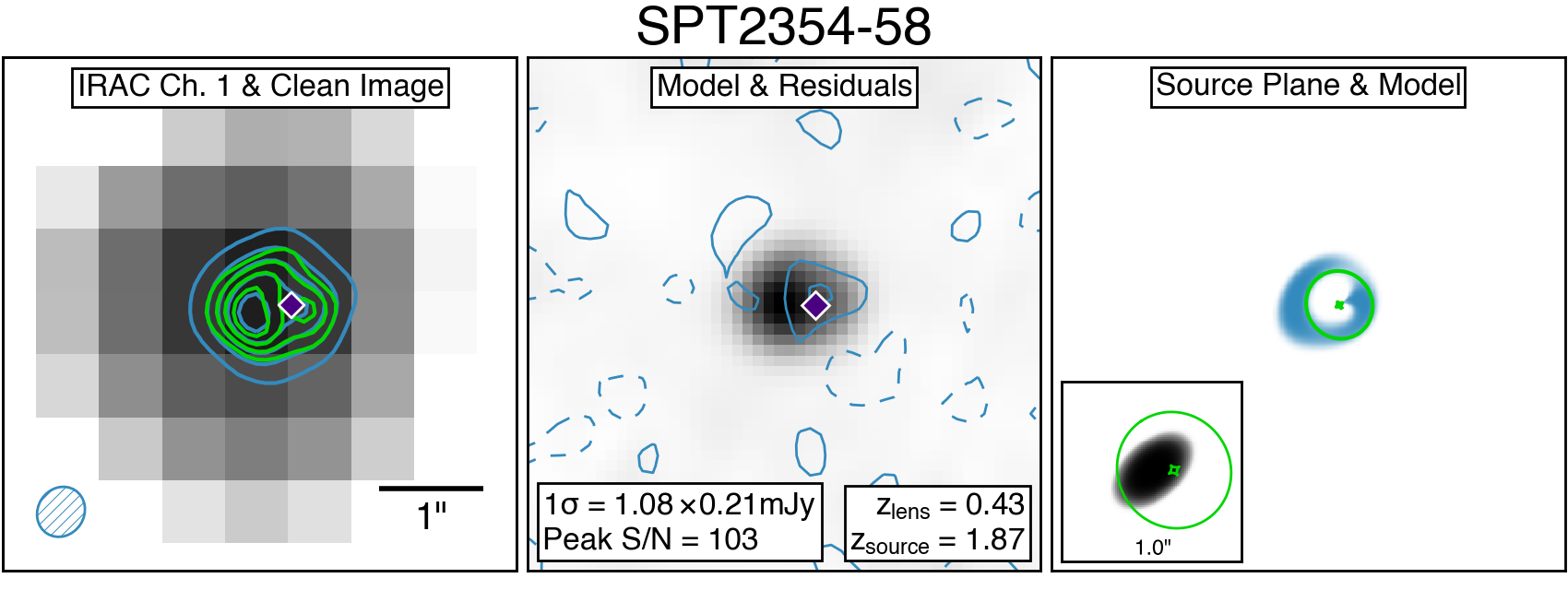}
\includegraphics[width=0.495\textwidth]{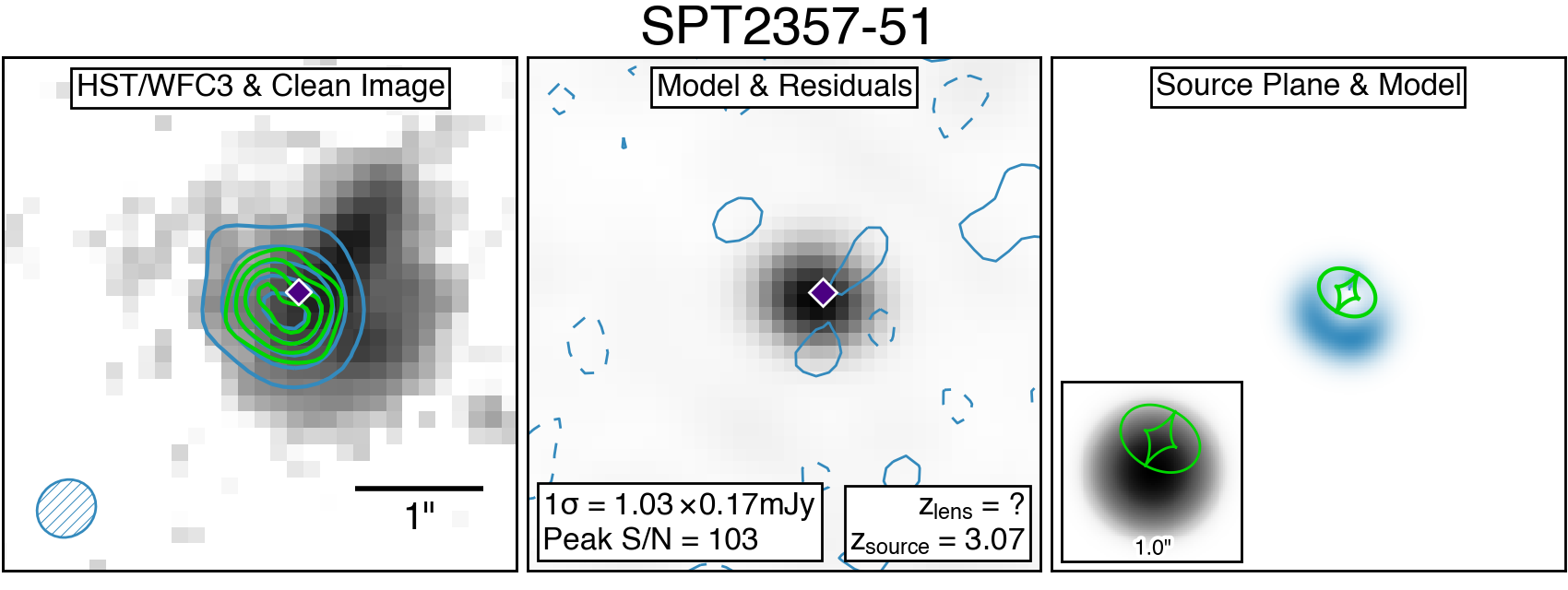}
\end{centering}
\caption{ Continued.  \label{fig:images3}}
\end{figure*}


\begin{deluxetable*}{lccccccc} 
\tablecaption{Modeled Properties of Foreground Gravitational Lenses \label{tab:lenses}} 
\startdata 
\tableline 
\\ 
Source & $x_L$ & $y_L$ & $\theta_{E,L}$ & $e_L$ & $\phi_L$ & $\gamma$ & $\phi_\gamma$ \\ 
 & arcsec & arcsec & arcsec &  & deg. E of N &  & deg. E of N \\ 
SPT0020-51 \hfill A & -1.26 $\pm$ 0.01 & -2.24 $\pm$ 0.01 & 0.614 $\pm$ 0.007 & 0.41 $\pm$ 0.03 & 12 $\pm$ 2 & 0.11 $\pm$ 0.01 & 47 $\pm$ 3\\ 
\hfill B & -0.27 $\pm$ 0.03 & -2.69 $\pm$ 0.01 & 0.171 $\pm$ 0.010 & 0.58 $\pm$ 0.08 & 94 $\pm$ 4 &  & \\ 
SPT0027-50 \hfill A & -2.49 $\pm$ 0.01 & -1.73 $\pm$ 0.01 & 0.638 $\pm$ 0.007 & 0.23 $\pm$ 0.03 & 3 $\pm$ 4 & 0.15 $\pm$ 0.01 & 62 $\pm$ 2\\ 
\hfill B & -3.98 $\pm$ 0.06 & -1.18 $\pm$ 0.07 & 0.316 $\pm$ 0.013 & 0.55 $\pm$ 0.05 & 14 $\pm$ 3 &  & \\ 
\hfill C & -2.48 $\pm$ 0.01 & -0.34 $\pm$ 0.01 & 0.119 $\pm$ 0.004 & 0.36 $\pm$ 0.05 & 77 $\pm$ 6 &  & \\ 
SPT0103-45 & -0.34 $\pm$ 0.01 & -2.44 $\pm$ 0.01 & 0.880 $\pm$ 0.003 & 0.11 $\pm$ 0.01 & 87 $\pm$ 1 & -- & -- \\ 
SPT0109-47 \hfill A & -3.61 $\pm$ 0.13 & -1.70 $\pm$ 0.04 & 1.304 $\pm$ 0.033 & 0.66 $\pm$ 0.06 & 119 $\pm$ 6 & 0.17 $\pm$ 0.01 & 54 $\pm$ 3\\ 
\hfill B & -1.81 $\pm$ 0.07 & -0.73 $\pm$ 0.03 & 0.930 $\pm$ 0.025 & 0.53 $\pm$ 0.03 & 6 $\pm$ 7 &  & \\ 
\hfill C & -7.65 $\pm$ 0.23 & -4.66 $\pm$ 0.18 & 0.839 $\pm$ 0.021 & 0.59 $\pm$ 0.04 & 96 $\pm$ 5 &  & \\ 
SPT0113-46 \hfill A & -0.14 $\pm$ 0.01 & 0.68 $\pm$ 0.01 & 1.157 $\pm$ 0.010 & 0.28 $\pm$ 0.01 & 84 $\pm$ 1 & -- & -- \\ 
\hfill B & -1.11 $\pm$ 0.02 & 1.63 $\pm$ 0.02 & 0.439 $\pm$ 0.012 & 0.08 $\pm$ 0.03 & 151 $\pm$ 8 &  & \\ 
\hfill C & -1.93 $\pm$ 0.10 & 3.07 $\pm$ 0.10 & 0.258 $\pm$ 0.019 & 0.50 $\pm$ 0.05 & 18 $\pm$ 5 &  & \\ 
SPT0125-47 & -1.76 $\pm$ 0.02 & -0.94 $\pm$ 0.01 & 1.011 $\pm$ 0.002 & 0.40 $\pm$ 0.01 & 23 $\pm$ 1 & 0.03 $\pm$ 0.00 & 97 $\pm$ 5\\ 
SPT0125-50 & -0.42 $\pm$ 0.02 & -4.26 $\pm$ 0.02 & 0.984 $\pm$ 0.005 & 0.40 $\pm$ 0.01 & 40 $\pm$ 0 & -- & -- \\ 
SPT0128-51 & -0.45 & 7.06 & 0.750 & 0.00 & 0 & -- & -- \\ 
SPT0202-61 & -0.04 $\pm$ 0.01 & 2.19 $\pm$ 0.02 & 0.758 $\pm$ 0.005 & 0.44 $\pm$ 0.03 & 70 $\pm$ 3 & 0.23 $\pm$ 0.01 & 178 $\pm$ 2\\ 
SPT0243-49 & -2.46 $\pm$ 0.01 & -2.01 $\pm$ 0.01 & 0.327 $\pm$ 0.003 & 0.55 $\pm$ 0.05 & 136 $\pm$ 1 & -- & -- \\ 
SPT0300-46 & 2.00 $\pm$ 0.01 & 1.20 $\pm$ 0.01 & 0.344 $\pm$ 0.008 & 0.53 $\pm$ 0.02 & 66 $\pm$ 1 & -- & -- \\ 
SPT0319-47 & -5.36 $\pm$ 0.01 & -0.77 $\pm$ 0.01 & 0.283 $\pm$ 0.009 & 0.48 $\pm$ 0.08 & 130 $\pm$ 5 & -- & -- \\ 
SPT0345-47 & -2.42 $\pm$ 0.01 & 1.26 $\pm$ 0.01 & 0.306 $\pm$ 0.002 & 0.45 $\pm$ 0.02 & 126 $\pm$ 1 & -- & -- \\ 
SPT0346-52 & -0.81 $\pm$ 0.01 & 3.04 $\pm$ 0.01 & 0.979 $\pm$ 0.007 & 0.52 $\pm$ 0.03 & 71 $\pm$ 1 & 0.12 $\pm$ 0.01 & 122 $\pm$ 3\\ 
SPT0348-62 & 10.90 & 3.10 & 1.002 & 0.00 & 0 & -- & -- \\ 
SPT0403-58 & 1.63 $\pm$ 0.11 & -3.23 $\pm$ 0.11 & 0.533 $\pm$ 0.047 & 0.59 $\pm$ 0.12 & 49 $\pm$ 5 & -- & -- \\ 
SPT0404-59 & -1.07 $\pm$ 0.05 & 9.75 $\pm$ 0.05 & 0.549 $\pm$ 0.027 & 0.49 $\pm$ 0.11 & 88 $\pm$ 7 & -- & -- \\ 
SPT0418-47 & 3.87 $\pm$ 0.01 & -2.71 $\pm$ 0.01 & 1.247 $\pm$ 0.003 & 0.11 $\pm$ 0.01 & 24 $\pm$ 1 & -- & -- \\ 
SPT0441-46 & -0.56 $\pm$ 0.01 & 4.24 $\pm$ 0.01 & 0.678 $\pm$ 0.006 & 0.42 $\pm$ 0.03 & 87 $\pm$ 1 & -- & -- \\ 
SPT0452-50 & 4.29 $\pm$ 0.19 & -3.73 $\pm$ 0.30 & 0.820 $\pm$ 0.140 & 0.27 $\pm$ 0.14 & 173 $\pm$ 13 & -- & -- \\ 
SPT0459-58 & -5.37 $\pm$ 0.03 & 2.64 $\pm$ 0.02 & 0.468 $\pm$ 0.015 & 0.34 $\pm$ 0.09 & 30 $\pm$ 6 & -- & -- \\ 
SPT0459-59 & -2.16 $\pm$ 0.03 & 0.85 $\pm$ 0.04 & 0.627 $\pm$ 0.018 & 0.40 $\pm$ 0.05 & 68 $\pm$ 6 & -- & -- \\ 
SPT0529-54 & -2.35 $\pm$ 0.01 & 0.02 $\pm$ 0.01 & 1.360 $\pm$ 0.008 & 0.17 $\pm$ 0.01 & 91 $\pm$ 3 & -- & -- \\ 
SPT0532-50 & -2.12 $\pm$ 0.02 & 1.84 $\pm$ 0.01 & 0.556 $\pm$ 0.003 & 0.41 $\pm$ 0.04 & 4 $\pm$ 2 & -- & -- \\ 
SPT0538-50 & -0.31 $\pm$ 0.02 & -0.09 $\pm$ 0.02 & 1.728 $\pm$ 0.004 & 0.13 $\pm$ 0.01 & 163 $\pm$ 0 & -- & -- \\ 
SPT2031-51 & -4.92 $\pm$ 0.01 & 1.58 $\pm$ 0.02 & 0.534 $\pm$ 0.005 & 0.45 $\pm$ 0.02 & 16 $\pm$ 1 & -- & -- \\ 
SPT2048-55 & -5.42 $\pm$ 0.01 & 2.47 $\pm$ 0.01 & 0.361 $\pm$ 0.006 & 0.07 $\pm$ 0.03 & 85 $\pm$ 10 & -- & -- \\ 
SPT2103-60 \hfill A & -4.25 $\pm$ 0.01 & 5.53 $\pm$ 0.01 & 0.455 $\pm$ 0.009 & 0.62 $\pm$ 0.02 & 41 $\pm$ 2 & -- & -- \\ 
\hfill B & -4.83 $\pm$ 0.04 & 7.38 $\pm$ 0.06 & 0.791 $\pm$ 0.022 & 0.11 $\pm$ 0.02 & 35 $\pm$ 9 &  & \\ 
\hfill C & -6.40 $\pm$ 0.06 & 5.19 $\pm$ 0.11 & 0.552 $\pm$ 0.020 & 0.81 $\pm$ 0.03 & 15 $\pm$ 3 &  & \\ 
SPT2132-58 & -2.64 $\pm$ 0.02 & 8.12 $\pm$ 0.02 & 0.335 $\pm$ 0.006 & 0.40 $\pm$ 0.03 & 144 $\pm$ 4 & -- & -- \\ 
SPT2134-50 & -4.76 $\pm$ 0.01 & 2.18 $\pm$ 0.01 & 0.518 $\pm$ 0.002 & 0.14 $\pm$ 0.01 & 37 $\pm$ 1 & -- & -- \\ 
SPT2146-55 & -0.50 $\pm$ 0.02 & -2.50 $\pm$ 0.02 & 0.858 $\pm$ 0.004 & 0.12 $\pm$ 0.02 & 48 $\pm$ 3 & -- & -- \\ 
SPT2147-50 & -6.18 $\pm$ 0.02 & 4.21 $\pm$ 0.02 & 1.195 $\pm$ 0.006 & 0.25 $\pm$ 0.02 & 14 $\pm$ 2 & -- & -- \\ 
SPT2311-54 & -2.89 $\pm$ 0.01 & 2.10 $\pm$ 0.02 & 0.209 $\pm$ 0.007 & 0.53 $\pm$ 0.06 & 83 $\pm$ 5 & -- & -- \\ 
SPT2319-55 & -4.74 $\pm$ 0.01 & -0.13 $\pm$ 0.02 & 0.430 $\pm$ 0.003 & 0.29 $\pm$ 0.03 & 117 $\pm$ 2 & -- & -- \\ 
SPT2340-59 & -2.16 $\pm$ 0.07 & -2.47 $\pm$ 0.03 & 1.581 $\pm$ 0.016 & 0.37 $\pm$ 0.06 & 19 $\pm$ 3 & -- & -- \\ 
SPT2349-50 & -4.38 $\pm$ 0.01 & 2.35 $\pm$ 0.02 & 0.244 $\pm$ 0.005 & 0.56 $\pm$ 0.06 & 4 $\pm$ 5 & -- & -- \\ 
SPT2354-58 & -2.50 $\pm$ 0.02 & -1.91 $\pm$ 0.01 & 0.321 $\pm$ 0.003 & 0.14 $\pm$ 0.02 & 124 $\pm$ 5 & -- & -- \\ 
SPT2357-51 & -0.30 $\pm$ 0.03 & -1.58 $\pm$ 0.02 & 0.215 $\pm$ 0.003 & 0.61 $\pm$ 0.03 & 151 $\pm$ 2 & -- & -- \\ 
\enddata 
\tablecomments{Lens positions ($x_L$, $y_L$) are relative to the ALMA phase center, shown for each source in Table~\ref{tab:targets}. Also listed are the lens Einstein radius ($\theta_{E,L}$, ellipticity ($e_L$), and position angle ($\phi_L$). For sources which require an external shear component, the shear strength ($\gamma$) and position angle ($\phi_\gamma$) are also given. Quantities without uncertainties have been fixed to the values shown during fitting.}  
\end{deluxetable*}

\begin{deluxetable*}{lcccccccc} 
\tablecaption{Intrinsic Properties of All Modeled Sources \label{tab:sources}} 
\startdata 
\tableline 
\\ 
Source & $x_S$ & $y_S$ & \ses & \reff & $n_S$ & $b_S/a_S$ & $\phi_S$ & \mues \\ 
 & arcsec & arcsec & mJy & arcsec & &  & deg. E of N & \\ 
SPT0020-51 \hfill A & 0.27 $\pm$ 0.01 & 0.37 $\pm$ 0.01 & 10.94 $\pm$ 0.42 & 0.104 $\pm$ 0.004 & 0.75 $\pm$ 0.06 & 0.51 $\pm$ 0.03 & 69 $\pm$ 3 & 4.2 $\pm$ 0.1 \\ 
\hfill B & 0.08 $\pm$ 0.01 & 0.05 $\pm$ 0.01 & 2.88 $\pm$ 0.13 & 0.043 $\pm$ 0.004 & 0.36 $\pm$ 0.06 & 0.52 $\pm$ 0.10 & 77 $\pm$ 6 & 10.3 $\pm$ 0.4 \\ 
SPT0027-50 \hfill A & -0.24 $\pm$ 0.01 & 0.67 $\pm$ 0.01 & 16.58 $\pm$ 0.56 & 0.142 $\pm$ 0.003 & 0.98 $\pm$ 0.06 & 0.64 $\pm$ 0.03 & 31 $\pm$ 3 & 5.1 $\pm$ 0.2 \\ 
\hfill B & -0.37 $\pm$ 0.01 & 0.37 $\pm$ 0.01 & 3.64 $\pm$ 0.21 & 0.057 $\pm$ 0.003 & 0.85 $\pm$ 0.09 & 0.50 $\pm$ 0.04 & 28 $\pm$ 5 & 11.2 $\pm$ 0.4 \\ 
\hfill C & -1.81 $\pm$ 0.03 & -0.16 $\pm$ 0.02 & 2.98 $\pm$ 0.15 & -- & -- & -- & -- & 1.0 \\ 
SPT0103-45 \hfill A & -0.05 $\pm$ 0.00 & 0.19 $\pm$ 0.01 & 1.06 $\pm$ 0.07 & 0.021 $\pm$ 0.003 & 1.04 $\pm$ 0.15 & 0.69 $\pm$ 0.15 & 128 $\pm$ 12 & 9.3 $\pm$ 0.5 \\ 
\hfill B & -0.41 $\pm$ 0.01 & 0.23 $\pm$ 0.00 & 19.66 $\pm$ 0.48 & 0.261 $\pm$ 0.005 & 0.83 $\pm$ 0.03 & 0.49 $\pm$ 0.01 & 117 $\pm$ 1 & 5.1 $\pm$ 0.1 \\ 
SPT0109-47 \hfill A & 0.14 $\pm$ 0.02 & 0.40 $\pm$ 0.02 & 3.37 $\pm$ 0.19 & 0.045 $\pm$ 0.005 & 0.43 $\pm$ 0.05 & 0.55 $\pm$ 0.10 & 148 $\pm$ 8 & 12.8 $\pm$ 3.7 \\ 
\hfill B & -0.06 $\pm$ 0.01 & 0.16 $\pm$ 0.02 & 3.91 $\pm$ 0.31 & 0.170 $\pm$ 0.023 & 1.23 $\pm$ 0.17 & 0.41 $\pm$ 0.15 & 50 $\pm$ 5 & 10.2 $\pm$ 1.0 \\ 
\hfill C & 0.23 $\pm$ 0.02 & 0.05 $\pm$ 0.01 & 0.21 $\pm$ 0.17 & 0.039 $\pm$ 0.006 & 0.74 $\pm$ 0.14 & 0.60 $\pm$ 0.07 & 34 $\pm$ 15 & 41.8 $\pm$ 17.1 \\ 
SPT0113-46 & -0.39 $\pm$ 0.02 & 0.40 $\pm$ 0.02 & 2.35 $\pm$ 0.06 & 0.075 $\pm$ 0.003 & 0.5 & 0.38 $\pm$ 0.02 & 53 $\pm$ 1 & 23.9 $\pm$ 0.5 \\ 
SPT0125-47 \hfill A & 0.31 $\pm$ 0.01 & 0.24 $\pm$ 0.00 & 21.42 $\pm$ 0.67 & 0.118 $\pm$ 0.003 & 0.21 $\pm$ 0.01 & 0.89 $\pm$ 0.03 & 74 $\pm$ 3 & 5.3 $\pm$ 0.1 \\ 
\hfill B & 0.37 $\pm$ 0.01 & -0.03 $\pm$ 0.01 & 3.91 $\pm$ 0.30 & 0.093 $\pm$ 0.009 & 0.78 $\pm$ 0.13 & 0.41 $\pm$ 0.09 & 48 $\pm$ 4 & 6.0 $\pm$ 0.2 \\ 
\hfill C & -0.18 $\pm$ 0.01 & 0.19 $\pm$ 0.01 & 0.82 $\pm$ 0.10 & 0.078 $\pm$ 0.016 & 0.95 $\pm$ 0.22 & 0.67 $\pm$ 0.13 & 112 $\pm$ 8 & 7.6 $\pm$ 0.7 \\ 
SPT0125-50 \hfill A & 0.13 $\pm$ 0.01 & 0.22 $\pm$ 0.01 & 4.39 $\pm$ 0.22 & 0.036 $\pm$ 0.002 & 0.72 $\pm$ 0.13 & 0.22 $\pm$ 0.02 & 39 $\pm$ 1 & 15.0 $\pm$ 0.5 \\ 
\hfill B & 0.03 $\pm$ 0.01 & 0.07 $\pm$ 0.01 & 1.39 $\pm$ 0.26 & 0.124 $\pm$ 0.031 & 1.5 & 0.49 $\pm$ 0.18 & 83 $\pm$ 13 & 11.7 $\pm$ 0.9 \\ 
SPT0128-51$^{\mathrm{a}}$ \hfill A & 3.30 $\pm$ 0.81 & -5.37 $\pm$ 1.49 & 9.79 $\pm$ 4.65 & 0.139 $\pm$ 0.015 & 0.5 & 1.0 & 0 & 1.1 $\pm$ 0.1 \\ 
\hfill B & 0.28 $\pm$ 0.13 & -5.53 $\pm$ 0.24 & 3.41 $\pm$ 0.38 & 0.122 $\pm$ 0.017 & 0.5 & 1.0 & 0 & 1.1 $\pm$ 0.0 \\ 
SPT0202-61 \hfill A & 0.09 $\pm$ 0.01 & -0.08 $\pm$ 0.01 & 3.29 $\pm$ 0.19 & 0.051 $\pm$ 0.004 & 0.26 $\pm$ 0.07 & 0.23 $\pm$ 0.03 & 130 $\pm$ 2 & 16.2 $\pm$ 0.8 \\ 
\hfill B & -0.03 $\pm$ 0.01 & 0.03 $\pm$ 0.01 & 4.19 $\pm$ 0.41 & 0.222 $\pm$ 0.026 & 1.50 $\pm$ 0.22 & 0.64 $\pm$ 0.09 & 88 $\pm$ 7 & 9.1 $\pm$ 0.7 \\ 
\hfill C & -1.24 $\pm$ 0.03 & -3.88 $\pm$ 0.03 & 4.06 $\pm$ 0.41 & 0.134 $\pm$ 0.029 & 0.5 & 1.0 & 0 & 1.0 \\ 
SPT0243-49 \hfill A & -0.06 $\pm$ 0.01 & -0.06 $\pm$ 0.01 & 6.23 $\pm$ 0.51 & 0.082 $\pm$ 0.005 & 0.5 & 1.0 & 0 & 6.7 $\pm$ 0.5 \\ 
\hfill B & -0.23 $\pm$ 0.01 & 0.21 $\pm$ 0.01 & 5.21 $\pm$ 0.52 & 0.145 $\pm$ 0.010 & 0.5 & 1.0 & 0 & 3.1 $\pm$ 0.1 \\ 
SPT0245-63 \hfill A & -1.21 $\pm$ 0.01 & 5.62 $\pm$ 0.01 & 20.73 $\pm$ 1.74 & 0.123 $\pm$ 0.017 & 0.5 & 0.29 $\pm$ 0.06 & 168 $\pm$ 3 & 1.0 \\ 
\hfill B & -0.85 $\pm$ 0.04 & 5.58 $\pm$ 0.03 & 19.28 $\pm$ 2.07 & 0.387 $\pm$ 0.022 & 0.5 & 0.67 $\pm$ 0.08 & 161 $\pm$ 9 & 1.0 \\ 
SPT0300-46 \hfill A & -0.10 $\pm$ 0.01 & -0.03 $\pm$ 0.01 & 4.07 $\pm$ 0.66 & 0.049 $\pm$ 0.007 & 0.5 & 1.0 & 0 & 9.0 $\pm$ 0.8 \\ 
\hfill B & -0.08 $\pm$ 0.03 & 0.16 $\pm$ 0.02 & 4.85 $\pm$ 1.10 & 0.143 $\pm$ 0.018 & 0.5 & 1.0 & 0 & 4.0 $\pm$ 0.3 \\ 
\hfill C & -7.68 $\pm$ 0.11 & 0.14 $\pm$ 0.06 & 9.65 $\pm$ 2.17 & 0.488 $\pm$ 0.132 & 0.5 & 1.0 & 0 & 1.0 \\ 
SPT0319-47 & 0.23 $\pm$ 0.01 & 0.09 $\pm$ 0.01 & 19.12 $\pm$ 1.93 & 0.162 $\pm$ 0.011 & 0.22 $\pm$ 0.02 & 0.74 $\pm$ 0.06 & 90 $\pm$ 6 & 2.9 $\pm$ 0.3 \\ 
SPT0345-47 & 0.02 $\pm$ 0.00 & -0.06 $\pm$ 0.01 & 9.71 $\pm$ 0.61 & 0.083 $\pm$ 0.005 & 0.5 & 1.0 & 0 & 7.9 $\pm$ 0.5 \\ 
SPT0346-52 & -0.22 $\pm$ 0.00 & 0.27 $\pm$ 0.01 & 19.64 $\pm$ 0.46 & 0.101 $\pm$ 0.003 & 0.81 $\pm$ 0.10 & 0.71 $\pm$ 0.03 & 121 $\pm$ 4 & 5.6 $\pm$ 0.1 \\ 
SPT0348-62$^{\mathrm{a}}$ \hfill A & -6.11 $\pm$ 0.02 & 1.22 $\pm$ 0.02 & 17.33 $\pm$ 1.11 & 0.173 $\pm$ 0.013 & 0.5 & 0.85 $\pm$ 0.09 & 58 $\pm$ 21 & 1.2 $\pm$ 0.0 \\ 
\hfill B & -3.48 $\pm$ 0.04 & 1.75 $\pm$ 0.04 & 4.58 $\pm$ 0.49 & 0.141 $\pm$ 0.029 & 0.5 & 0.55 $\pm$ 0.11 & 7 $\pm$ 27 & 1.3 $\pm$ 0.0 \\ 
\hfill C & -5.16 $\pm$ 0.09 & 1.65 $\pm$ 0.06 & 1.81 $\pm$ 0.29 & 0.080 $\pm$ 0.019 & 0.5 & 0.65 $\pm$ 0.11 & 72 $\pm$ 10 & 1.2 $\pm$ 0.0 \\ 
\hfill D & -5.88 $\pm$ 0.21 & 1.36 $\pm$ 0.10 & 3.38 $\pm$ 0.60 & 0.180 $\pm$ 0.030 & 0.5 & 0.53 $\pm$ 0.09 & 148 $\pm$ 45 & 1.2 $\pm$ 0.0 \\ 
SPT0403-58 \hfill A & 1.11 $\pm$ 0.10 & 0.82 $\pm$ 0.09 & 24.14 $\pm$ 1.48 & 0.486 $\pm$ 0.020 & 0.5 & 1.0 & 0 & 1.6 $\pm$ 0.1 \\ 
\hfill B & 1.16 $\pm$ 0.10 & 0.90 $\pm$ 0.09 & 3.64 $\pm$ 0.47 & 0.051 $\pm$ 0.022 & 0.5 & 1.0 & 0 & 1.7 $\pm$ 0.2 \\ 
SPT0404-59 & 0.02 $\pm$ 0.03 & -0.29 $\pm$ 0.03 & 2.87 $\pm$ 0.57 & 0.126 $\pm$ 0.029 & 0.5 & 1.0 & 0 & 4.1 $\pm$ 0.6 \\ 
SPT0418-47 & 0.01 $\pm$ 0.00 & 0.05 $\pm$ 0.01 & 2.58 $\pm$ 0.23 & 0.092 $\pm$ 0.008 & 0.74 $\pm$ 0.18 & 0.86 $\pm$ 0.03 & 134 $\pm$ 10 & 32.7 $\pm$ 2.7 \\ 
SPT0441-46 & -0.01 $\pm$ 0.00 & 0.16 $\pm$ 0.01 & 6.88 $\pm$ 0.57 & 0.076 $\pm$ 0.007 & 2.30 $\pm$ 0.32 & 1.0 & 0 & 12.7 $\pm$ 1.0 \\ 
SPT0452-50 & -0.63 $\pm$ 0.11 & 1.22 $\pm$ 0.21 & 30.74 $\pm$ 2.32 & 0.265 $\pm$ 0.015 & 1.50 $\pm$ 0.22 & 1.0 & 0 & 1.7 $\pm$ 0.1 \\ 
SPT0459-58 & -0.15 $\pm$ 0.02 & 0.08 $\pm$ 0.02 & 11.82 $\pm$ 1.51 & 0.217 $\pm$ 0.023 & 0.5 & 1.0 & 0 & 5.0 $\pm$ 0.6 \\ 
SPT0459-59 \hfill A & -0.08 $\pm$ 0.02 & 0.29 $\pm$ 0.03 & 11.26 $\pm$ 1.11 & 0.323 $\pm$ 0.023 & 0.5 & 0.60 $\pm$ 0.07 & 171 $\pm$ 6 & 4.2 $\pm$ 0.4 \\ 
\hfill B & -1.17 $\pm$ 0.06 & -0.98 $\pm$ 0.07 & 3.22 $\pm$ 0.41 & 0.205 $\pm$ 0.048 & 0.5 & 1.0 & 0 & 1.5 $\pm$ 0.1 \\ 
SPT0529-54 & -0.02 $\pm$ 0.01 & 0.03 $\pm$ 0.01 & 7.02 $\pm$ 0.58 & 0.248 $\pm$ 0.019 & 0.96 $\pm$ 0.21 & 0.22 $\pm$ 0.02 & 122 $\pm$ 1 & 13.2 $\pm$ 0.8 \\ 
SPT0532-50 & -0.05 $\pm$ 0.01 & 0.00 $\pm$ 0.00 & 13.23 $\pm$ 0.85 & 0.134 $\pm$ 0.007 & 0.35 $\pm$ 0.12 & 0.90 $\pm$ 0.05 & 102 $\pm$ 16 & 10.0 $\pm$ 0.6 \\ 
SPT0538-50 \hfill A & 0.08 $\pm$ 0.01 & 0.15 $\pm$ 0.02 & 3.77 $\pm$ 0.43 & 0.060 $\pm$ 0.006 & 0.5 & 1.0 & 0 & 18.8 $\pm$ 2.3 \\ 
\hfill B & 0.15 $\pm$ 0.02 & -0.00 $\pm$ 0.02 & 1.50 $\pm$ 0.20 & 0.152 $\pm$ 0.016 & 0.5 & 1.0 & 0 & 23.4 $\pm$ 1.9 \\ 
SPT2031-51 & 0.10 $\pm$ 0.01 & -0.25 $\pm$ 0.01 & 15.33 $\pm$ 0.78 & 0.170 $\pm$ 0.008 & 1.44 $\pm$ 0.17 & 0.88 $\pm$ 0.05 & 99 $\pm$ 12 & 3.9 $\pm$ 0.2 \\ 
SPT2048-55 & 0.05 $\pm$ 0.01 & 0.09 $\pm$ 0.01 & 9.27 $\pm$ 1.06 & 0.111 $\pm$ 0.010 & 1.71 $\pm$ 0.32 & 0.96 $\pm$ 0.05 & 125 $\pm$ 19 & 6.3 $\pm$ 0.7 \\ 
SPT2052-56 & 4.99 $\pm$ 0.03 & 0.73 $\pm$ 0.02 & 10.92 $\pm$ 0.67 & 0.183 $\pm$ 0.011 & 0.5 & 1.0 & 0 & 1.0 \\ 
SPT2103-60 & -0.86 $\pm$ 0.02 & 0.80 $\pm$ 0.02 & 2.29 $\pm$ 0.16 & 0.073 $\pm$ 0.004 & 0.5 & 0.78 $\pm$ 0.08 & 4 $\pm$ 13 & 27.8 $\pm$ 1.8 \\ 
SPT2132-58 & 0.05 $\pm$ 0.01 & -0.15 $\pm$ 0.01 & 8.72 $\pm$ 0.90 & 0.095 $\pm$ 0.009 & 0.39 $\pm$ 0.08 & 0.40 $\pm$ 0.06 & 134 $\pm$ 3 & 5.7 $\pm$ 0.5 \\ 
SPT2134-50 & -0.01 $\pm$ 0.00 & -0.05 $\pm$ 0.01 & 4.31 $\pm$ 0.50 & 0.033 $\pm$ 0.004 & 0.95 $\pm$ 0.14 & 0.82 $\pm$ 0.12 & 95 $\pm$ 23 & 21.0 $\pm$ 2.4 \\ 
SPT2146-55 & -0.23 $\pm$ 0.01 & 0.09 $\pm$ 0.00 & 8.34 $\pm$ 0.61 & 0.136 $\pm$ 0.010 & 2.29 $\pm$ 0.28 & 0.75 $\pm$ 0.05 & 59 $\pm$ 6 & 6.6 $\pm$ 0.4 \\ 
SPT2146-56 & 0.94 $\pm$ 0.05 & 1.03 $\pm$ 0.03 & 2.80 $\pm$ 0.30 & -- & -- & -- & -- & 1.0 \\ 
SPT2147-50 & 0.27 $\pm$ 0.01 & 0.27 $\pm$ 0.02 & 11.00 $\pm$ 0.78 & 0.145 $\pm$ 0.008 & 1.17 $\pm$ 0.18 & 0.63 $\pm$ 0.04 & 93 $\pm$ 5 & 6.6 $\pm$ 0.4 \\ 
SPT2311-54 & 0.09 $\pm$ 0.01 & 0.24 $\pm$ 0.01 & 21.58 $\pm$ 1.04 & 0.143 $\pm$ 0.005 & 0.61 $\pm$ 0.12 & 0.84 $\pm$ 0.04 & 160 $\pm$ 8 & 1.9 $\pm$ 0.1 \\ 
SPT2319-55 \hfill A & 0.05 $\pm$ 0.00 & 0.12 $\pm$ 0.01 & 4.38 $\pm$ 0.49 & 0.043 $\pm$ 0.007 & 2.21 $\pm$ 0.30 & 0.40 $\pm$ 0.13 & 122 $\pm$ 7 & 6.9 $\pm$ 0.6 \\ 
\hfill B & -0.00 $\pm$ 0.01 & -0.05 $\pm$ 0.01 & 0.72 $\pm$ 0.15 & 0.059 $\pm$ 0.016 & 0.90 $\pm$ 0.32 & 1.0 & 0 & 13.9 $\pm$ 1.8 \\ 
SPT2340-59 & -0.92 $\pm$ 0.07 & 0.21 $\pm$ 0.02 & 10.90 $\pm$ 0.95 & 0.121 $\pm$ 0.006 & 1.26 $\pm$ 0.17 & 0.58 $\pm$ 0.04 & 76 $\pm$ 5 & 3.4 $\pm$ 0.3 \\ 
SPT2349-50 & -0.22 $\pm$ 0.01 & 0.00 $\pm$ 0.01 & 21.01 $\pm$ 0.99 & 0.159 $\pm$ 0.006 & 1.02 $\pm$ 0.14 & 0.76 $\pm$ 0.05 & 16 $\pm$ 6 & 2.1 $\pm$ 0.1 \\ 
SPT2349-56 \hfill A & -0.33 $\pm$ 0.01 & -0.46 $\pm$ 0.01 & 8.76 $\pm$ 0.27 & 0.176 $\pm$ 0.013 & 0.5 & 0.36 $\pm$ 0.06 & 67 $\pm$ 3 & 1.0 \\ 
\hfill B & 0.61 $\pm$ 0.01 & -4.91 $\pm$ 0.01 & 7.85 $\pm$ 0.26 & 0.103 $\pm$ 0.011 & 0.5 & 0.53 $\pm$ 0.07 & 88 $\pm$ 10 & 1.0 \\ 
\hfill C & 1.06 $\pm$ 0.02 & -6.13 $\pm$ 0.02 & 4.60 $\pm$ 0.27 & 0.114 $\pm$ 0.019 & 0.5 & 0.65 $\pm$ 0.08 & 60 $\pm$ 22 & 1.0 \\ 
\hfill D & 0.21 $\pm$ 0.02 & -6.19 $\pm$ 0.03 & 1.53 $\pm$ 0.16 & -- & -- & -- & -- & 1.0 \\ 
\hfill E & -4.73 $\pm$ 0.05 & -6.83 $\pm$ 0.05 & 1.67 $\pm$ 0.22 & -- & -- & -- & -- & 1.0 \\ 
\hfill F & -10.83 $\pm$ 0.23 & -3.59 $\pm$ 0.18 & 1.59 $\pm$ 0.27 & -- & -- & -- & -- & 1.0 \\ 
SPT2354-58 & 0.13 $\pm$ 0.01 & -0.01 $\pm$ 0.00 & 10.23 $\pm$ 0.61 & 0.100 $\pm$ 0.004 & 0.22 $\pm$ 0.02 & 0.61 $\pm$ 0.03 & 133 $\pm$ 4 & 6.3 $\pm$ 0.4 \\ 
SPT2357-51 & 0.05 $\pm$ 0.02 & -0.18 $\pm$ 0.01 & 13.80 $\pm$ 0.66 & 0.152 $\pm$ 0.004 & 0.5 & 1.0 & 0 & 2.9 $\pm$ 0.1 \\ 
\enddata 
\tablecomments{For lensed sources ($\mues > 1$), source positions are relative to the first lens listed for each source in Table~\ref{tab:lenses}. For unlensed sources, positions are relative to the ALMA phase center given in Table~\ref{tab:targets}. Quantities without uncertainties have been fixed during fitting to the values listed. Sources without a listed size are unresolved (pointlike) in the ALMA data.}  
\tablenotetext{a}{Parameters derived under the assumption of fixed lens Einstein radius; 
see Appendix~\ref{app:notes}.} 
\end{deluxetable*}

\subsection{Basic Lens Model Properties} \label{lensstats}

As expected, a large fraction of the 47 fields observed by ALMA are consistent
with strongly lensed systems -- for 38 sources (81\%), strong gravitational
lensing is the most plausible explanation for the ALMA emission. Of these,
4 sources (11\% of the strongly lensed sources) appear to be lensed by large 
groups or clusters of galaxies.  An additional 8 sources (17\%) appear to
be unlensed or weakly lensed.  Of these sources, 
2 are co-located ($<0.5$\arc)
with objects also detected in the optical or near-infrared but do not appear
to be lensed, two more are within 3\arc of optical/NIR 
counterparts and are likely either weakly lensed background sources or
unlensed sources with undetected optical counterparts, while
the remaining 4 sources do not appear to be closely associated with any
objects detected in the best-available optical/NIR imaging.  The final source,
SPT2300-51, was undetected by ALMA at $>5\sigma$ significance within
the ALMA primary beam half-power radius and was determined to
be a spurious detection in the LABOCA follow-up of SPT sources; this source is
shown in Appendix~\ref{app:notes}.  

Figure~\ref{fig:lensstats} summarizes some key parameters of the lens models. The
left panel shows the distribution of Einstein radii for the strongly lensed
sources, where we have added the Einstein radii of systems with multiple lenses
in quadrature. We find a median Einstein radius of 0.64\arc, with the distribution
rising until approximately the half-resolution radius of our data.  This may indicate
that higher resolution observations will reveal that some of the sources which
are unresolved in the current data may also be gravitationally lensed,
because the multiple images of strongly lensed sources are generally
separated by $\sim$2 Einstein radii. A similar median Einstein radius of $\sim0.6$\arc
was found for the \textit{Herschel}-selected sample of \citet{bussmann13}. This may
indicate that the two surveys probe a similar population of lens galaxies,
in spite of the difference in background source properties (e.g., 
Fig.~\ref{fig:selection}). We defer a more thorough discussion of the lens
galaxies to a future work.

The center
panel of Fig.~\ref{fig:lensstats} shows the distribution of \mues for the SPT
sources, with a median magnification of 5.5 for all sources, or 6.3 for the
strongly lensed subset alone.  This is somewhat higher than the median 
magnification of 4.6 found by \citet{bussmann13} in a study of 
\textit{Herschel}-selected objects.  The magnification distribution for the SPT
sources also appears to contain a tail to higher magnifications compared to the
\textit{Herschel} sample;
for approximately 30\% of the strongly lensed
sources, the best-fit magnifications are $\mues>10$. 

The fraction of strongly lensed sources is expected to vary with the
flux density threshold used to create the source catalogs.  Lower flux density 
limits will include a higher proportion of unlensed sources.  
Equivalently, the median magnification of an
observed sample of objects is a function of the flux density threshold. The right
panel of Fig.~\ref{fig:lensstats} illustrates this effect: on average, apparently
brighter sources are magnified more highly.
This effect is also apparent in the brighter
\textit{Herschel} sample studied by \citet{bussmann13}, in which at least 21 of
30 sources are strongly lensed, compared to a fainter sample described in
\citet{bussmann15}, in which only 6 of 29 sources are strongly lensed.  This 
difference is likely due to the shape of the submillimeter number counts, which
drop steeply for sources with intrinsic $\ses\gtrsim8.5$\,mJy \citep{karim13,simpson15b}.

\begin{figure*}[thb]%
\centering
\includegraphics[width=\textwidth]{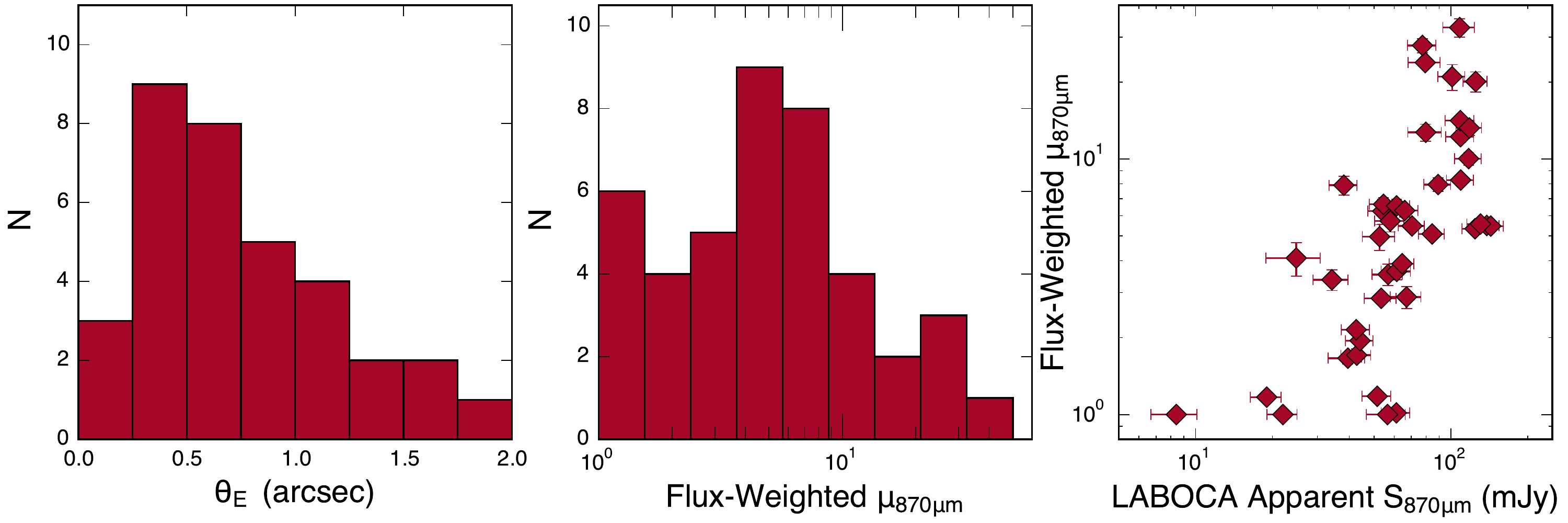}%
\caption{
\textit{Left:} Distribution of Einstein radii for the strongly lensed SPT
sources. For objects with multiple lenses, the Einstein radii of the individual
lens galaxies have been added in quadrature.
\textit{Middle:} Distribution of \mues for all modeled sources. For sources with
multiple components, the flux density-weighted mean magnification is shown.
\textit{Right:} Source magnification as a function of apparent LABOCA flux
density. }
\label{fig:lensstats}
\end{figure*}

\subsection{Flux Recovery} \label{fluxcomp}

Every source targeted was detected, with the exception of SPT2300-51
(this source was determined to be false after it was included in the ALMA sample).
Each source had previously been observed at 
the same frequency using LABOCA on APEX, a single-dish bolometer camera with the same
primary beam size as the ALMA data. By comparing
the 870\,\um flux density measured by LABOCA to that recovered in the ALMA data,
we can test whether significant flux has been resolved out by ALMA due to
limited coverage of the $uv$-plane.  This could occur if the sources have 
structure extended on scales greater than the largest scale recoverable by
the data, or if additional sources are present in the maps which are too faint
to have been detected individually or are outside the primary beam.

Almost all of the sources in our sample are significantly resolved in the ALMA data. To
estimate the total flux density present in the ALMA maps, we first image the
data using a taper in the $uv$-plane at 50\,k$\lambda$, corresponding to a
resolution of $\gtrsim4$''.  This ensures that we measure a value as close
as possible to the true single-dish ``zero-spacing'' flux density. We then
CLEAN the images to a 3$\sigma$ threshold and correct for the 
response of the primary beam. The total
ALMA flux is then defined as the sum of the CLEAN components, in order to avoid
the need to define an aperture over which to measure the total flux density.  If our
sources were unresolved on scales $<50$\,k$\lambda$, this would be equivalent
to reporting the maximum pixel value in the images; in practice, many of our
sources still show some structure on these scales. 

\begin{figure*}[thb]%
\centering
\includegraphics[width=\textwidth]{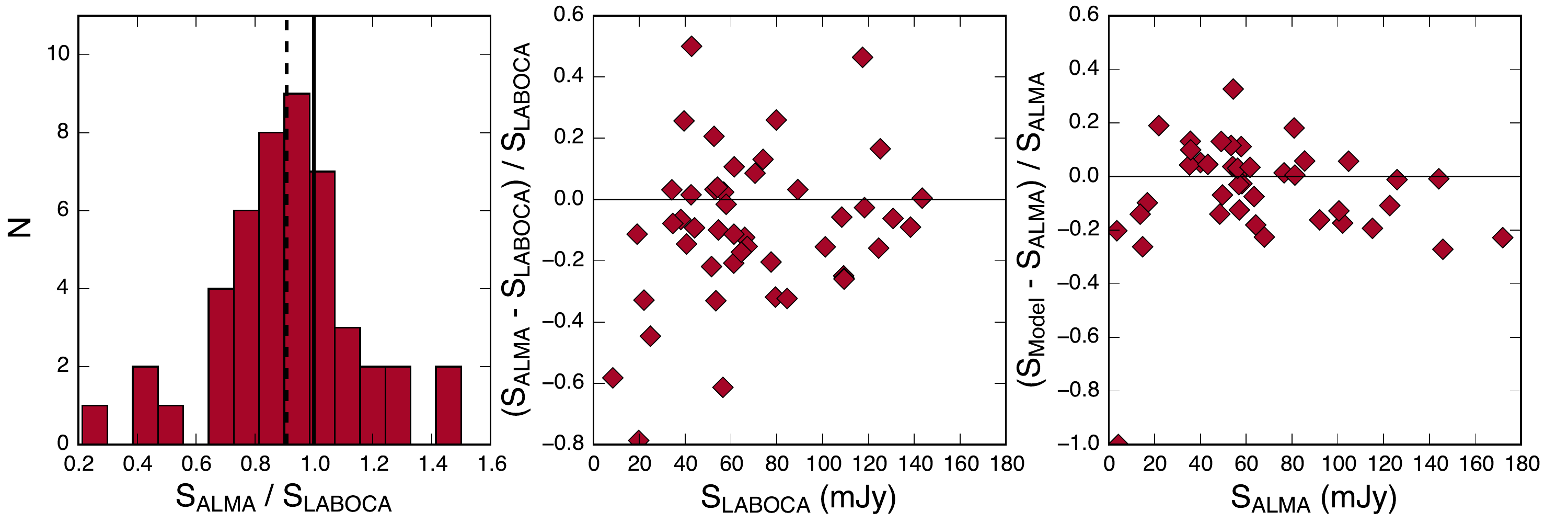}%
\caption{
Extent to which flux densities derived from ALMA, LABOCA, and the lens models
agree; see Sec.~\ref{fluxcomp} for details. 
\textit{Left:} The ALMA data recover a median of 91\% of the single-dish
flux density measured by LABOCA, indicated by the dashed line. 
\textit{Middle:} No clear trend is seen in the
fraction of flux detected by ALMA as a function of LABOCA flux density.
\textit{Right:} The lens models of all sources contain a median of 102\% of
the total ALMA flux density.}
\label{fig:fluxcomp}
\end{figure*}

In the left panel of Fig.~\ref{fig:fluxcomp}, we compare the total flux
densities of the ALMA sources determined in this way to the LABOCA measurements
\citep{weiss13}.  Note that we have made no effort
to correct for the different bandwidths of the two instruments (8 vs. $\sim$60\,GHz).
We recover a median of ($91\pm24$)\% of the LABOCA flux density, consistent within the 
mutual absolute flux scaling uncertainties ($\sim$10\% for both instruments).  
Meanwhile, the middle panel of
Fig.~\ref{fig:fluxcomp} shows no clear trend in the fraction of flux recovered
as a function of LABOCA flux density.  These plots suggest that, in general, 
the ALMA data do not resolve out significant extended emission or hide a large
population of sources too faint to detect individually. \citet{hodge13} reached
a similar conclusion using ALMA to image a large sample of unlensed
870\,\um--selected sources discovered by LABOCA in the Extended \textit{Chandra} Deep
Field-South. The sample of SPT DSFGs observed in this work shows a better
degree of consistency between the ALMA and LABOCA flux densities, which may be
due to the fact that the SPT-selected sources are apparently brighter. Indeed, 
the brightest sources studied by Hodge \etal correspond to the faintest sources
in the present sample.

We also test the extent to which the total ALMA flux densities agree
with the total flux densities inferred from the lens models. In this case,
we define the total model flux density as the sum over all components of
$\ses \times \mues$.  This is shown in the right panel of Fig.~\ref{fig:fluxcomp}.
The models contain a median of 102\% of the total ALMA flux densities, indicating
that no significant sources of emission remain unaccounted for by the models.
As the residual maps generated by the best-fit models shown in Fig.~\ref{fig:images}
show no significant remaining peaks, this is unsurprising.

\subsection{Multiplicity in the SPT Sample} \label{multiplicity}

Several high-resolution ALMA follow-up studies of submillimeter sources originally
detected in low-resolution single-dish surveys have concluded that a significant
fraction of the sources break up into multiple components when observed at
higher resolution.  In the ALESS program, \citet{hodge13} find that at least 35\% of their
sources contain multiple components, but that these components are consistent
with being distributed randomly on the sky.  In contrast, \citet{bussmann15} report
a multiplicity fraction of 69\%, with the multiple sources strongly concentrated
at separations $\lesssim3$\arc.  Similarly, \citet{simpson15b} report that 61\% of
SCUBA-2 sources contain multiple components.

Our ability to determine multiplicity fractions from the follow-up of SPT-selected
sources is hampered by two potential issues.  First, the large majority of the
sources considered here are strongly lensed. This makes finding close-in multiple
components difficult, as any faint nearby companions will be overwhelmed by the
much brighter lensed emission.  Second, the SPT sources have a much higher
apparent brightness compared to the unlensed single-dish sources observed in other
follow-up campaigns.  This reduces our ability to detect faint sources, as
low-level phase errors can create spurious ``companions.'' 
For this reason, we use a higher (5$\sigma$, with $\siges \sim 0.18-0.5$\,mJy) 
threshold for source detection than
other source catalogs.  We also refrain from counting sources which require
multiple source-plane components to reproduce the lensed emission as multiples,
because these components are generally separated by $<0.5$\arc, and the source-plane
components are likely an approximation of complex underlying structure within a
single galaxy.

In the ALMA data presented here, only 13\% (6/47) of sources contain multiple 
components at $>5\sigma$ significance.  
This fraction is significantly lower than the high multiplicity
rates reported by other ALMA follow-up programs. While obviously dependent on the 
depth of the follow-up
observations, the high reported multiplicity fractions in other programs 
come from data with roughly comparable depth and resolution to the ALMA data 
presented here (ALESS detection threshold $\sim1.1-2.1$\,mJy, compared to 
$\sim0.9-2.5$\,mJy here).  After considering our lack of sensitivity to close-in
sources and a higher source detection threshold, the ALESS sample is the most
natural comparison sample -- the higher detection threshold in our data is
balanced by the increased depth of our observations, and both samples are 
insensitive to multiples at separations of $\lesssim1.5$\arc.  While the
overall multiplicity fraction does appear to be lower in the SPT sample,
the few multiples in our data are consistent with being uniformly distributed
within the fields, as in the ALESS data.

\section{Discussion} \label{discussion}
We are now in a position
to take advantage of the comprehensive followup programs we have been conducting
to revisit a number of topics of interest which may be investigated further using
our new knowledge of source sizes.

\subsection{Reliability of Lens Models} \label{reliability}

For the four sources studied by \citet{hezaveh13} using low-resolution ($\sim$1.5\arc)
data, we find generally good agreement with the updated models. The differences
between the previous and updated models can be entirely explained by the
difference in background source parameterization -- that is, fitting only the
data used by Hezaveh \etal with elliptical source-plane components recovers
the models presented here, while fitting all of the data used in this work with
the circularly symmetric Gaussian components assumed by Hezaveh \etal recovers the
models shown there. This indicates that the model uncertainties on
the properties of the background sources are dominated by systematic, rather
than statistical, uncertainty.  We have attempted to counter this issue by use
of the DIC for model selection, which effectively penalizes models with more
degrees of freedom unless they reproduce the data
significantly better.

ALMA is now capable of resolutions as fine as a few tens of
milliarcseconds. To what extent can we expect that the model properties (e.g., \mues)
derived here would agree with the properties
derived from observations with the $\sim$20$\times$
better resolution now possible? Given that the true source structure of DSFGs
is expected to be clumpy and irregular \citep[e.g.,][]{swinbank11,dye15}, 
in contrast to the smooth source
parameterization assumed here, this question is difficult to answer. This irregular
structure means that different regions of a given source will be magnified by
different amounts, as opposed to the magnifications derived here, which are
averaged over the assumed-elliptical source profile.

One instructive comparison comes via the ALMA Long Baseline Campaign
observations of the lensed DSFG SDP.81 at $\sim$0.023\arc
resolution \citep{almapartner15}.
This source was also included in the sample studied by \citet{bussmann13}, who
used $\sim$0.5--1\arc SMA observations to construct the lens models,
comparable to the resolution of the ALMA data used in this work. Note,
however, that the SMA observations reached only a peak signal-to-noise of 12
for SDP.81, far less than the typical significance of our detections (median
peak signal-to-noise of 62). The lens model, which also represented the 
background source as an elliptical S\'{e}rsic profile, as in this work,
yielded a magnification
of $\mues \sim 11$. Several authors have constructed lens models of the 
continuum emission using the high-resolution ALMA observations of SDP.81,
finding magnifications $\mues \sim 16-22$ using pixelated
source-plane reconstructions \citep{rybak15,dye15,tamura15}. It is difficult 
to know whether these $\sim$50\% variations are to be generally expected, 
or whether the differences arise chiefly from data resolution, data
signal-to-noise, or modeling approach. In at least this single
case, however, shrinking the beam area by $\gtrsim$100$\times$ 
leads to less than factor of two changes in source-averaged magnification.

\subsection{Size Distribution of Background Sources} \label{sourcesizes}

Gravitational lensing allows us to study the background sources at effective
resolutions higher than the instrumental resolution of our observations. It is
worth considering, however, the biases which may be present when comparing lensed
and unlensed samples.  For example, numerous authors have explored a potential
size bias of lensed samples \citep[e.g.,][]{serjeant12,hezaveh12a,wardlow13}, in which 
sources with high magnification are preferentially smaller than sources with
lower magnification factors.  This effect is due to the angular extent of the
background source in comparison to the relatively small region near caustics
over which high magnification is possible -- small sources near caustics can
experience a higher net magnification compared to more extended sources.  Different
regions of a given background source experience different magnifications, depending
on the lensing geometry, an effect known as differential magnification.

\begin{figure*}[htb]%
\centering
\includegraphics[width=\textwidth]{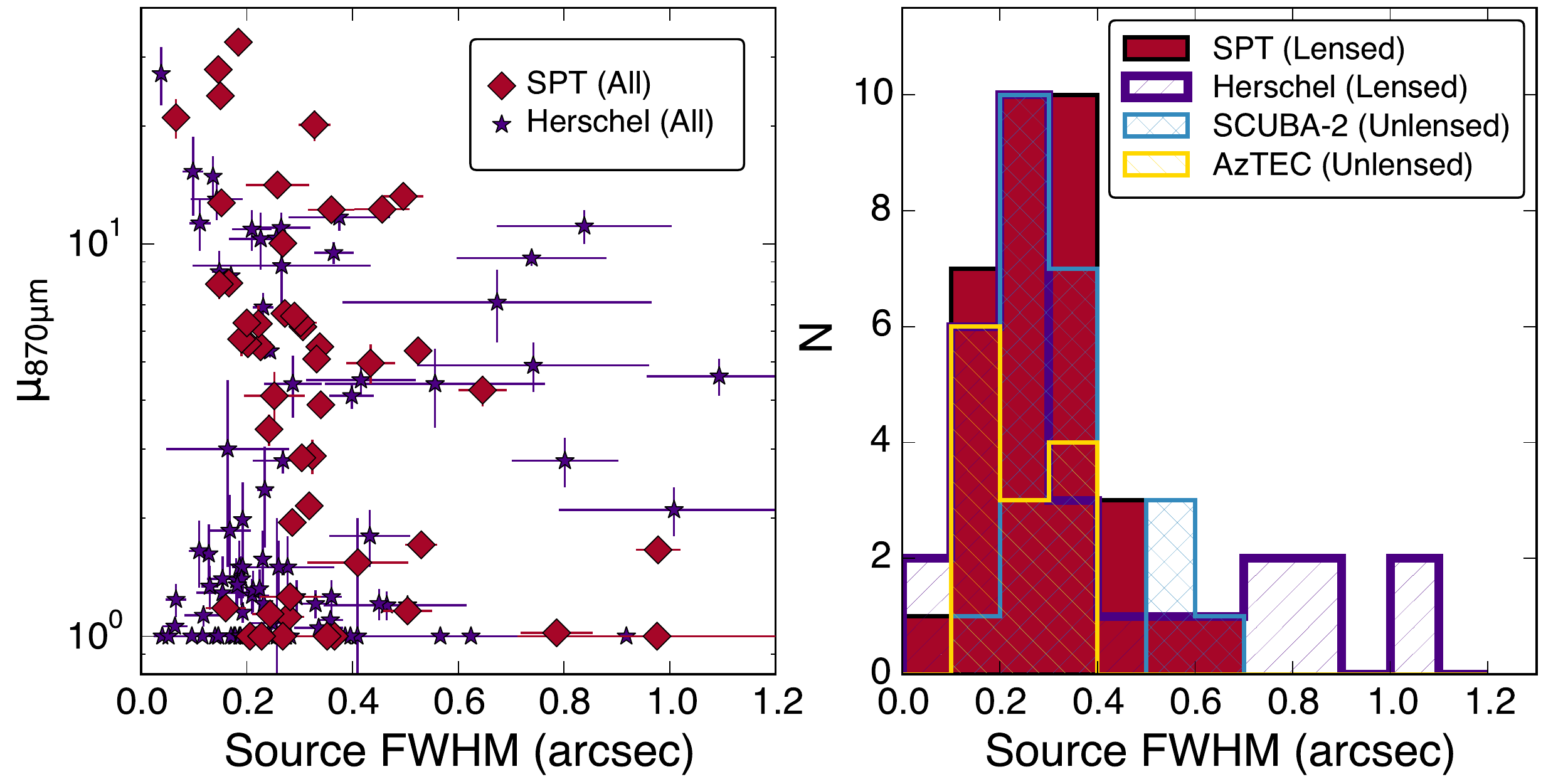}%
\caption{
\textit{Left:} Intrinsic source size plotted as a function of 
870\,\um magnification, 
for all sources in the SPT and \textit{Herschel} DSFG samples. Sources with the 
highest magnifications are preferentially more compact than the full sample.
\textit{Right:} Size distributions of strongly lensed ($\mues>2$) SPT and
\textit{Herschel} sources \citep{bussmann13,bussmann15}, 
compared to samples of unlensed DSFGs observed by
ALMA from the 850\,\um-selected SCUBA-2 Cosmology Legacy Survey 
\citep{simpson15b} and AzTEC 1.1\,mm-selected sources
\citep{ikarashi15}. For all samples, we plot the circularized FWHM; Simpson
\etal report only source major axes (priv. comm.), so we have circularized their 
measurements assuming an average axis ratio of 0.8. This figure indicates that
the lensed samples recover the same size distribution as the unlensed samples,
despite the potential size bias shown in the left panel.
}
\label{fig:sourcesizes}
\end{figure*}

We explore this effect in the left panel of Fig.~\ref{fig:sourcesizes}.  Here,
we show the source magnification as a function of its size for both the SPT sample
and the \textit{Herschel}-selected samples of \citet{bussmann13,bussmann15}. Many
of the lens models reproduce the complex background source morphology by invoking
multiple S\'{e}rsic components.  These components are likely to be physically
associated, so we show the total flux-weighted magnification and the total source
area of all related components (so the two components of, e.g., SPT0103-45 are shown
as a single point, while the two components of, e.g., SPT0128-51 are shown
separately).  We find a median intrinsic FWHM of 0.28''.
This figure shows no clear correlation between the two
parameters. However, in agreement with the size bias mentioned, it does appear
that the sources with the highest magnifications are preferentially smaller
than sources with lower magnifications.  In other words, small size appears
to be a necessary but not sufficient criterion for the highest magnifications.

A separate but related question is whether selecting strongly lensed sources
results in a biased measurement of the true size distribution of DSFGs 
\citep[e.g.,][]{hezaveh12a}.  Even
though high-magnification sources are preferentially compact, the size distribution
of lensed samples is not necessarily biased, depending on the true underlying
brightness and size distributions (for example, if the true size distribution
were a delta function, no bias would exist).  The presence of such a bias can
be investigated by comparing size distributions measured from lensed and unlensed
samples. In the right panel of Fig.~\ref{fig:sourcesizes},
we compare the size distribution measured from the strongly lensed ($\mues>2$) sources
in the SPT and \textit{Herschel} samples with two unlensed DSFG samples. 
\citet{simpson15b} measure sizes of 22 sources based on 870\,\um ALMA imaging of 
objects selected from the 850\,\um SCUBA-2 Cosmology Legacy Survey \citep{geach13}. 
Only one source was unresolved by these data, with a FWHM$\lesssim0.18$'',
although the sample is restricted to sources with $\ses \sim 5-12$\,mJy to 
ensure sufficient signal-to-noise to measure an accurate source size.
\citet{ikarashi15} report 1.1\,mm sizes from ALMA observations of 13 AzTEC 
1.1\,mm-selected objects spanning $\sopo \sim 1.2-3.5$\,mJy.  Assuming
a dust emissivity index $\beta = 2$, this corresponds to $\ses \sim 3-9$\,mJy.
Even after accounting for the gravitational magnification, the SPT sources are 
typically brighter than many of the unlensed comparison sources, although no
significant correlations between source flux density and size are seen in either
unlensed sample or our own.
Both unlensed
samples have sizes measured from the dust continuum emission, eliminating possible
confusion in comparing to sizes measured with alternative methods (e.g., from the
radio continuum; \citealt{biggs11}).

Given the present sample sizes, both samples of 
strongly lensed sources have size distributions consistent with the distribution
of unlensed sources.  The two-sample K--S test confirms that we cannot reject
the hypothesis that both distributions are drawn from the same parent
distribution ($p=0.84$).
Few of the unlensed sources have robust spectroscopic 
redshifts, which hinders our ability to infer whether the consistent angular
size distributions correspond to differing physical size distributions.  As
detailed in \citet{bethermin15b} and \citet{strandet16}, we expect more
sources at higher redshifts in the SPT sample due to its long selection wavelength
and preferential selection of lensed sources.  We note, however, that the 
angular size scale evolves slowly for $z>2$; the difference in the size
scale between the SPT median redshift and the median redshift of the unlensed
DSFGs of \citealt{chapman05} is $<$15\%.

The lensed samples appear to recover the ``true'' unlensed size distribution 
in spite of the bias discussed above.  This seems to indicate
one of two possibilities.  First, it may be that neither the lensed 
nor unlensed samples are
sufficiently complete for differences to be noticeable. The lensed samples
effectively select sources based on the product of intrinsic flux density and
magnification, while the unlensed samples would not measure the true size
distribution if faint sources are preferentially more extended, precluding size
measurements from the current ALMA data.  Alternatively, the underlying DSFG 
size distribution may lack
sufficient dynamic range for the size bias to become noticeable without a very
large number of sources.  The true size distribution may have few objects
at both very small and very large sizes, making the magnification bias unimportant.
Both scenarios are testable from deeper
observations of a larger sample of unlensed sources.

\subsection{Constraining the Dust Opacity} \label{dusttau}

The size information we have determined affords us additional constraints on
other fitted parameters which would be difficult to determine from unresolved
observations. 
One of the most common fitting functions used to describe the dust emission of 
galaxies is the ``modified blackbody'' function,
\begin{equation} \label{mbb}
S_{\nu_r} = \frac{\Omega_{\mathrm{source}}}{(1+z_S)^3} (B_{\nu_r}(\Tdust) - 
B_{\nu_r}(T_{\mathrm{CMB}})) (1 - e^{-\tau_{\nu_r}})
\end{equation}
where $B_{\nu_r}(T)$ is the Planck function evaluated at rest-frame frequency 
$\nu_r$ and temperature
$T$.  This blackbody is ``modified'' by the dust optical depth term, and the overall
normalization of the SED is related to the intrinsic source solid angle
$\Omega_{\mathrm{source}} = \pi \reff^2 / D_A^2$. 
At long wavelengths,
the dust optical depth can be parameterized as a power-law in frequency
\citep[e.g.,][]{draine06}, with
$\tau_\nu = (\nu/\nu_0)^\beta = (\lambda_0/\lambda)^\beta$, and the 
optical depth reaching unity at wavelength \Lz.
The value of $\beta$ governs the slope of the Rayleigh-Jeans tail of the dust emission,
while the combination of \Tdust and \Lz governs the peak wavelength and width 
of the peak of the dust emission.  The value of $\beta$ is generally in the range
1.5--2, while the value of \Lz is commonly assumed to be 100--200\,\um
(3--1.5\,THz) \citep[e.g.,][]{blain03,casey14}.  

For sources without size measurements, the source solid angle is unknown, in addition
to the other parameters which control the shape of the dust SED.  Even with sizes
derived from the lens models, we are forced to assume a single dust temperature
and value of \Lz averaged over the source for each object. Improvements on this
scenario require spatially resolved continuum measurements at several
widely spaced frequencies, especially those which straddle the SED peak. While this
may one day be possible, at present we assume that the source emission is uniform,
mirroring the assumptions which must be made with unresolved photometry.

The spatially unresolved long-wavelength SED alone 
is usually insufficient to constrain the 
value of \Lz, as degeneracies with the other parameters (particularly the
dust temperature \Tdust) allow for good matches to the data for a wide range
of \Lz.  The inferred \Tdust, in turn, has a large effect on other
inferred quantities, such as the total dust mass \citep[e.g.,][]{casey12}.  Our
new knowledge of the intrinsic size of the SPT DSFGs offers an alternative avenue
for constraining an effective \Lz.  For those sources with spectroscopic redshifts, we fit
the photometry at rest wavelengths $> 50$\,\um with the modified blackbody function
given above, assuming $\beta = 2$
and allowing \Lz to be a free parameter, although allowing $\beta$ as a free
parameter does not alter our results. The cutoff
at short wavelengths is used because neither the modified blackbody function nor
our lens models are expected to capture the emission from hot dust which dominates
the short-wavelength side of the SED.  This assumption ignores any possible
contribution of a hot dust component to the long wavelength photometry, but 
\textit{Herschel}/PACS photometry indicates that this component is negligible
at the relevant wavelengths (Strandet \etal, in prep.). We have verified that 
neither a hot dust component nor a short-wavelength power-law significantly
affect our conclusions.

We perform the fitting described using the source photometry in 
\citet{weiss13,strandet16}, and an MCMC fitting routine.  The free
parameters are the SED normalization (which stands in for the source solid angle
at wavelengths without size measurements; see below), \Tdust, and \Lz. At each MCMC step, 
we calculate the log-likelihood of producing the spatially unresolved continuum
measurements given the proposed
combination of parameters, and add to this the log-likelihood of the proposed
\Tdust and \Lz reproducing the intrinsic source flux density determined from
the lens models, after marginalizing over the uncertainty in source size.  The reason
the contributions from the spatially unresolved and resolved measurements
must be calculated 
separately is that, as shown in
Fig.~\ref{fig:fluxcomp}, there is a median 10\% offset and large scatter 
between the total flux density
measured in our (resolved) ALMA images compared to the (unresolved) LABOCA images
at the same wavelength; presumably this scatter would also be present if we 
resolved the sources at all other wavelengths.  In order to avoid biases
introduced by this scatter, we use the exact form in Eq.~\ref{mbb} at
observed-frame 870\,\um only, and allow a normalization at other wavelengths.
This normalization effectively allows us to match flux captured by the large
single-dish beams (primarily from galaxies associated with the foreground lensing
haloes; \citealt{welikala16}) not included in the lens modeling, as well as general
measurement and calibration errors.
We have verified that this method does not give unphysical results, and that the
inclusion of the lens model sizes merely shrinks the allowable parameter space
without driving the solutions to otherwise unfavored values.

\begin{figure*}[htb]%
\centering
\includegraphics[width=0.95\textwidth]{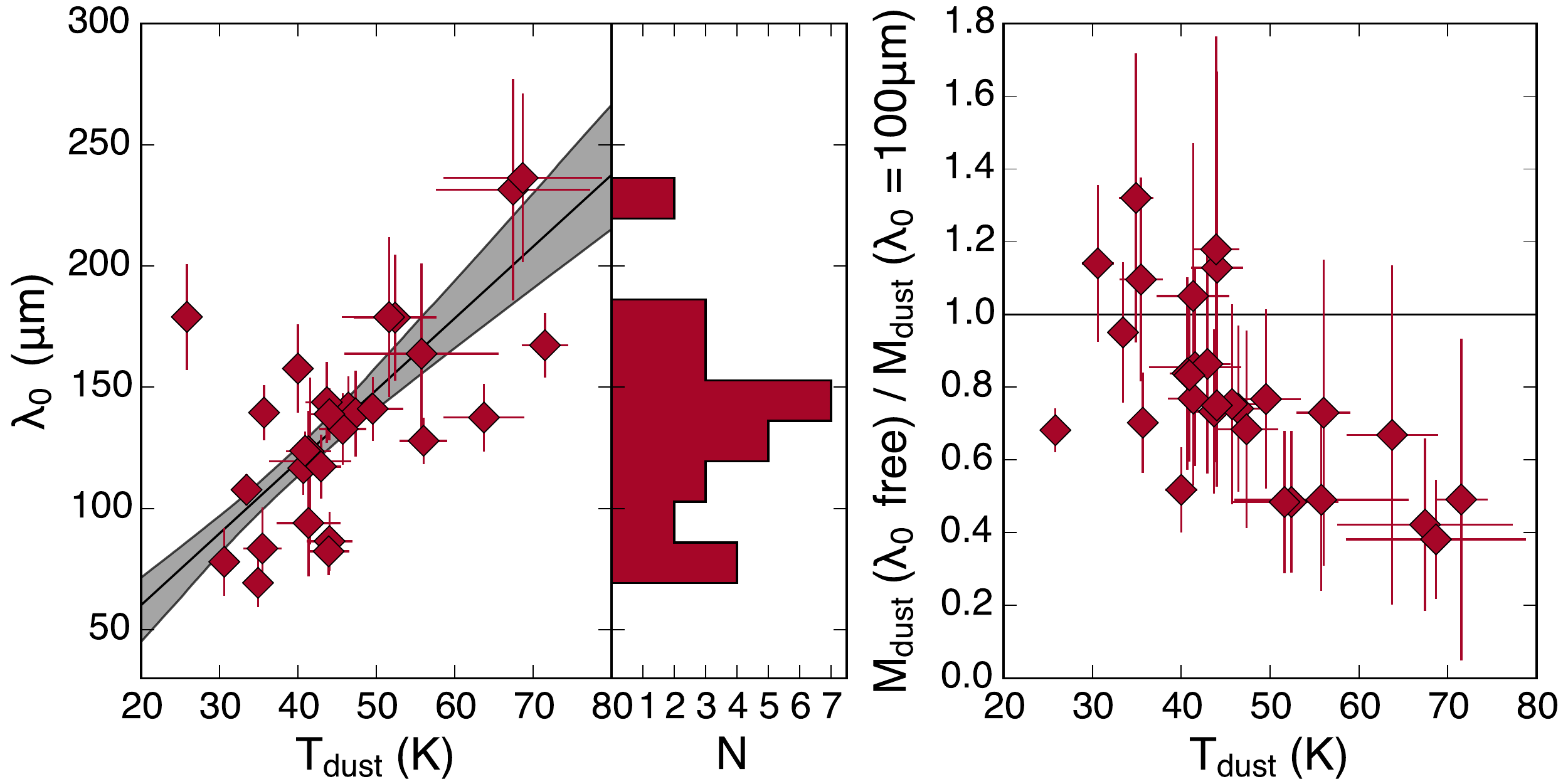}%
\caption{
\textit{Left: } Correlation between the inferred dust temperature and \Lz,
the wavelength where the dust optical depth is unity, derived from a joint fit
to the FIR photometry and the source properties inferred from the lens models for
sources with spectroscopic source redshifts. 
The solid line and grey region indicate the relation in Eq.~\ref{Lzeq}
and its associated 68\% credibility interval, respectively.
The histogram of the inferred values of \Lz is also shown. The median and standard
deviation for the SPT DSFGs is 140 $\pm$ 40\,\um.
\textit{Right: } Ratio of the dust mass inferred by allowing
\Lz to be a free parameter in the joint fit of the SED and derived source properties
over the dust mass inferred by fixing \Lz to 100\,\um.  Fixing $\Lz = 100$\,\um
over-predicts the dust mass by more than a factor of 2 for the sources with the
highest dust temperatures.
}
\label{fig:dustsed}
\end{figure*}

The results of this fitting are shown in the left panel of Fig.~\ref{fig:dustsed}.
We find a median value of \Lz = 140 $\pm$ 40\,\um, somewhat larger than the
canonically assumed value of 100\,\um \citep[e.g.,][]{greve12}.  
Moreover, as previously mentioned, this
wavelength is correlated with the inferred dust temperature.  Fitting a
line to the points shown in Fig.~\ref{fig:dustsed} using orthogonal distance
regression (marginalizing over the probability of points being outliers;
e.g., \citealt{hogg10}) yields
\begin{equation} \label{Lzeq}
\Lz = (3.0 \pm 0.7) \times (\Tdust - 40) + (118 \pm 12) \um.
\end{equation}
Using this relation provides a better alternative to assuming a single
value for \Lz when the available photometry cannot constrain both \Lz and \Tdust --
this relation can be easily inserted into likelihood functions when fitting the dust SED.
This correlation may manifest in part from the relationship between star formation and
molecular gas -- at a simplistic level, the star formation rate of dusty
galaxies is related to \lfir, which in turn is related to \Tdust; meanwhile, the gas
mass is related to the dust mass, which, as we discuss further below, is related
to the dust emissivity encapsulated in \Lz.

The impact of this correlation has little effect on the integrated \lfir. This is
as expected, since our photometric coverage fully samples the SED peak.
The dust mass \Mdust, on the other hand, is strongly influenced.  In the 
optically thin limit, \Mdust is related to the source flux density and \Tdust via
\begin{equation} \label{mdust}
\Mdust = \frac{S_{\nu_\mathrm{obs}}{D_L^2}}{\kappa_{\nu_r} (1+z_S) (B_{\nu_r}(\Tdust) - 
B_{\nu_r}(T_{\mathrm{CMB}}))},
\end{equation}
\citep{greve12} where $\kappa_\nu$ is the dust mass absorption coefficient.  
At present, we are
concerned only with the relative difference in the dust mass determined under
various assumptions, so the form and normalization of $\kappa_\nu$
are irrelevant (as it is related to the source flux density at one frequency
and \Tdust).  The right-hand panel of Fig.~\ref{fig:dustsed} shows the 
ratio of the dust mass determined through our SED fitting when
leaving \Lz as a free parameter compared to the dust mass inferred by assuming
$\Lz = 100$\,\um, effected through the changes in the fitted \Tdust. 
A similar range of inferred dust masses are seen for other assumed values, although
the range of temperatures with reasonable agreement shifts higher for higher \Lz. 
For dust temperatures $\lesssim 45$\,K, the difference
is relatively small. However, as dust temperature increases, the dust mass is
increasingly over-predicted under the assumption that $\Lz = 100$\,\um, reaching
more than a factor of 2 for the hottest sources. A similar result, ignoring the
dust optical depth and instead framed in terms of \Tdust, was obtained by
\citet{magdis12}, who showed that single-temperature fits underestimated \Mdust
compared to more complex models.
This demonstrates that the
assumption of a single, constant value of \Lz can cause a severe distortion in
other derived quantities, especially those which rely on \Tdust.

\subsection{Revisiting the \cii/FIR Ratio} \label{cii}

The 158\,\um \cii
line has long been known as a powerful coolant of the ISM \citep[e.g.,][]{crawford86},
radiating about 0.1--1\% of the total IR luminosity \citep[e.g.,][]{stacey91,stacey10}. 
Unfortunately, \cii can be emitted by gas under a wide variety of conditions, which
makes its physical interpretation challenging.  

One challenge in interpreting \cii manifests as the ``\cii deficit,'' in which
the \cii/\lfir ratio can fall rapidly for $\lfir \gtrsim 10^{11}$\,\lsol
\citep[e.g.,][]{malhotra97,luhman98,graciacarpio11}.  A variety of physical
mechanisms for this deficit have been proposed, including AGN contributions
to \lfir \citep[e.g.,][]{sargsyan12}, increased ionization parameter
\citep[e.g.,][]{malhotra01,graciacarpio11}, collisional de-excitation of
\cii \citep{appleton13}, and differences in emitting column 
\citep{goicoechea15}.  The \cii emission of a sample of 20 SPT DSFGs was
studied in detail by \citet{gullberg15}, who noted that nearly saturated \cii 
emission (via, e.g., excitation or optical depth effects)
could cause much of the \cii/\lfir variation to
be controlled by variations in \lfir alone. This is tentatively supported by 
photodissociation models which attempt to simultaneously explain both the 
\cii and CO(1--0) emission.

\begin{figure}[htb]%
\centering
\includegraphics[width=\columnwidth]{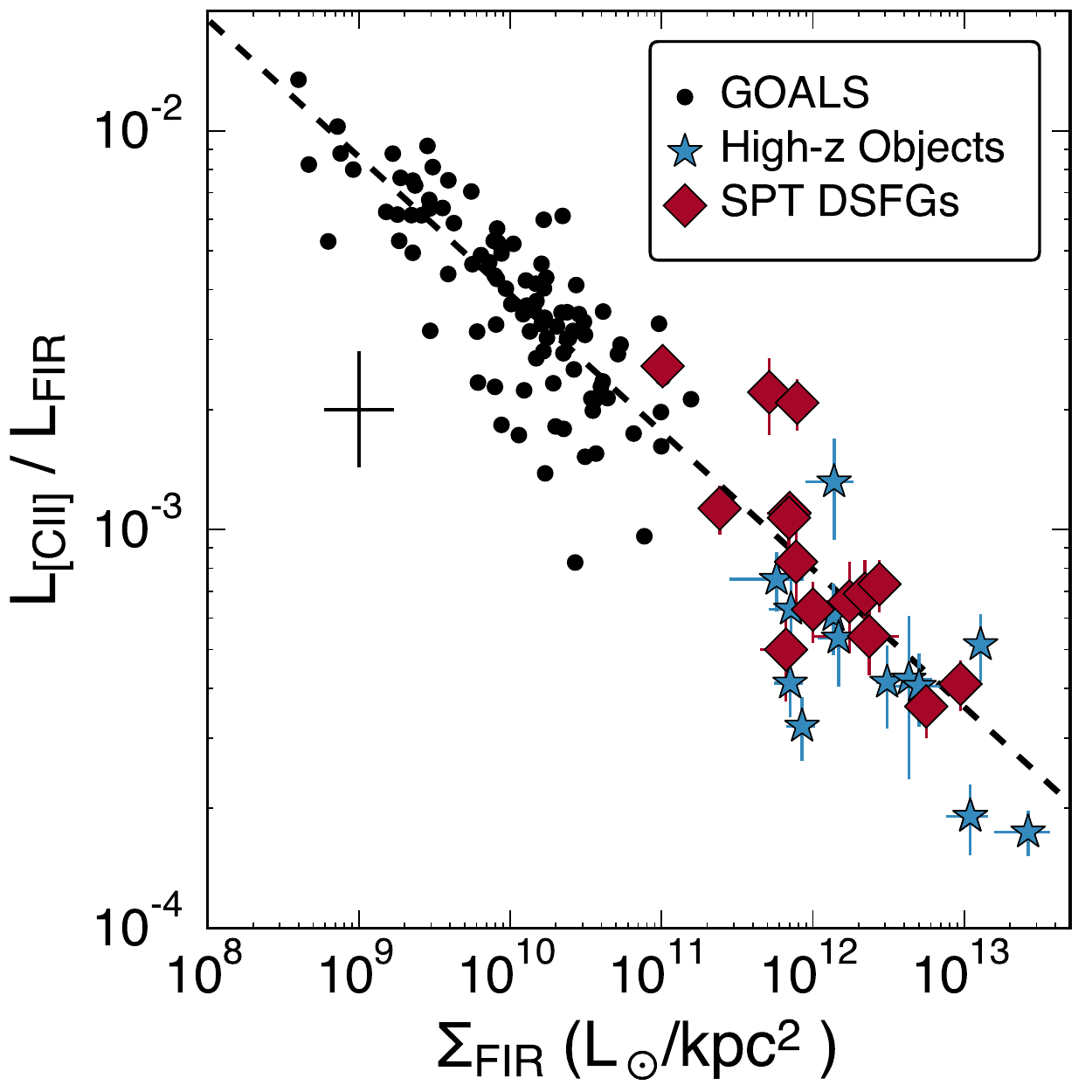}%
\caption{
The \cii/FIR luminosity ratio as a function of \sigfir for low-redshift
star-forming sources with a resolved mid-IR size
from the GOALS survey \citep{diazsantos13}, a collection of high-redshift 
sources from the literature, and the SPT DSFGs \citep{gullberg15}. The remarkably
tight relation (dashed line) noted by \citet{diazsantos13} continues for at least
another two orders of magnitude. A typical uncertainty for the GOALS objects 
is shown as a black cross. The high-redshift objects are drawn from 
\citet{walter09,carniani13,riechers13,wang13,debreuck14,neri14,riechers14,yun15,
diazsantos16}. Note that we have
re-fit the photometry of all objects in a consistent manner, as described in
\citet{gullberg15}.
}
\label{fig:ciifir}
\end{figure}

In their studies of a large sample of local IR-luminous galaxies from the
Great Observatories All-Sky LIRG Survey (GOALS), \citet{diazsantos13} find that 
the \cii/\lfir ratio is also correlated with the FIR luminosity surface
density \sigfir.  This correlation held for both purely star-forming galaxies
as well as objects with significant AGN activity (although many of the 
AGN-dominated sources were spatially unresolved, resulting in lower limits
on \sigfir). A similar result was obtained for galaxies at $z<0.2$ by
\citet{ibar15}, who additionally noted that the spiral galaxies in their
sample had higher \cii/\lfir ratios than irregular and elliptical galaxies.
Using our new measurements of the size of the dust continuum
emitting regions of the SPT DSFGs, and drawing on a compilation of 
high-redshift objects from the literature, we can extend this work two orders of
magnitude higher in \sigfir. The result is shown in Fig.~\ref{fig:ciifir}.
We have re-fit the photometry of all sources to ensure a uniform determination
of \lfir.

The dashed line in Fig.~\ref{fig:ciifir} represents the best-fit relation
determined by \citet{diazsantos13}. We have shifted their relation
vertically to match our re-determination of \lfir, but the slope is exactly
as determined by \citet{diazsantos13}, i.e., 
$\cii/\lfir \propto \sigfir^{-0.35}$.  
The decline continues unabated another two
orders of magnitude beyond the limits of the GOALS survey, to at least
$\sigfir \sim 10^{13}$\,\lsol/kpc$^2$.  This lends further support to the
claim that the compactness of the IR-emitting region drives the relationship
between \cii and \lfir.  A similar correlation can be seen by comparing
the \cii/\lfir ratio with the dust temperature \Tdust, since, to
first order, $\sigfir \propto \lfir/\reff^2 \propto \Tdust^4$. This
correlation was first shown by \citet{malhotra97} and further explored by
\citet{gullberg15}, who determined that most of the variation could
indeed by explained by the Stefan-Boltzmann law, with a small residual
dependence on $T_\mathrm{dust}$.  Formulating the correlation in terms of
\sigfir itself, however, leads to a dispersion approximately a factor
of 2 smaller than formulating it in terms of \Tdust 
\citep{diazsantos13}. While the nature of the \cii emission is still
uncertain, it is clear that the compactness of the IR-emitting region
plays a vital role in determining the coupling of the \cii-emitting gas
with the warm dust.

\section{Conclusions} \label{conclusions}

We have used ALMA 870\,\um observations of 47 gravitationally
lensed dusty, star-forming galaxies to model the effects of gravitational lensing. Using
a visibility-based modeling routine which accounts for several calibration
uncertainties, we can recover the intrinsic properties of the background sources.
At least 33 of the sources are confirmed to undergo galaxy-scale strong lensing
($\mues > 2$), while the remaining sources are lensed by galaxy clusters, or 
are weakly- or un-lensed ($\mues < 2$).  The background sources are magnified
by a median factor of 5.5 for all sources, or 6.3 
for the strongly lensed subset alone, with a tail that extends to $\mues > 30$.

The sources have a median intrinsic angular FWHM of 0.28''. In spite of a
potential size bias of lensed systems, in which compact background sources can be 
magnified more highly than extended sources, we find no significant differences
between the size distributions of existing strongly lensed and unlensed samples
of DSFGs.  Increasing the number of unlensed sources with spectroscopic redshifts
will indicate whether this corresponds to a difference in physical size scale, 
though this effect is small over the plausible range of redshifts. If the 
similarity in size distributions is not a chance effect owing to the limited number
of sources with size measurements, we argue that this may indicate that the 
intrinsic size distribution of DSFGs is sufficiently narrow that the effect of 
the size bias is not detectable.

We use the sizes derived from the lens models together with the extensive 
FIR/submillimeter photometric coverage to constrain \Lz, the wavelength where
the dust opacity is unity.  The size information from the lens models allows
us to overcome parameter degeneracies which limit our ability to constrain 
this wavelength from the SED alone.  We find a median transition wavelength
of $\Lz = 140 \pm 40$\,\um, somewhat longer than the generally assumed 100\,\um.
We provide a fitting formula between \Lz and the dust temperature \Tdust which
can be used for sources without size measurements.  We show that assuming
a single, fixed value for \Lz leads to variations of a factor of 2 in the
inferred dust mass which can be propagated forward to, e.g., the gas
mass under overly simplified assumptions.

Finally, we make use of our extensive follow-up program targeting the
158\,\um FIR fine structure line of \cii.  We show that high-redshift
galaxies (over half of them from the SPT DSFG sample) 
follow the same relationship
between \cii/\lfir and \sigfir as the $z \sim 0$ IR-luminous galaxies in the
\textit{Herschel} GOALS sample, extending this correlation another two
orders of magnitude higher in \sigfir.  This agrees with the claim that the
controlling parameter in the ``\cii deficit'' is the compactness of the
IR-emitting region, regardless of the dust heating source.  Future 
spatially resolved observations of the \cii line at high redshifts will 
indicate whether this global correlation is also present on sub-galactic scales.

The high-resolution images and lens models in this work, along with the high
spectroscopic completeness of our sample, provide a wealth of information useful
for future follow-up programs.  The sensitivity and resolution afforded by ALMA
in full operation indicate that the future of DSFG studies is bright, and the
models we have presented should help prioritize the best sources to help
answer questions of interest.

\nocite{bethermin16} 

\acknowledgements
{
J.S.S., D.P.M., K.C.L. and J.D.V. acknowledge support from the U.S. National Science Foundation under grant No. AST-1312950 and through award SOSPA2-012 from the NRAO.
M.A. acknowledges partial support from FONDECYT through Grant 1140099.
This material has made use of the El Gato high performance computer, supported by the U.S. National Science Foundation under grant No. 1228509.
This paper makes use of the following ALMA data: ADS/JAO.ALMA \#2011.0.00957.S, \#2011.0.00958.S, \#2012.1.00844.S, and \#2012.1.00994.S.  ALMA is a partnership of ESO (representing its member states), NSF (USA) and NINS (Japan), together with NRC (Canada) and NSC and ASIAA (Taiwan), in cooperation with the Republic of Chile. The Joint ALMA Observatory is operated by ESO, AUI/NRAO and NAOJ.  The National Radio Astronomy Observatory is a facility of the National Science Foundation operated under cooperative agreement by Associated Universities, Inc. 
The SPT is supported by the National Science Foundation through grant PLR-1248097, with partial support through PHY-1125897, the Kavli Foundation and the Gordon and Betty Moore Foundation grant GBMF 947.
This work makes use of an extensive optical/NIR follow-up campaign, including \textit{Hubble Space Telescope} programs \#12659 and \#13614, Very Large Telescope programs 085.A-0608, 086.A-0797, 088.A-0902, 090.A-0503, 092.A-0480, 284.A-5029, and 285.A-5034, Gemini Observatory programs GS-2013B-Q-5 and GS-2013A-Q-33, and data gathered with the Magellan Telescopes and the \textit{Spitzer Space Telescope}.
This research has made use of NASA's Astrophysics Data System.
}

\bibliographystyle{apj}

\clearpage

\appendix

\section{Notes on the Lens Models of Individual Sources}\label{app:notes}
\textit{SPT0020-51} -- This source is lensed by a small group of galaxies,
and the source emission can be adequately described by two S\'{e}rsic profiles.
Oddly, the brightest, northernmost galaxy seems to have virtually no effect on
the lensing geometry. We have verified the astrometry of the images shown in
Fig.~\ref{fig:images}. It may be that the northern lensing galaxy is at a different
redshift from the other two, such that its influence on the overall deflection
is small. The complex lensing environment is likely responsible for the
$\sim6\sigma$ peak residual structure seen, given the limitations of our relatively
simple model.

\textit{SPT0027-50} -- This source presents a complex lensing environment. We find
that modeling three of the lensing galaxies and an external shear produces a model 
which describes the data reasonably well.

\textit{SPT0103-45} -- The lens galaxy in this system appears to have a tidal
feature or spiral arm outside of the lensed emission seen by ALMA. Two S\'{e}rsic
components are required to model the emission.  Interestingly, the CO(3--2) line
observed by \citealt{weiss13} also shows an asymmetric double-peaked profile.

\textit{SPT0109-47} -- This source is reasonably well-fit by a model accounting
for three lensing galaxies. Even still, the flux ratios of the four images are
not entirely reproduced by the model. This likely indicates that the complexities
of the lensing environment are not fully accounted for by the model.

\textit{SPT0113-46} -- This source is also lensed by a group of galaxies.
We model the three closest galaxies, which is sufficient to reproduce the 
observed emission.  The best-fit model with the positions of all three galaxies
as free parameters shows some tension with the locations of the galaxies in our
\textit{HST} imaging. This may be due to the gravitational potential in which the
galaxies reside or the effects of other group members we have not modeled.
Fixing the relative positions of the galaxies to their separations in the
\textit{HST} imaging results in significantly higher residuals but changes
the inferred source size and magnification by $\lesssim 10$\%.  Since the effect
on these quantities of greatest interest is small, we show the best-fit model
with the positions of all three galaxies as free parameters.

\textit{SPT0125-47} -- This source is detected with a peak significance of nearly
150, and required three source-plane components to model the data. Even this model
leaves $\sim6\sigma$ residuals, which may indicate that an SIE is an imperfect
representation of the lens mass profile. The structure of the residuals indicates
that additional angular structure in the lens in the form of multipoles may
be necessary to accurately capture the complexity of the lens mass profile.

\textit{SPT0125-50} -- We model this source using two S\'{e}rsic components. The
faint counterimage to the southwest of the lens is only moderately well-reproduced
by this model, which may indicate further structure in the source and/or lens
planes.

\textit{SPT0128-51} -- This source appears to consist of two components, at 
best mildly lensed by a galaxy $\sim7.5$'' to the north. We estimate the magnification
experienced by these two components by assuming the lens galaxy has an Einstein
radius of 0.75'', approximately the median of all the lens galaxies in our
sample. Under this assumption, the eastern and western components have 
magnifications of $\mues = $1.16 and 1.20, respectively.

\textit{SPT0202-61} -- This source is well-represented by two source-plane 
components.  We also detect a second source $\sim7$'' south of the main
lens. This second component has $\ses = 4.1 \pm 0.4$\,mJy. As we do
not know the redshift(s) of the background sources, it is unclear whether
they are physically related or simply a chance projection.

\textit{SPT0243-49} -- This is the highest-redshift source we consider here,
at $z=5.70$. The lens appears highly elongated, with a possible tidal tail
extending east.  

\textit{SPT0245-63} -- This source appears to be a rare case in which the 
emission we detect is associated with a large foreground galaxy.  We 
consider a lensing origin of the 870\,\um emission unlikely, as the implied
lens mass is implausibly small ($<10^{10}$\,\msol for $z_L = 0.3$) given
the brightness of the putative lens (K-band magnitude 16.3). Additional 
optical/NIR and millimeter spectroscopy is needed to determine the nature
of this source conclusively.

\textit{SPT0300-46} -- This source is well-fit by two source-plane
components, and the CO(4--3) line clearly shows two velocity components 
\citep{gullberg15}.  
While the source is not clearly resolved into multiple images,
imaging the visibilities with baselines $>$100\,k$\lambda$ clearly shows
the arc-like structure reproduced in the best-fit model.  The field also
includes a 9\,mJy source $\sim$8\arc west of the lensed source.

\textit{SPT0319-47} -- This source is also moderately resolved at 0.5''
resolution, and is adequately fit by a single S\'{e}rsic component.
Imaging the visibilities with baselines $>$100\,k$\lambda$ confirms
that the source splits into two lensed images as predicted by the
best-fit model.

\textit{SPT0345-47} -- This source is unambiguously lensed, as we have
measured distinct lens and source redshifts. Imaging only baselines
$>$100\,k$\lambda$ confirms that the arc and counterimage structure seen
in the best-fit model is correct.

\textit{SPT0346-52} -- This source was one of four resolved at $\sim$1.5''
resolution and studied by \citet{hezaveh13}. Combining those data with
the higher-resolution observations presented here largely confirms
the previous model. The data also favor the existence of an external
shear component at an angle roughly in line with the galaxy $\sim$3''
east of the main lens.

\textit{SPT0348-62} -- This object appears to be a collection of several
weakly lensed sources. We assume an Einstein radius of 1'' for the galaxy
approximately 6'' east of the bulk of the emission. Much of the 870\,\um emission
is not well-represented by simple Gaussian components, although four
elliptical Gaussian components provide an adequate fit. The four components
are magnified by factors of 1.16--1.26, assuming the Einstein radius above.

\textit{SPT0403-58} -- This source appears to be marginally lensed by
the nearby galaxy, with a flux-weighted total magnification of $\mues = 1.7$.

\textit{SPT0404-59} -- This object is the most weakly detected source
in this sample, and the images we do detect straddle the ALMA primary
beam half-power radius.  A single source-plane component represents
the data well, although deeper observations will be necessary to place
better constraints on the lens and source geometry.

\textit{SPT0418-47} -- This source was also considered by \citet{hezaveh13}.
Our model, which incorporates higher-resolution data available after
publication of Hezaveh \etal, finds a higher magnification,
$\mues = 32$, as the higher-resolution data show a nearly-perfect
Einstein ring. This source is the most highly magnified galaxy-scale
lens in the current sample.

\textit{SPT0441-46} -- This source is well-modeled by a single 
S\'{e}rsic background component.

\textit{SPT0452-50} -- With distinct lens and source redshifts, this
source appears to be only mildly lensed by the faint galaxy $\sim$3''
to the south of the ALMA emission. It is the highest-redshift lens
with a confirmed redshift in the current sample.

\textit{SPT0459-58} -- This source is marginally resolved into multiple
images by ALMA.  Imaging those visibilities with baselines $>$75\,k$\lambda$
confirms that the multiple images seen in the best-fit model are real.

\textit{SPT0459-59} -- In addition to a strongly lensed component,
this source also contains a fainter source weakly lensed, by a factor
of $\mues = 1.5$.

\textit{SPT0529-54} -- This source was also studied by \citet{hezaveh13},
and is the only source for which the model which incorporates the
higher-resolution data now available to us significantly differs from
the previous model. The model presented here allows ellipticity in the 
background source, which entirely accounts for the model differences.
That is, fitting an elliptical component to the data available at the
time of publication of \citet{hezaveh13} recovers the model shown here,
while fitting a circularly symmetric Gaussian source to the data from
both array configurations recovers the model shown in \citet{hezaveh13}.
The higher-resolution data now available clearly resolve the source
into four images, which was not apparent in the lower-resolution data. The full
data strongly prefer the elliptical source-plane model over a
circularly symmetric model.

\textit{SPT0532-50} -- This source shows a clear hole in the center of 
a ring of emission, and is magnified by a factor of 10.

\textit{SPT0538-50} -- This source has been the subject of detailed
studies by \citet{bothwell13b} and \citet{spilker15}, and was included 
in the sample of \citet{hezaveh13}. Our model confirms the need for
two source-plane components, which Spilker \etal associate with
the two velocity components seen in observations of CO(1--0) and CO(3--2).

\textit{SPT2031-51} -- A single S\'{e}rsic component is sufficient to
accurately reproduce the ALMA data. The source is magnified by a 
factor of $\mues = 3.9$.

\textit{SPT2048-55} -- The lens in this system appears to have a tidal
feature extending to the west of the galaxy; nevertheless, a single
SIE lens and background S\'{e}rsic component fit the data quite well.

\textit{SPT2052-56} -- This source is not pointlike at 0.5'' resolution,
but appears completely unlensed. A symmetric Gaussian with $\ses = $10.9\,mJy
reproduces the data well. 

\textit{SPT2103-60} -- This source presents an unusual lensing geometry,
caused by the combined effects of the three foreground galaxies. We place
loose priors on the locations of the three lensing galaxies but do not
require that they have the same relative locations as seen in the VLT/ISAAC
image. Remarkably, a single source-plane component adequately reproduces the data.

\textit{SPT2132-58} -- The background source in this system is moderately
resolved by ALMA. Imaging only the visibilities with baselines
$>$75\,k$\lambda$ confirms the arc-like structure with faint counterimage
predicted by the best-fit model. This source was studied in detail by
\citet{bethermin16}.

\textit{SPT2134-50} -- A single source-plane component with half-light
radius of just $\sim$270\,pc is magnified by a factor of 21 in this source.

\textit{SPT2146-55} -- This source, at $z=4.5$, is well-described by a
single SIE lens and background S\'{e}rsic component.

\textit{SPT2146-56} -- This source appears pointlike in the ALMA data,
and lies on top of a galaxy identified at $z=0.67$.  The unresolved
SED of this source (Strandet \etal, in prep.) is also atypical, showing
what appears to be two dust peaks. Whatever the true nature of this
source, it does not appear to be a galaxy-scale lensed DSFG.

\textit{SPT2147-50} -- A simple SIE lens and single S\'{e}rsic 
background component are sufficient to reproduce the ALMA data.

\begin{figure}[hbt]%
\centering
\includegraphics[width=0.3\textwidth]{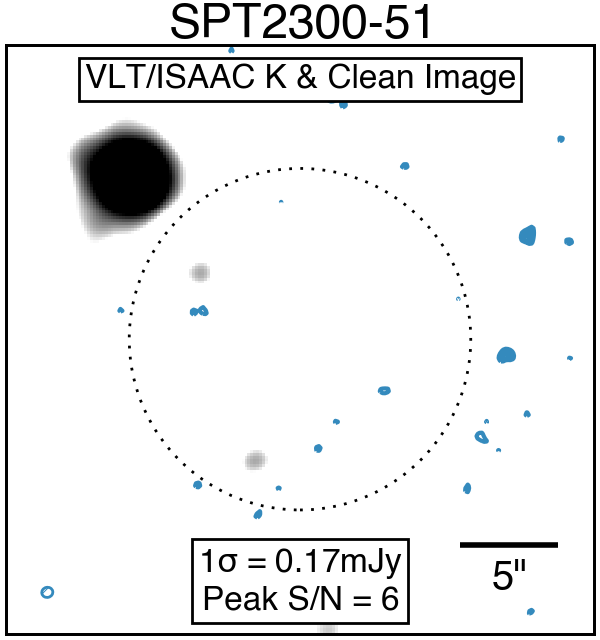}
\caption{
ALMA 870\,\um image of SPT2300-51 (blue contours) overlaid on a VLT/ISAAC
K-band image (greyscale), centered on the ALMA phase center. 
Contours are drawn at 50, 70, and 90\% of the image
peak value, and the ALMA primary beam half-power radius is indicated with a dotted
line. The synthesized beam is indicated in the lower left corner.
The two highest peaks in the image are located at approximately
38\% and 25\% of the peak primary beam response, and both are likely false.
}
\label{fig:nondetect}
\end{figure}

\textit{SPT2300-51} -- No sources were detected at $>5\sigma$ significance
within the ALMA primary beam half-power radius, and this object was determined to
be a spurious detection in the LABOCA follow-up of SPT sources. This source
is shown in Fig.~\ref{fig:nondetect}. Two potential
sources exist at approximately the 38\% and 25\% primary beam response levels,
with nominal significance of 4 and $6\sigma$, respectively; we consider both
likely to be false.

\textit{SPT2311-54} -- This source is only marginally resolved
by ALMA, but we have measured distinct lens and source redshifts which
confirm the lensed nature of this source. Imaging only those visibilities
with baselines $>$150\,k$\lambda$ reveals structure consistent with the
best-fit model.

\textit{SPT2319-55} -- This source at $z_S = 5.3$, is well-fit by two
source-plane components.

\textit{SPT2340-59} -- The nature of this source is unclear. While the 870\,\um
data appear to show a standard doubly imaged background source, the two
images were also spatially resolved in an ALMA 3\,mm project to determine
its redshift. These data show a weak line, identified as CO(4--3), in the eastern
component only \citep{strandet16}. 
This suggests the two images seen at 870\,\um may in fact
be two unlensed or weakly lensed sources at different redshifts. In 
Fig.~\ref{fig:images}, we show the model assuming the two sources seen by 
ALMA are lensed images of the same source, in which case the source is
magnified by a factor of $\mues = 4.0$ and has $\ses =$9.1\,mJy. If, on the
other hand, the two images are distinct sources, we derive their intrinsic
properties by fixing the position of the foreground lens galaxy to the location
in the VLT/FORS2 image. This yields $\mues=1.3$, $\ses =$15.2\,mJy and $\mues=1.7$,
$\ses =$8.7\,mJy for the western and eastern components, respectively.

\textit{SPT2349-50} -- This source is doubly imaged by the foreground lens.
Imaging the visibilities on baselines $>$125\,k$\lambda$ clearly resolves the
two lensed images separated by $\sim$0.5''.

\textit{SPT2349-56} -- We detect six sources at $\geq5\sigma$ significance
in this source, none of which appear to be significantly lensed. ALMA 3\,mm
data \citep{strandet16} resolve the bright northern component from the
three more southern components. A line identified as CO(4--3) in these data,
along with \cii confirmation using APEX, places this source at $z_S = 4.30$.
It is unclear whether all six components detected at 870\,\um are at the
same redshift or whether they represent a chance alignment of sources
at different redshifts. Only three of the six components are spatially
resolved at the $\sim$0.5'' resolution of these data.

\textit{SPT2354-58} -- This source likely has the lowest confirmed redshift of any
object in the SPT sample, at $z_S = 1.87$ \citep{strandet16}. 
Imaging the visibilities with
baselines $>$125\,k$\lambda$ confirms the arc and counterimage predicted
by the best-fit model, separated by $\sim$0.5''.

\textit{SPT2357-51} -- This source, at $z_S = 3.07$, does not obviously
separate into multiple images in the ALMA data. If the source were unlensed,
however, its implied flux density would be $\ses \sim 40$\,mJy, the 
brightest of any DSFG. Given this and the proximity of foreground galaxies
along the line of sight, we consider the lensing hypothesis more probable.
The current data are well-fit by a single Gaussian component magnified by
a factor of $\sim$2.9.

\section{Cluster-Lensed Sources}\label{app:clusters}

In this appendix we show the ALMA images of the four sources which appear to
be lensed by large groups or clusters of galaxies.  The ALMA data show only
single images of the background sources, making lens models impossible to constrain
from the limited data available.

\begin{figure}[htb]%
\centering
\includegraphics[width=0.245\textwidth]{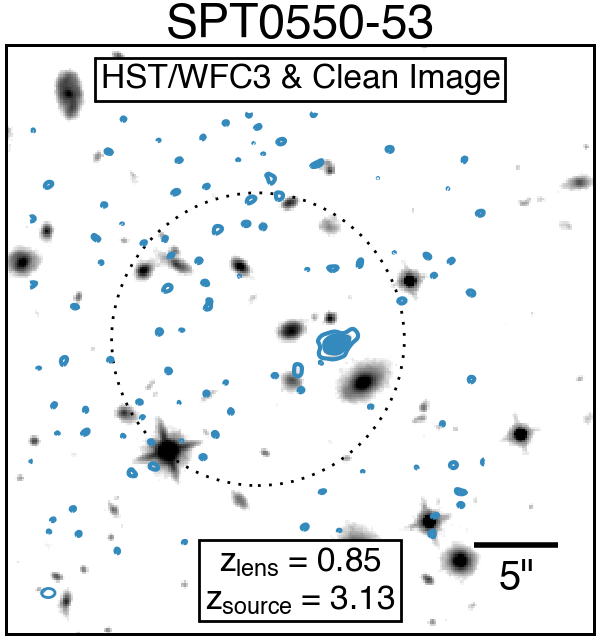}
\includegraphics[width=0.245\textwidth]{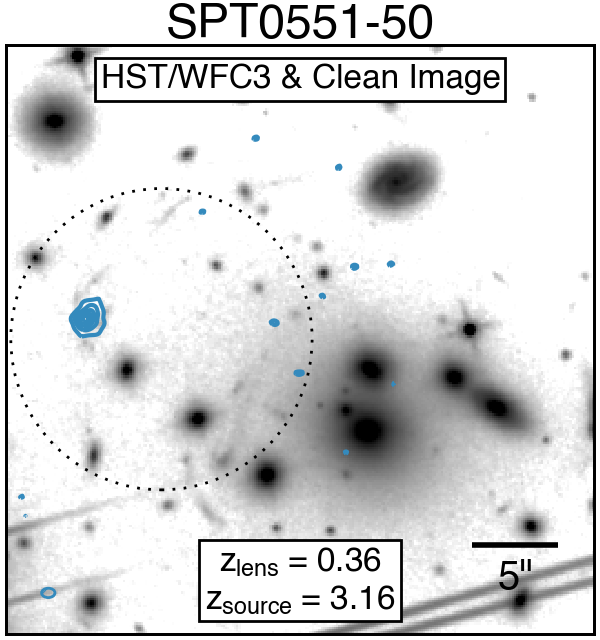}
\includegraphics[width=0.245\textwidth]{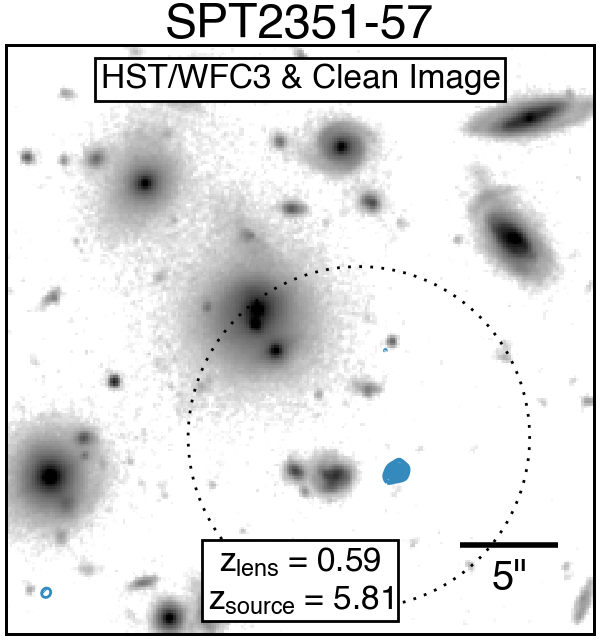}
\includegraphics[width=0.245\textwidth]{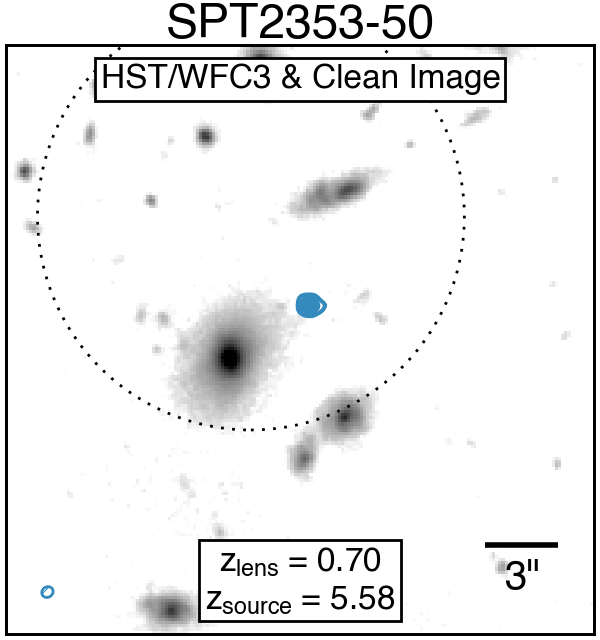}
\caption{
ALMA 870\,\um emission (blue contours) overlaid on the HST/WFC3 images
(greyscale) for the four cluster-lensed galaxies in the current sample. 
Contours are drawn at 10, 30, ... percent of the peak 
value, and the ALMA primary beam half-power radius is indicated with a dotted
line. The synthesized beam is indicated in the lower left corner.
Greyscale images are logarithmically scaled to emphasize the 
relevant objects detected.
}
\label{fig:clusters}
\end{figure}

\end{document}